\documentclass[a4paper,11pt]{article}

\usepackage[margin=2.5cm]{geometry}

\usepackage[utf8]{inputenc}
\usepackage[T1]{fontenc}
\usepackage{microtype}
\usepackage{lmodern}
\usepackage{exscale}
\usepackage{upgreek}

\usepackage{float} 

\usepackage{amsmath}
\usepackage{amssymb}
\usepackage{mathtools}

\numberwithin{equation}{section}

\allowdisplaybreaks

\usepackage[linktocpage]{hyperref}
\usepackage[dvipsnames]{xcolor}
\hypersetup{
    colorlinks,
    linkcolor={purple},
    citecolor={purple},
    urlcolor={purple}
}

\usepackage{orcidlink}
\usepackage{cite}

\usepackage{mathrsfs}

\usepackage{relsize,exscale}
\usepackage{tensor}

\usepackage[low-sup]{subdepth}

\newcommand{\scr}{\scriptscriptstyle}

\newcommand{\longsim}{\scalebox{1.8}[1]{$\sim$}}

\usepackage[labelfont=bf]{caption}

\makeatletter
\newcommand{\dalembertian}{\mathop{\mathpalette\dalembertian@\relax}}
\newcommand{\dalembertian@}[2]{%
  \begingroup
  \sbox\z@{$\m@th#1\square$}%
  \dimen0=\fontdimen8
    \ifx#1\displaystyle\textfont\else
    \ifx#1\textstyle\textfont\else
    \ifx#1\scriptstyle\scriptfont\else
    \scriptscriptfont\fi\fi\fi3
  \makebox[\wd\z@]{%
    \hbox to \ht\z@{%
      \vrule width \dimen0
      \kern-\dimen0
      \vbox to \ht\z@{
        \hrule height \dimen0 width \ht\z@
        \vss
        \hrule height 2\dimen0
      }%
      \kern-2.5\dimen0
      \vrule width 2.5\dimen0
    }%
  }%
  \endgroup
}
\makeatother

\usepackage{graphicx}
\graphicspath{ {./AccretionImages/} }

\usepackage{cancel}

\begin{document}

\begin{center}

{\bf \Large On the Bondi accretion of a self-interacting complex scalar field}

\bigskip

\renewcommand{\thefootnote}{\fnsymbol{footnote}}

Dra\v{z}en Glavan\,${\orcidlink{0000-0002-1983-0448}}^{\,a,}$\footnote[1]{email: 
	\href{mailto:glavan@fzu.cz}{\tt glavan@fzu.cz}},
Alexander Vikman\,${\orcidlink{0000-0003-3957-2068}}^{\,a,}$\footnote[2]{email:
	\href{mailto:vikman@fzu.cz}{\tt vikman@fzu.cz}}
and Tom Zlosnik\,${\orcidlink{0000-0001-7715-5842}}^{\,b,}$\footnote[3]{email: 
	\href{mailto:thomas.zlosnik@ug.edu.pl}{\tt thomas.zlosnik@ug.edu.pl}}

\setcounter{footnote}{0} 

\bigskip

{\it
${}^a$\,CEICO, FZU --- Institute of Physics of the Czech Academy of Sciences,
\\
Na Slovance 1992/2, 182 00 Prague 8, Czech Republic

\smallskip

${}^b$\,Institute of Theoretical Physics and Astrophysics,  University of Gda\'{n}sk, \\
ul.~Wita Stwosza 57, 80-308 Gda\'{n}sk, Poland
}

\bigskip

\smallskip

\parbox{0.9\linewidth}{
Scalar fields with a global $U(1)$ symmetry often appear in cosmology 
and astrophysics. We study the spherically-symmetric, stationary accretion 
of such a classical field onto a Schwarzschild black hole in the test-field 
approximation. Thus, we consider the relativistic Bondi accretion beyond a 
simplified perfect-fluid setup. We focus on the complex scalar field with 
canonical kinetic term and with a generic quartic potential that either 
preserves the~$U(1)$ symmetry or exhibits  spontaneous symmetry breaking. It is 
well known that in the lowest order in gradient expansion the dynamics of such 
a scalar field is well approximated by a perfect superfluid. We demonstrate 
that going beyond this approximation systematically reduces the accretion rate 
with respect to the perfect fluid case. Hence, black holes can provide a way to 
distinguish a perfect fluid from its ultraviolet completion in form of the 
complex scalar field.
}

\end{center}

\bigskip

\hrule

\tableofcontents

\bigskip

\hrule

\bigskip
\bigskip

\section{Introduction}
\label{sec: Introduction}

Observations from precision cosmology~\cite{Planck:2018vyg} constrain dark matter 
(DM) to behave approximately as a dust-like perfect fluid on large
scales~\cite{Kopp:2018zxp}. A wide range of DM models are consistent with this 
behavior; see~\cite{Mukhanov:2005sc} for a textbook discussion 
and~\cite{Cirelli:2024ssz} for a recent review. This dust-like behavior could 
arise from gravity itself, in the form of primordial black holes (PBHs) --- 
see~\cite{Carr:2020xqk} for a recent review --- or more traditionally from 
relatively heavy, collisionless particle candidates;
see~\cite{Bertone:2004pz,Roszkowski:2017nbc} for reviews.
Most relevant to the present work, however, this dynamics can also arise from 
light, oscillating, coherent semiclassical fields that behave as fluid-like DM, 
in which the underlying degrees of freedom form a 
condensate~\cite{Peebles:2000yy,Hu:2000ke,Hui:2016ltb}; see 
also~\cite{Ferreira:2020fam} for a recent review. DM could also be a mixture of 
all of the above, and could even involve modifications of gravity --- particularly 
interesting on galactic scales, see e.g. review~\cite{Clifton:2011jh}.
Given that all viable DM models must satisfy cosmological constraints, additional 
astrophysical phenomena must be considered in order to discriminate between them. 
Here, we examine the potential for spherically-symmetric black hole (BH) accretion 
as a process that may serve to distinguish between different fluid-like DM models.

In this paper, we focus on two classes of fluid-like DM models that have 
attracted particular interest. The first class comprises the so-called 
\emph{$P(X)$ models}, which form a \emph{purely kinetic} subclass of more general 
\emph{k-essence} models~\cite{Armendariz-Picon:2000nqq,Armendariz-Picon:2000ulo,Chiba:1999ka} --- single-scalar field theories with derivative 
self-interactions. 
In particular, these purely kinetic models~\cite{Arkani-Hamed:2003pdi,Chimento:2003ta,Scherrer:2004au,Armendariz-Picon:2005oog} 
are non-linear in the kinetic term $X$ and possess a global shift symmetry in 
field space. The second class describes DM in terms of a \emph{canonical complex 
scalar field} with a global $U(1)$ symmetry of the action. The latter provides 
the simplest relativistic model of a Bose--Einstein condensate and, at zero temperature, of a 
superfluid~\cite{Greiter:1989qb,Son:2000ht,Son:2002zn,Alford:2012vn,Glodkowski:2025tnv}. 
Such models are of interest as candidates for cold, fluid-like\footnote{It is worth mentioning that $U(1)$ gauge singlet scalar field extensions of the Standard Model can also play the role of a cosmological particle-like DM, see e.g. \cite{McDonald:1993ex}. This particle DM may still condense around BHs.} DM; see
e.g.~\cite{Arbey:2001qi,Arbey:2001jj,Li:2013nal,Li:2016mmc,Fan:2016rda,Suarez:2017mav,Chavanis:2021jwr,Chavanis:2022vzi}. 
In particular, it has been proposed that DM may undergo a phase transition 
within galaxies such that, while on cosmological scales it behaves as a dilute 
gas of particles, on galactic scales it condenses into a coherent 
fluid~\cite{Sin:1992bg,Berezhiani:2015bqa,Berezhiani:2025maf}. Furthermore, both 
$P(X)$ models~\cite{Arkani-Hamed:2003pdi,Chimento:2003ta,Scherrer:2004au} and 
complex scalar fields have also been considered as candidates for dark energy 
(DE)~\cite{Boyle:2001du,Gu:2001tr,Anguelova:2021jxu}. In addition, motivated by 
the suggestion that QCD matter inside neutron stars may be in a superfluid 
state~\cite{Ginzburg:1965cpw,Baym:1969fiw}, complex scalar fields also arise in 
this context as an effective description.

Crucially for our work, a canonical complex scalar field provides a natural 
ultraviolet (UV) completion of $P(X)$ 
theories~\cite{Colpi:1986ye,Bilic:2001cg,Bilic:2008zk,Tolley:2009fg,Babichev:2018twg,Mukohyama:2020lsu}, 
making the latter an effective description of the former in the low-energy regime.\footnote{Note that $P(X)$ models, and more generally models with 
non-canonical kinetic terms such as k-essence, can also arise as effective 
descriptions capturing quantum corrections, 
e.g.~\cite{Ai:2021gtg,Joyce:2022ydd,Hung:2023bxc}.}
While $P(X)$ models, similar to other fluids, are known to evolve towards singular 
classical dynamics such as the formation of caustics~\cite{Babichev:2016hys}, 
their corresponding complex scalar UV completions are free of these 
issues~\cite{Babichev:2017lrx}. We therefore examine spherically-symmetric 
accretion not only as a process that can discriminate between different DM 
models, but also as a potential means of distinguishing between effective field 
theories (EFTs) and their UV completions.

Since DM interacts predominantly through gravity, it is natural to investigate 
its behavior in the vicinity of the strongest gravitational sources, namely 
BHs. This is particularly relevant in light of the rapid progress and promising 
prospects in gravitational-wave astronomy. Due to the no-hair theorems, see 
e.g.~\cite{Bekenstein:1971hc,Bekenstein:1995un,Hui:2012qt}, the most natural 
configuration for matter around a BH is accretion. The steady-state, 
spherically-symmetric accretion of a perfect fluid---known as Bondi 
accretion---is a classical and well-studied problem in astrophysics~\cite{Bondi:1952ni,Michel:1972oeq,Moncrief} (for a pedagogical 
discussion, see~\cite{Beskin:2002si}).

Relativistic accretion of a perfect-fluid-like DE component described by 
a real, shift-symmetric scalar field has been investigated in~\cite{Babichev:2004yx,Babichev:2005py,Babichev:2006vx,Babichev:2007wg,Babichev:2008dy,Babichev:2008jb,Babichev:2012sg} and reviewed in~\cite{Babichev:2013vji}. It is worth noting that shift symmetry 
in field space is required for an exact stationary flow~\cite{Akhoury:2008nn}. Furthermore, 
accretion beyond the perfect-fluid approximation was explored in~\cite{Babichev:2010kj}, 
while departures from the steady-state regime were analysed in~\cite{Akhoury:2011rc}, 
where it was shown that the steady-state configuration acts as a late-time attractor. 
In the context of fluid-like dark matter modelled by a 
\emph{ghost condensate}~\cite{Arkani-Hamed:2003pdi}, accretion was studied in two distinct 
regimes in~\cite{Frolov:2004vm} and~\cite{Mukohyama:2005rw}, and the stability analysis 
was performed in~\cite{RivasplataPaz:2014gng}.

Accretion of a real scalar field without shift symmetry was investigated 
in~\cite{Jacobson:1999vr,Bean:2002kx,Frolov:2002va} and, more recently, 
in~\cite{Hui:2019aqm,Brax:2019npi,Brax:2020tuk,Boudon:2022dxi,Ravanal:2023ytp,Gomez:2024ack}. 
In these works, the authors found non-stationary solutions that describe a 
steady-state accretion flow only when averaged over many field oscillations. 
In addition,~\cite{deCesare:2022aoe} examined accretion of a real self-interacting 
scalar field beyond the test-field approximation.

Finally, accretion of a complex scalar field has recently been considered in~\cite{Hui:2019aqm,Bamber:2020bpu,Hui:2022sri,Feng:2021qkj}. In~\cite{Hui:2019aqm,Bamber:2020bpu,Hui:2022sri}, 
the field was taken to be non-self-interacting, while~\cite{Feng:2021qkj} studied the 
system in the limit of $P(X)$ theories, i.e., to leading order in the gradient expansion. 
Moreover, the non-stationary behavior of a self-interacting complex scalar field was 
numerically investigated in~\cite{Aguilar-Nieto:2022jio}.

The main purpose of our work is to determine the Bondi accretion of a self-interacting complex scalar field~$\Psi$ and compare it with that of the 
corresponding~$P(X)$ model, with the goal of quantifying the extent to which 
the differences between the two models manifest themselves. 
We work in the 
regime where backreaction is negligible~\cite{Babichev:2012sg,Dokuchaev:2011gt}, 
treating the scalar as a test field. While~$P(X)$ models generally admit an ideal 
fluid description for timelike gradients of the scalar 
field,\footnote{Note that \cite{Armendariz-Picon:2005oog} considers spacelike gradients and therefore does not correspond to a perfect fluid.}
their complex scalar UV completions in general do not. Because of this, it is 
necessary to consider the field equations directly rather than the ideal-fluid 
treatment customarily employed in accretion studies. We compute the profiles of the complex scalar 
modulus by solving the full equations of motion while retaining the gradient 
terms, and use them to determine the accretion rate and its dependence on the 
model parameters. In particular, unlike~\cite{Feng:2021qkj}, we do not assume 
that the gradients of the phase are timelike everywhere. This condition is imposed
only at spatial infinity, where it is motivated by cosmology.
\\
\\
Our work is relevant for the phenomena including:  
i) accretion of self-interacting  complex scalar field playing the role of DM onto BH; ii) accretion of neutron star QCD condensates onto a PBH inside; iii) accretion of boson star material (usually taken to be a complex scalar field \cite{Kaup:1968zz,Ruffini:1969qy,Colpi:1986ye}) onto a PBH inside; iv) accretion of the quantum liquid\footnote{AV is thankful to Iggy Sawicki for recalling this 
relevant physical setup.} in a class of dwarf stars \cite{Gabadadze:2008mx,Gabadadze:2009jb} onto PBH inside.
\\
\\
For concreteness, we restrict our attention to renormalizable potentials, 
comprised of the quartic self-interaction and the quadratic mass term. We 
allow for spontaneous symmetry breaking --- another crucial novelty of our work ---
and consider both signs of the mass term, without restricting its magnitude. 
We also do not restrict the magnitude of the quartic coupling, but require it to be positive 
to avoid unbounded potentials. Nevertheless, we aim to keep the formulas 
applicable to more general potentials whenever possible. 
This generality is motivated by the possibility that DM (or DE) may be described 
by an effective, non-fundamental complex scalar field within galaxies, while 
behaving as a particles on cosmological scales. Furthermore, unknown heavy degrees 
of freedom could form an effective scalar condensate~$\Psi$ in earlier epochs, 
when BHs (or PBHs) had already formed. On the other hand, we do not want to delve 
into specific details of neutron star condensates and QCD and consider general 
parameters.

The paper is organized as follows. In Section~\ref{sec: Complex scalar generalities}, 
we present the complex scalar model, its general equations of motion, the associated 
Noether current, and the energy–momentum tensor (EMT), together with its decomposition in the local Eckart rest frame~\cite{Eckart:1940te} used in the relativistic hydrodynamics of imperfect fluids. The relation 
between complex scalar models and~$P(X)$ models --- namely, how the latter arise as an 
EFT of the former in the gradient expansion --- is discussed in 
Section~\ref{sec: EFT of the complex scalar}. In Section~\ref{sec: Steady state accretion}, 
we outline the formalism of steady-state accretion. Section~\ref{sec: Accretion for P(X)} reviews the Bondi accretion for a particular~$P(X)$ model, providing a useful point of 
comparison for the results obtained later in Section~\ref{sec: Accretion for complex scalar field}. The latter section examines accretion for the complex scalar field and clarifies in what sense the complex field
can be regarded as a UV completion of the~$P(X)$ model. In Section~\ref{sec: Equation of state}, 
we further analyze the EMT of the complex scalar field, focusing on terms going beyond the perfect fluid structure. Finally, 
Section~\ref{sec: Discussion} summarizes our results, 
presents our conclusions, and discusses further perspectives.

\section{Complex scalar generalities}
\label{sec: Complex scalar generalities}

The dynamics of a canonical complex scalar~$\Psi$
with a~$U(1)$-invariant potential~$V$ is given by the action\footnote{Unless constants appear explicitly, as in $M_{Pl}^{2}=1/G$, we work in
Planck units~$\hbar \!=\! G \!=\! c \!=\!1$. Throughout the 
paper, we use $(+,-,-,-)$ signature convention.}  
\begin{equation}
S_{\rm cs} \bigl[ \Psi,\Psi^*,g_{\mu\nu} \bigr] 
    =
    \int\! d^{4}x \, \sqrt{-g} \, \biggl[
	\frac{1}{2} g^{\mu\nu} \partial_\mu \Psi \partial_\nu \Psi ^* - V(|\Psi|)
	\biggr] \, ,
\label{csAction}
\end{equation}
where $g_{\mu\nu}$ is the spacetime metric in the mostly negative signature convention,
and its determinant denoted by $g \!\equiv\! \text{det}(g_{\mu\nu})$. Investigating accretion 
and establishing connections to~$P(X)$ models and hydrodynamics are best accomplished using the polar field decomposition,
\begin{equation}
\Psi = \rho \, e^{i\varphi}\, ,
\label{polar decomposition}
\end{equation}
in which the action above reads
\begin{equation}
S_{\rm cs} \bigl[ \rho,\varphi,g_{\mu\nu} \bigr]
    =
    \int\! d^{4}x \, \sqrt{-g} \, \biggl[
	\frac{1}{2} ( \partial\rho )^2
	+\frac{1}{2}\rho^2 ( \partial\varphi )^2
	- V(\rho)
	\biggr]
	\, ,
	\label{CS action}
\end{equation}
with~$( \partial\rho )^2 \!\equiv\! g^{\mu\nu} \,\partial_\mu \rho \,\partial_\nu \rho$
and~$( \partial\varphi )^2 \!\equiv\! g^{\mu\nu} \,\partial_\mu \varphi \,\partial_\nu \varphi$. 
Note that the modulus field~$\rho$ carries the canonical 
scalar-field dimension, while the 
phase field~$\varphi$ is dimensionless. 

The equation of motion for the phase field takes the form of a conservation 
equation,
\begin{equation}
\nabla_\mu J^\mu = 0 \, ,  \qquad \text{where} \qquad 
J_{\mu} = -\frac{i}{2}\left(\Psi^{*}\partial_{\mu}\Psi - \Psi\partial_{\mu}\Psi^{*}\right)
= \rho^2 \,\partial_\mu \varphi \, ,
\label{current}
\end{equation}
and $\nabla_{\mu}$ is the standard covariant derivative. It encodes the 
conservation of the Noether current $J^{\mu}$ associated with the $U(1)$ global 
symmetry. The equation of motion for the modulus field is
\begin{equation}
\dalembertian \rho + V'(\rho) - \rho X = 0 \, ,  
\qquad \text{where} \qquad 
X = g^{\mu\nu} \partial_\mu \varphi \,\partial_\nu \varphi \, .
\label{EOMs}
\end{equation}
Here primes denote derivatives with respect to~$\rho$,
and~$\dalembertian \!=\! g^{\mu\nu}\nabla_{\mu}\nabla_{\nu}$ is the d'Alembertian operator.

For timelike gradients of the phase field,~$J^{\mu}$ defines\,\footnote{We assume 
that~$\partial_\mu \varphi$ is future-directed, i.e.~$\partial_t \varphi>0$ for the relevant choice of
time coordinate $t$.} 
a local rest frame comoving with the $U(1)$ charge, with four-velocity 
\begin{equation}
u_{\mu} = \frac{\partial_\mu \varphi}{\sqrt{X}}\, .
\label{velocity Eckart}
\end{equation}
This is the so-called Eckart local rest frame~\cite{Eckart:1940te} in the terminology of 
the hydrodynamics of imperfect fluids (for a recent pedagogical discussion see 
e.g.~\cite{Kovtun:2012rj,Andersson:2020phh}). 
It is convenient to introduce notation for the convective derivative,
denoted by an overdot, e.g.
\begin{equation}
\mathring{\varphi} \equiv u^{\lambda}\nabla_{\lambda}\varphi = \sqrt{X}\,,
\label{eq:convection}
\end{equation}
and analogously for all other quantities. It is well known, see e.g. \cite{Son:2000ht,Son:2002zn}, that \eqref{eq:convection} defines the \emph{chemical potential}, so that 
\begin{equation}
\upmu=\mathring{\varphi}\,.
\end{equation}
The $U(1)$ charge density in
this frame is given by
\begin{equation}
n=u_{\mu}J^{\mu}=\rho^{2}\mathring{\varphi}\,,
\label{eq:charge_density}
\end{equation}
which gives the modulus field~$\rho$ a physical meaning in 
terms of the charge density and the chemical potential. Charge conservation equation \eqref{current}, written for $J^{\mu}=n u^{\mu}$, gives 
\begin{equation}
\mathring{n}+\Theta n=0\,,
\label{eq:charge_density_evolution}
\end{equation}
where, $\Theta\equiv\nabla_{\lambda}u^{\lambda}$, denotes the expansion parameter of the Eckart frame. As $\Theta\equiv\mathring{V}/V$, where $V$ denotes the comoving volume, see, e.g. \cite{Poisson:2009pwt}, one obtains $n\propto 1/V$.  

It is also useful to define the projector
\begin{equation}
{\perp}_{\mu\nu} \equiv g_{\mu\nu} - u_{\mu}u_{\nu}\,,
\label{eq:Projector}
\end{equation}
which projects all tensor quantities onto purely spatial 
directions orthogonal to $u^{\mu}$. Using this projector, 
one introduces a purely
spatial covariant derivative\footnote{This derivative is not compatible with the transverse purely spatial
metric $-{\perp}_{\mu\nu}$, as $D_{\mu}{\perp}_{\alpha\beta}\neq0$ for
four-velocities with nonvanishing four-acceleration.}, 
\begin{equation}
D_{\mu} \equiv {\perp}_{\mu}^{\nu}\, \nabla_{\nu}
	\, .
\label{eq:D}
\end{equation}
Ordinary covariant derivatives can then be decomposed in terms of convective and purely spatial covariant derivatives, e.g.
\begin{equation}
\partial_{\mu}\rho = \mathring{\rho}\,u_{\mu} + D_{\mu}\rho\,.
\label{eq:radial_decomposition}
\end{equation}

It is important to note that the classical complex scalar field, as a continuous medium, does 
not have an energy–momentum tensor (EMT) of the perfect-fluid form, even for timelike 
currents $J^{\mu}$. Indeed, the complex scalar EMT is given by
\begin{equation}
T_{\mu\nu}
	\equiv
	\frac{2}{\sqrt{-g}} \frac{\delta S_{\rm cs}}{\delta g^{\mu\nu} } 
	=
	\partial_{\mu}\rho\,\partial_{\nu}\rho
	+
	\rho^{2}\partial_{\mu}\varphi\,\partial_{\nu}\varphi
	-
	g_{\mu\nu}
		\left[
		\frac{1}{2}\left(\partial\rho\right)^{2}
		+
		\frac{1}{2}\rho^{2}\left(\partial\varphi\right)^{2}
		-
		V\left(\rho\right)
		\right]\, ,
\label{cs emt}
\end{equation}
which can be decomposed in the Eckart local rest frame~\eqref{velocity Eckart} using~\eqref{eq:radial_decomposition} as 
\begin{equation}
T_{\mu\nu}
	=
	\epsilon \, u_{\mu}u_{\nu}
	-
	p \, {\perp}_{\mu\nu}
	+
	\Pi_{\mu\nu}
	+
	q_{\mu}u_{\nu} + q_{\nu}u_{\mu}\,,
\label{T decomp}
\end{equation}
where the introduced quantities\footnote{
There is no physical heat flow in this system, as we work at vanishing 
temperature. Nevertheless, energy is still being transported orthogonally to the 
shift-charge current, which is captured by the term in~(\ref{heat transfer}).
} are
\begin{align}
\label{energy density}
\text{energy density:}
	\qquad & 
	\epsilon \equiv u^{\mu}u^{\nu} T_{\mu\nu}
		= \frac{1}{2} \left(
			\mathring{\rho}^{2} + \rho^{2}\mathring{\varphi}^{2} + |D\rho|^{2}
			\right)
			+
			V(\rho)
			\, , 
\\ 
\label{pressure}
\text{pressure:}
	\qquad & 
	p \equiv - \frac{1}{3} {\perp}^{\mu\nu} T_{\mu\nu}
		= \frac{1}{2}\left(
			\mathring{\rho}^{2}
			+
			\rho^{2}\mathring{\varphi}^{2}
			-
			\frac{1}{3}|D\rho|^{2}
			\right)
			-
			V\left(\rho\right)
			\, ,
\\
\label{anisotropic stress}
\text{anisotropic stress:}
	\qquad & 
	\Pi_{\mu\nu} \equiv
		{\perp}_{\mu}^\alpha {\perp}_{\nu}^\beta \,T_{\alpha\beta}
			+ p {\perp}_{\mu\nu}
		=
		D_{\mu}\rho \, D_{\nu}\rho
		+
		\frac{1}{3} {\perp}_{\mu\nu} |D\rho|^{2}
		\, ,
\\
\label{heat transfer}
\text{``heat flux'':}
	\qquad & 
	q_{\mu} \equiv {\perp}_{\mu}^{\alpha} \, u^\beta T_{\alpha\beta}
	=
	\mathring{\rho} D_{\mu}\rho\,,
\end{align}
and where we denote $ |D\rho|^{2} \!\equiv\! -g^{\mu\nu}D_{\mu}\rho \, D_{\nu}\rho$. 
The energy density and pressure are perfect-fluid properties, while
the anisotropic stress and heat flux account for dissipative effects.
Since the pressure and energy density generally depend on two fields and their derivatives, 
an equation of state of the form~$p(\epsilon)$ does not,
in general, exist. Nonetheless, it is useful to introduce 
the equation-of-state parameter $w$ as the usual ratio,
\begin{equation}
w \equiv p/\epsilon\,.
\label{w}
\end{equation}

For general configurations of the complex scalar field, including accretion onto a black hole, 
the heat flux~\eqref{heat transfer} and anisotropic stress~\eqref{anisotropic stress} 
do not vanish, indicating deviations of the energy–momentum tensor from 
the perfect fluid form. Later in the paper, we provide 
expressions~\eqref{heat_transfer_spec}
and~\eqref{anisotropic_stress_spec} for these quantities for 
the Bondi accretion. Their absolute values for this case are plotted in 
Fig.~\ref{q plots} and Fig.~\ref{Pi plots}. While there are special circumstances where the 
perfect fluid description, including the equation of state~$p(\epsilon)$, might 
constitute an accurate approximation, there is no a priori 
reason to  expect that this is the 
case for the steady-state accretion setup used in~\cite{Feng:2021qkj}. Indeed, we show here 
that the full field-theoretic description is necessary to obtain the correct field profiles and 
accretion rate in the general setting. 

In this paper, we consider the standard renormalizable potential for the complex scalar,
\begin{equation}
V(\rho) = \frac{m^2}{2} | \Psi |^2 + \frac{\lambda}{4} | \Psi |^4 + V_0
    = \frac{m^2}{2} \rho^2 + \frac{\lambda}{4} \rho^4 + V_0 \, ,
\label{eq:Potential}
\end{equation}
where~$\left|m\right|$ is the mass of excitations around vacuum,~$\lambda$ is the dimensionless quartic self-coupling constant, and
\begin{equation}
\label{V_0}
V_0 = \theta(-m^2) \frac{m^4}{4\lambda}\equiv \theta(-m^2)\varrho_{m} \, ,
\end{equation}
is a constant added to $V(\rho)$, to ensure that the potential vanishes also in the spontaneously broken vacuum and to avoid negative energy densities. Here $\theta$ is the unit step function which is only non-vanishing in the symmetry-breaking case $m^2<0$. 
Note that~\cite{Feng:2021qkj} considered Bondi accretion of the complex scalar only for $m^2\!>\!0$, i.e., without spontaneous symmetry breaking, for which the potential minimum is at $\rho\!=\!0$.

\section{EFT of the complex scalar}
\label{sec: EFT of the complex scalar}

The connection between the complex scalar model and $P(X)$ models becomes apparent in the regime of a slowly varying modulus field. When we can treat derivatives
of the modulus as small perturbations, we can derive an effective field theory (EFT) 
for the phase field. We first present this procedure, also considered in~\cite{Aoki:2021ffc}, for arbitrary potentials,
and then apply it to the quartic potential of interest here.

\subsection{General gradient expansion}
\label{subsec: General gradient expansion}

In the limit of small modulus field gradients, we should organize the equations
of motion~(\ref{EOMs}) such that the kinetic term is treated as a perturbation. 
This assumption can conveniently be tracked by introducing a bookkeeping
parameter~$\varepsilon$ that counts the number of derivatives,
\begin{equation}
\rho X - V'(\rho) = \varepsilon \dalembertian \rho \, ,
\label{EOMepsilon}
\end{equation}
which is set to unity at the end.
By requiring the modulus~$\rho$
to take the form of a power series solution,
\begin{equation}
\label{gradient expansion}
\rho = \rho_0 + \varepsilon \rho_1 + \varepsilon^2 \rho_2 + \dots
\end{equation}
equation~(\ref{EOMepsilon}) essentially changes character from a differential 
equation into an algebraic one. This is because the leading order is determined only by the
potential and the phase field kinetic term. This algebraic character is then 
maintained at all higher orders.

\paragraph{Leading order.}
The leading order equation contains no derivatives,
\begin{equation}
X \rho_0 - V'(\rho_0) = 0 \, ,
\label{rho0eq}
\end{equation}
and defines the leading-order approximation for the modulus as a function of the phase field kinetic 
term,~$\rho_0\!=\!\rho_0(X)$. 
For flat space vacua, the phase-field kinetic term
is constant, and~(\ref{rho0eq}) constitutes an exact solution for the modulus field.
Thus, we can consider that~$X \!=\! \text{const.}$ labels the flat-space vacua.
For a slowly varying~$X$, this picture remains qualitatively valid, with
all quantities receiving gradient corrections, captured perturbatively.

Truncating the action~(\ref{CS action}) at leading order in the gradient expansion
corresponds to the auxiliary-field formulation of {\it $P(X)$ theory},
\begin{equation}
S_{0} \bigl[ \rho,\varphi,g_{\mu\nu} \bigr]
    = \int\! d^4x \, \sqrt{-g} \, \biggl[ \frac{\rho^2X}{2} - V(\rho) \biggr]
    \, .
\end{equation}
The modulus field is an auxiliary field satisfying Eq.~(\ref{rho0eq})
on-shell. Solving for it and substituting it back into the action produces
the more standard formulation of the~$P(X)$ theory,
\begin{equation}
S_{0} \bigl[ \varphi, g_{\mu\nu} \bigr]
    =
    \int\! d^4x \, \sqrt{-g} \, P(X) \, ,
\qquad
\text{where}
\qquad
P(X) = \frac{\rho^2_0(X)X}{2} -  V\bigl( \rho_0(X) \bigr)
\, .
\label{PXaction}
\end{equation}

\paragraph{Subleading orders.}
The subleading order of Eq.~(\ref{EOMepsilon}) is linear in~$\rho_1$, 
which can then be solved readily,
\begin{equation}
\rho_1 = \frac{ \dalembertian \rho_0 }{ X - V''(\rho_0) } \, .
\end{equation}
The first order correction~$\rho_1$ is a function of~$X$ and its
two covariant derivatives. 
For higher order corrections, the structure of the equations essentially does not 
change. In fact, it is not difficult 
to see that the solution of the~$n$-th order 
equation descending from~(\ref{EOMepsilon}) is
\begin{equation}
\rho_n = \frac{ \dalembertian \rho_{n-1} + V_{(n)} }
        { X - V''(\rho_0) } \, ,
\qquad \text{where} \qquad
V_{(n)} = \frac{1}{n!} \biggl[ \frac{\partial^n}{\partial \varepsilon^n}
    V' \bigl( \rho_0 + \varepsilon \rho_1 + \dots + \varepsilon^{n-1} \rho_{n-1} \bigr)
    \biggr]_{\varepsilon \to 0}
    \, ,
\end{equation}
so that the~$n$-th order correction contains~$2n$ covariant derivatives of~$X$.

Equation~(\ref{EOMepsilon}) effectively plays the role 
of an algebraic constraint, upon interpreting the 
kinetic term perturbatively. That is why its solution can be 
plugged directly into the action~(\ref{CS action}) to eliminate the modulus field,
and to obtain an effective field theory description for the phase field,
\begin{equation}
S_{\rm eff} \bigl[ \varphi, g_{\mu\nu} \bigr] =
\int\! d^{4}x \, \sqrt{-g} \, \biggl[
    \frac{\rho_0^2 X}{2}
	-
    V(\rho_0 )
    +
    \frac{\varepsilon}{2} (\nabla^\mu \rho_0) (\nabla_\mu \rho_0)
    -
    \frac{\varepsilon^2}{2} 
            \frac{ (\dalembertian \rho_0)^2 }{ X - V''(\rho_0) }
        +
        \mathcal{O}(\varepsilon^3)
	\biggr]
	\, ,
\label{Seff}
\end{equation}
which maintains shift symmetry at all orders.
We should note that the leading order term in this effective action is precisely
the~$P(X)$ theory, parametrized in terms of~$\rho_0(X)$. This is how the two
theories are connected: at leading
order in gradient expansion, the canonical complex scalar theory reduces to the
non-canonical~$P(X)$ theory~\cite{Babichev:2017lrx}. 
The higher order terms in 
the effective action~(\ref{Seff}) represent gradient corrections.
Note that higher-derivative terms, arising 
in the field equations that descend from (\ref{Seff}), should be treated 
perturbatively in order to avoid introducing spurious degrees of 
freedom~(see e.g.~\cite{Glavan:2017srd}).

\subsection{Application to quartic self-coupling}
\label{subsec: Application to quartic self-coupling}

For the potential~(\ref{eq:Potential}) of interest to us, the leading
order solution of Eq.~(\ref{rho0eq}) is
\begin{equation}
\rho_0^2 = \frac{X \!-\! m^2}{\lambda} \, ,
\label{eq:Rho_X_quartic}
\end{equation}
which exists only for $X \!\ge\! m^2$.
Consequently, the effective action~(\ref{Seff}) takes the form
\begin{equation}
S_{\rm eff} \bigl[ \varphi, g_{\mu\nu} \bigr]
=
\frac{1}{4\lambda}\int\! d^{4}x \, \sqrt{-g} \, 
    \biggl[
    ( X \!-\! m^2 )^2
    +
    \frac{(\partial X )^2}{2 (X \!-\! m^2) }
    +
            \biggl(
                \frac{2(X \!-\! m^2){\dalembertian} X \!-\! (\partial X )^2}
                    {4(X \!-\! m^2)^2}
            \biggr)^{\!2}
        \!+
        \mathcal{O}\bigl(\partial^6\bigr)
	\biggr]
    ,
\label{EFTexample}
\end{equation}
where, for brevity, we omitted the constant term \eqref{V_0}, 
and we also dropped the bookkeeping parameter~$\varepsilon$, as 
it is clear
that the expansion is organized in terms of the number of derivatives\footnote{Note that~(\ref{EFTexample}) 
is the tree level 
effective action; for the one-loop-corrected effective action in the regime of negligible gradients of~$X$, see~\cite{Joyce:2022ydd}.} acting on~$X$.
We see that the leading order term in this effective action is a particular~$P(X)$ 
theory~(\ref{PXaction}) with
\begin{equation}
P(X) = \frac{( X \!-\! m^2 )^2}{4\lambda} - V_0 \, .
\label{specific P}
\end{equation}
For $m^2 \!>\! 0$, this action is known as the \emph{ghost condensate}~\cite{Arkani-Hamed:2003pdi}, while for $m \!=\! V_0 \!=\! 0$, it represents effective radiation. 

Note that the~$P(X)$ theory in~(\ref{specific P}) 
permits exploration of the ghostly region around $X\!=\!0$, 
whereas the original action from which it descends definitely forbids all ghostly configurations $X \!<\! m^2$ on account of~(\ref{eq:Rho_X_quartic})
being satisfied. For $m^2 \!<\! 0$, the action in~\eqref{specific P} is known as purely kinetic k-essence~\cite{Scherrer:2004au}, which has been proposed as a framework for unifying DM and DE.
It is worth noting that the gradient expansion defining the 
EFT in~(\ref{EFTexample}) is not the same as the EFT 
expansion obtained by integrating out the heavy modulus field~\cite{Burgess:2014lwa,Burgess:2020tbq}. However, 
expanding~(\ref{EFTexample}) in inverse powers of~$m^2$
and specializing to flat space does produce the latter, at 
least for the lowest orders of the expansion.

\bigskip

The complex scalar model in~(\ref{CS action}) with the potential in~(\ref{eq:Potential}) 
is therefore virtually indistinguishable from the~$P(X)$ model in~(\ref{specific P}) in
situations where no large gradients of $X$ develop. 
The equivalence in this regime holds not only for background evolution, but also for perturbations~\cite{Babichev:2018twg}, provided the description is limited to frequencies parametrically lower than the chemical potential $\upmu=\mathring{\varphi}$ of the background. 
For timelike gradients of the phase field, 
any $P(X)$ theory describes a perfect superfluid, see e.g. \cite{Greiter:1989qb,Son:2000ht,Son:2002zn,Alford:2012vn,Glodkowski:2025tnv}. In the leading order in gradient expansion, for model \eqref{specific P} (or for potential \eqref{eq:Potential}) the non-diagonal terms are vanishing,  $q_{\mu}=0$ and $\Pi_{\mu\nu}=0$, while the perfect-fluid quantities\footnote{Here, for simplicity, we omit the superscripts $0$, but all quantities $p$, $\epsilon$, $n$, $\upmu$ and $c_s$ are calculated at zeroth order in the gradient expansion~\eqref{gradient expansion}, i.e., for the theory defined by the Lagrangian~\eqref{specific P}.} are given by 
\begin{align}
&\epsilon=\frac{1}{2}\rho V'+V=m^{2}\rho^{2}+\frac{3}{4}\lambda\rho^{4}+V_{0}\,, \\
&p=\frac{1}{2}\rho V'-V=\frac{\lambda}{4}\rho^{4}-V_{0}\,,\\
&n=\rho\sqrt{\rho V'}=\rho\sqrt{m^{2}\rho^{2}+\lambda\rho^{4}}\,,\\
&\upmu=\sqrt{\frac{V'}{\rho}}=\sqrt{m^{2}+\lambda\rho^{2}}\,.
\end{align}
From the definitions above, for general potential $V$ one recovers the standard thermodynamical expressions involving the chemical potential: the Euler relation
\begin{equation}
\epsilon=\upmu\,n-p\,, 
\end{equation}
along with
\begin{equation}
\label{UPmu}
\upmu=\frac{d\epsilon}{dn}=\frac{\epsilon+p}{n}\,.
\end{equation}
Note that the last relation implies the adiabatic flow (no change in entropy per particle) of the fluid, due to the second law of thermodynamics, written as\footnote{See, e.g. equation (2.10.18) on page 50 from \cite{Weinberg:1972kfs}.}
\begin{equation}
pd\left(\frac{1}{n}\right)+d\left(\frac{\epsilon}{n}\right)=0\,.
\end{equation}
Thus, when studying the accretion in the leading order in gradient expansion, the flow is always \emph{adiabatic}. 
The adiabatic processes are often approximately described as $pV^{\gamma}=const$ or $p\propto n^{\gamma}$, where $\gamma$ is the so-called adiabatic index.  In the leading order in gradient expansion 
for potential \eqref{eq:Potential} one obtains 
\begin{equation}
n^2=\frac{8}{\sqrt{\lambda}}\left(\varrho_{m}^{1/2}\left(p+V_{0}\right)+\left(p+V_{0}\right)^{3/2}\right)\,,
\end{equation}
 which defines $p(n)$ as an implicit function. From where one can read off the asymptotic values of the  adiabatic index. Namely, without spontaneous symmetry breaking, ($m^2>0$ and $V_0=0$), for $p\gg\varrho_{m}$ one has $\gamma\approx4/3$, similarly to radiation, while for $p\ll\varrho_{m}$ the adiabatic index is stiffer: $\gamma\approx2$. On the other hand, with spontaneous symmetry breaking, ($m^2<0$ and $V_0>0$), for $p\gg\varrho_{m}$ the index is again similar to radiation: $\gamma\approx4/3$. However, in this case, $\rho>\sqrt{\left|m^{2}\right|/\lambda}$,
implying that $p+V_{0}>\varrho_{m}$, so that for small pressures, or $\rho$ just slightly larger than the expectation value in the spontaneously broken vacuum $\sqrt{\left|m^{2}\right|/\lambda}$, the adiabatic index, $\gamma=\left(n/p\right)\left(dp/dn\right)$, is formally ultra-stiff, $\gamma\rightarrow\infty$, because $p\rightarrow0$ by the constant charge density, $n$, and constant derivative, $dn/dp$.

Furthermore, one can write an implicit equation of state $p(\epsilon)$
\begin{equation}
\epsilon - V_0
	= 
	3 (p + V_0)
	+ \frac{2m^2}{ \sqrt{\lambda} } \sqrt{p + V_0}
	\, ,
    \label{eq:eospx}
\end{equation}
valid for both signs of $m^2$.

The implicit equation of state~(\ref{eq:eospx}) yields rather different solutions depending on the sign of  $m^2$: 
\begin{equation}
\label{equation of state_symm}
p\left(\epsilon\right)_{[+]}
	=
	\frac{4}{9}\varrho_{m}
	\bigg( \sqrt{1+\frac{3}{4}\frac{\epsilon}{\varrho_{m}}}-1 \bigg)^{\!2}\,,\qquad 
\text{for}
\qquad 
m^2>0\,,
\end{equation}
where the characteristic energy-density scale, $\varrho_{m}$, is defined in \eqref{V_0} (see \cite{Colpi:1986ye}), and
\begin{equation}
\label{equation of state_SSB}
p\left(\epsilon\right)_{[-]}
	=
	\frac{4}{9}\varrho_{m}
	\bigg[
	\bigg( 1+\frac{1}{2}\sqrt{1+3\frac{\epsilon}{\varrho_{m}}} \, \bigg)^{\!2} - \frac{9}{4}
	\bigg]
	\,,
\qquad
\text{for}
\qquad
m^2<0\,.
\end{equation}
Note that $p\left(\epsilon\right)_{[-]} \!\geq\! p\left(\epsilon\right)_{[+]}$. 
Moreover, for $\epsilon\gg \varrho_{m}$, both equations of state give $p\!\simeq\!\frac{1}{3}\epsilon$ (which is also obtained in the formal limit $\varrho_{m}\!\rightarrow \!0$). 
However, for $\epsilon\ll \varrho_{m}$ the two equations of state are radically different: 
in the latter case one has an ultra-hard equation of state, 
$p\left(\epsilon\right)_{[-]} \!\simeq\! \epsilon$, whereas in the former case one finds a dust-like equation of state, $p\left(\epsilon\right)_{[+]}\!\simeq\! \epsilon^{2}/(16\varrho_{m})$. 

Equation of state~(\ref{equation of state_symm}) was employed in~\cite{Feng:2021qkj}
to study the steady-state accretion of a complex scalar field in the hydrodynamic limit, which 
was claimed to be realized when
\begin{equation}
\label{eq:lambda_condition_Feng_Colpi}
\lambda\gg Gm^{2}\,,
\end{equation}
where we restored the Newton constant $G$. This condition just implies that the scalar self-interaction is stronger than the gravitational interaction at the scale $m$. Note that, surprisingly, this condition involves neither the mass of the BH nor the chemical potential associated with the $U(1)$-symmetry (or the energy density of the scalar field) at spatial infinity. In fact, $p\left(\epsilon\right)_{[+]}$, along with the above condition on $\lambda$, were adopted by~\cite{Feng:2021qkj} from the study of 
boson stars in~\cite{Colpi:1986ye}. In the latter work this approximation was shown to hold 
for \emph{self-gravitating} complex scalar configurations with $m^2\!>\!0$. As we will demonstrate later, see \eqref{dimless profile eq}, the applicability of the gradient expansion and of the hydrodynamic limit given by \eqref{specific P} requires a different condition, at least in the test field approximation neglecting self-gravity.  
To the best of our knowledge, no prior work studied accretion for $p\left(\epsilon\right)_{[-]}$ which may be relevant for 
the growth of a PBH immersed in the superfluid nuclear matter of a neutron-star interior.

In the hydrodynamical case, the EFT in~\eqref{specific P} propagates
small perturbations with the sound speed
\begin{equation}
c_{s}^{2}
    =
    \frac{\partial p}{\partial\epsilon}
    =
    \frac{\rho V'' \!-\! V'}{\rho V'' \!+\! 3V'}
    =
    \frac{1}{3} \Bigl( \frac{3\lambda\rho^{2}}{3\lambda\rho^{2} \!+\! 2m^{2}} \Bigr)
    =
    \frac{X \!-\! m^{2}}{3X \!-\! m^{2}} \, .
\label{C_s}
\end{equation}
For $m^2>0$, this formula gives the speed of sound 
only for~$X \!>\! m^2$ because of~\eqref{eq:Rho_X_quartic}. 
However, for $m^2\!<\!0$, it only works 
for $\rho^{2} \!>\! |m^{2}|/\lambda$, thus limiting the speed of sound 
to $c_{s}^2\!<\!1$.
The above formula implies that for $m^2>0$ the EFT sound speed is $0<c_{s}^{2}<1/3$, while for $m^2<0$ one has $1/3<c_{s}^{2}<1$.

\bigskip

Rather than considering the effective theory and studying only small corrections, 
we focus on systems that show large deviations between
the two models that look the same in the limit of small gradients. The effective-theory 
description outlined in this section does not apply in such circumstances. That is why 
we mostly study the equation of motion for the modulus field without assuming the derivatives of $X$
to be small. 
Moreover, we do not assume that the condition~$X\!>\!0$ holds everywhere, as required in the hydrodynamic picture of accretion. We impose it only at spatial infinity, and find that our solutions preserve this inequality throughout the flow.

\section{Steady state accretion}
\label{sec: Steady state accretion}

We now consider the accretion of a complex scalar field onto a Schwarzschild black hole. Most often, the line element for this spacetime is expressed in 
Schwarzschild coordinates~($t,r,\theta,\phi$),
\begin{equation}
ds^2 = f(r) dt^2 - \frac{dr^2}{f(r)} - r^2 d\Omega^2 \, ,
\qquad \qquad
f(r) = 1 - \frac{r_{\scr S}}{r} \, ,
\label{SchwCoord}
\end{equation}
where~$d\Omega^2 \!=\! d\theta^2 \!+\! \sin^2(\theta) d\phi^2$ is the line element 
of a 2-sphere, and the Schwarzschild radius~$r_{\scr S} \!=\!2GM$ is expressed in 
terms of Newton’s constant~$G$ and the black hole mass~$M$. However, for our 
purposes, the ingoing Eddington–Finkelstein coordinates~($v,r,\theta,\phi$) are
more appropriate. They are related to the Schwarzschild ones via
\begin{equation}
v = t + r_{*} \, ,
\qquad \qquad
r_{*} = r + r_{\scr S} \ln\Bigl|  \frac{r}{r_{\scr S}} - 1 \Bigr| \, ,
\label{tortoise}
\end{equation}
where~$r_{*}$ is known as the tortoise coordinate. The line element in these 
coordinates reads
\begin{equation}
ds^2 = f(r) dv^2 - 2 dv \, dr - r^2 d\Omega^2 \, ,
\end{equation}
and is not singular at the Schwarzschild horizon~$r\!=\!r_{\scr S}$.

\medskip

We consider the complex scalar field as a test field in the Schwarzschild black 
hole spacetime, which is a reasonable approximation in the limit where 
backreaction can be neglected. We briefly discuss the conditions for the validity of this approximation at the end of this section. 

We assume {\it steady-state accretion}, in which 
the infalling scalar field respects the isometries of the Schwarzschild spacetime. 
We formulate the accretion in ingoing Eddington–Finkelstein coordinates, which are 
well adapted to this problem as they remain finite at the Schwarzschild horizon. 
The most general ansatz for the complex scalar field
respecting Schwarzschild isometries is
\begin{equation}
\varphi = \varphi(v,r) =  \dot{\varphi}_0 \biggl[ v + \int^r \! \! dr' \, W(r') \biggr] \, ,
\qquad \qquad
\rho = \rho(r) \, ,
\label{ansatz}
\end{equation}
where~$\dot{\varphi}_0$ is the constant velocity of the phase field,
$\partial_v \varphi \!=\! \partial_t \varphi \!=\! \dot{\varphi}_0$,
in either Eddington–Finkelstein or Schwarzschild coordinates.
Our main goal is solving the equations of 
motion~(\ref{EOMs}) and~(\ref{current}) assuming the ansatz~(\ref{ansatz}). 
The strategy is to first integrate the conservation equation~\eqref{current}. In this way, we find~$W$
in terms of~$\rho$, thus obtaining a single profile equation for the modulus.

\medskip

The shift-charge current~(\ref{current}) has only two non-vanishing components
for the ansatz in~(\ref{ansatz}),
\begin{equation}
J^r =
	- \dot{\varphi}_0 \rho^2 \bigl( 1 + f W \bigr) 
	\, ,
\qquad \qquad
J^v =
	- \dot{\varphi}_0 \rho^2 W
	\, .
\end{equation}
Only the radial component contributes to the conservation 
equation in~(\ref{current}), which is readily integrated to give
\begin{equation}
J^r = - \frac{ \dot{\varphi}_0 \mathcal{F} }{ r^2 } \, .
\label{Jr}
\end{equation}
Here~$\mathcal{F}\!>\!0$ is a constant of integration representing the 
incoming~$U(1)$ charge flux per unit solid angle, normalized to the chemical 
potential~$\dot{\varphi}_0$ at infinity. 
The radial function~$W$ from the 
ansatz~(\ref{ansatz}), and consequently the phase field kinetic term from~(\ref{EOMs}) 
immediately follow,
\begin{equation}
W = \frac{1}{f} \biggl[ \frac{ (1\!-\!f)^2 \mathcal{F} }{ r_{\scr S}^2 \rho^2 } - 1 \biggr] \, ,
\qquad \qquad
X = \frac{ \dot{\varphi}_0^2 }{f} \biggl[
	1 - \frac{( 1\!-\!f )^4 \mathcal{F}^2 }{ r_{\scr S}^4 \rho^4 }
	\biggr]
	\, .
\label{WandX}
\end{equation}
Since $X$ is a scalar and an observable quantity --- the square of the chemical 
potential for $X\!>\!0$ --- it must remain finite at the horizon. This requirement 
immediately fixes the relation between the flux and the modulus field value at the 
horizon,
\begin{equation}
\rho(r_{\scr S} ) \!=\! \frac{\sqrt{\mathcal{F}}}{r_{\scr S}} \!\equiv\! \rho_{\scr S}\,. 
\label{modulus_on_Horizon}
\end{equation}
Thus, the $U(1)$ charge of the BH increases at a constant rate,
\begin{equation}
\dot{Q}=-4\pi r^{2}J^{r}=-4 \pi r^2 n\,u^{r}=4\pi r_{\scr S}^{2}\,\dot{\varphi}_{0}\,\rho_{\scr S}^{2}
	\, .
\label{Charge_Accretion}
\end{equation}

Substituting the expression for $X$ from~(\ref{WandX}) into the first equation of 
motion in~(\ref{EOMs}) yields an ordinary differential equation for the scalar modulus profile,
\begin{equation}
f \frac{d^2 \rho}{dr^2}
	+ \frac{(1 \!-\! f^2)}{r_{\scr S}} \, \frac{d \rho}{dr}
	- U'(\rho)
	=
	0
	\, ,
\label{rhoEOM}
\end{equation}
We regard this as the \emph{master equation}, since its solutions fix 
all background profiles and derived observables. The effective potential $U(\rho)$ 
entering this equation,
\begin{equation}
U(\rho) \equiv V(\rho) - \int^\rho\!\! d\rho \, \rho \, X(\rho)
	= \frac{m^2}{2} \rho^2 + \frac{\lambda}{4} \rho^4 
		+ V_0
		- \frac{ \dot{\varphi}_0^2 \rho^2 }{2f} \biggl[ 1 + \frac{ (1\!-\!f)^4 \mathcal{F}^2 }{ r_{\scr S}^4 \rho^4 } \biggr] \, .
\end{equation}
receives contributions both from the complex scalar potential in the action~(\ref{CS action})
and from the nonvanishing phase-field velocity. Note that this effective potential also
depends explicitly on the radial coordinate through~$f\!=\!f(r)$, and not only on the modulus 
field~$\rho$, and that the prime denotes a derivative of the effective potential with respect 
to~$\rho$ only. Finding solutions of the master equation~(\ref{rhoEOM}) is the focus of the remainder of 
the paper. 

The task of solving~(\ref{rhoEOM}) is facilitated and made more transparent by introducing 
dimensionless variables. It is natural to measure radius in terms of the 
Schwarzschild radius, and it is useful to define the following quantities,
\begin{equation}
x = \frac{r}{r_{\scr S}} \, ,
\qquad \qquad
f = 1 - \frac{1}{x} \, .
\end{equation}
It is convenient to express the remaining dimensionful quantities in 
units of~$\dot{\varphi}_0$. Furthermore, it is advantageous to absorb the 
quartic coupling constant~$\lambda$ into the modulus profile; this makes it 
transparent that the profile equation does not depend on all the parameters
independently. 
Moreover, the limit~$\lambda\!=\!0$ is not accessible in the steady-state accretion
scenario, as it corresponds to the case of dust, for which steady-state accretion is not possible.
All of this suggests introducing the following dimensionless quantities, in part already introduced in~\cite{Aguilar-Nieto:2022jio}, for the modulus profile, mass, potential, Schwarzschild radius, flux, and phase-field kinetic term, respectively:
\begin{align}
&
\sigma = \frac{\sqrt{\lambda} \, \rho}{ \dot{\varphi}_0 } \, , 
&&
\mu^2 = \frac{m^2}{\dot{\varphi}_{0}^2} \, ,
&&
\mathcal{V}_0 = \frac{\lambda V_0}{\dot{\varphi}_0^4} 
    = \theta(-\mu^2) \frac{ \mu^4}{4} \, ,
\nonumber\\
&
\xi = r_{\scr S} \dot{\varphi}_0 \, , 
&&
\beta^2 = \frac{ \lambda \mathcal{F} }{ \xi^2 } \, ,
&&
Y = \frac{X}{\dot{\varphi}_0^2}
\, .
\label{dimensionless def}
\end{align}
Thus, the profile equation takes the form
\begin{equation}
\frac{1}{\xi^2} \mathcal{D}^2 \sigma
	- \mathcal{U}' \bigl( \mu^2, \beta^2; f, \sigma \bigr)
	=
	0
	\, ,
\label{dimless profile eq}
\end{equation}
where we have introduced a shorthand notation for the gradient terms,
\begin{equation}
\mathcal{D}^2 = f \frac{d^2 }{dx^2}
	+ (1 \!-\! f^2) \frac{d }{dx} 
	=
	(1\!-\!f)^4 \biggl[
		f \frac{d^2}{df^2} + \frac{d}{df}
		\biggr]
	\, ,
\end{equation}
and where the dimensionless effective potential is
\begin{equation}
\mathcal{U} \bigl( \mu, \beta; f, \sigma \bigr)
	= \frac{\mu^2 \sigma^2}{2} + \frac{ \sigma^4 }{4} 
	 + \mathcal{V}_0 
		- \frac{ \sigma^2 }{2f} \biggl[ 1 + \frac{(1\!-\!f)^4 \beta^4}{\sigma^4} \biggr] 
		\, .
\label{eff potential}
\end{equation}
The prime on the potential in~(\ref{dimless profile eq}) stands for the derivative 
with respect to~$\sigma$. Formally, in the limit~$\xi\!\to\!\infty$, the profile 
equation reduces\footnote{It is worth mentioning that $\xi^{-2}$ plays the role of the formal bookkeeping parameter $\varepsilon$ in Eq.~\eqref{EOMepsilon}. Note that a $P(X)$ ideal fluid has only one independent equation of motion, which is simultaneously the current conservation equation \eqref{current}. Thus, \eqref{dimless profile eq} only changes what is transported by the current, providing in the limit of $\xi\!\to\!\infty$ a relation between charge density $n$ and chemical potential $\upmu$.} to that descending from the corresponding~$P(X)$ model. This 
is useful when studying the profile equation in itself. However, the 
limit~$\xi\!\to\!\infty$ should not be conflated with the~$P(X)$ limit in general. This is because, even in the exact~$P(X)$ limit, the mass of the black hole and the 
phase-field velocity can be chosen independently, and hence so can~$\xi$. The 
accretion rate, defined below, is a quantity that exemplifies this point.

The profile equation~(\ref{dimless profile eq}) is singular both at the event 
horizon at~$f\!=\!0$, and at radial infinity~$f\!=\!1$. If we require the boundary values to be finite and nonvanishing, they are uniquely fixed:
\begin{equation}
\sigma_{-}\equiv \sigma(f\!=\!0) = \beta \,,
\qquad \qquad
\sigma_{+}\equiv \sigma (f\!=\!1) = \sqrt{1 \!-\! \mu^{2}} \,.
\label{boundary conditions}
\end{equation}
The precise values of the boundary conditions are dictated
by the effective 
potential~(\ref{eff potential}), as the kinetic term becomes irrelevant in the 
two limits. This is why the boundary conditions for steady-state accretion 
for the~$P(X)$ model are the same as for the corresponding complex scalar model,
though we stress that the value of~$\beta$ corresponding to a particular value 
of~$\mu$ will generally differ when the parameter~$\xi$ is finite.

Another useful quantity is the {\it physical} velocity~$v$ associated with the shift charge. 
Following Eq.~(2.3.2) of~\cite{Frolov:1998wf}, 
this velocity is defined as
\begin{equation}
v =
    \frac{\sqrt{-g_{rr}}dr}{\sqrt{g_{tt}}dt}
    =
    \frac{\sqrt{-g_{rr}}u^{r}}{\sqrt{g_{tt}}u^{t}}
    =
    -f\frac{\partial_{r}\varphi}{\dot{\varphi}}
    =
    - \Bigl( \frac{r_{\scr S}}{r} \Bigr)^{\!2}
        \Bigl( \frac{\rho_{\scr S}}{\rho} \Bigr)^{\!2}
    \, ,
\label{definition_vc}
\end{equation}
so that its square 
\begin{equation}
v^{2}
    = \Bigl( \frac{f\partial_r \varphi}{\dot{\varphi}} \Bigr)^{\!2}
    = 1-fY
    \, ,
\label{definition vc2}
\end{equation}
is equal to the square of the shift-charge {\it relative three-velocity} 
as measured by an observer at rest in Schwarzschild coordinates.

\bigskip 

The time-translation symmetry, manifest in the Schwarzschild coordinates~(\ref{SchwCoord}), 
corresponds to the Killing vector given by $\zeta^{\mu}\!=\!\delta_{t}^{\mu}$. This vector 
defines the \emph{conserved} energy-momentum current $T_{\nu}^{\mu}\zeta^{\nu}$. In turn, 
similarly to the charge accretion rate~\eqref{Charge_Accretion}, 
the flux of this current defines the accretion rate, and for the spherically-symmetric 
setup it gives\footnote{\label{footnote_rate}This definition of accretion rate is general-relativistic and also used e.g. in \cite{Frolov:2004vm, Babichev:2013vji}, but it is different from the one used in \cite{Feng:2021qkj} with which it coincides only in the limit of a non-relativistic speed of sounds at spatial infinity. This mismatch is caused by different interpretations of the "mean baryon
mass" from Appendix G of \cite{Shapiro:1983du}. Indeed, \cite{Feng:2021qkj} interprets it as a mass $m$ from \eqref{eq:Potential}, whereas \eqref{AccretionGenRel} implies that the role of the "mean baryon
mass" is played by $\dot\varphi_0$ – the chemical potential at the spatial infinity. Note that this "mean baryon
mass" is absent in (G.2) or (G.6), Appendix G of \cite{Shapiro:1983du} and appears in (G.21) purely as an integration constant. Of course, for $m^2>0$ and non-relativistic speeds of sound at spatial infinity, $\dot\varphi_0\approx m$, see Eq.~\eqref{C_s} and discussion in \cite{Hui:2019aqm}. However, \cite{Shapiro:1983du} considers the baryon number to be conserved, whereas here it is the $U(1)$-charge which is conserved and not the number of particles. Yet, $\left|m\right|$ is the mass per particle around vacuum, whereas $\dot\varphi_0$ is the mass per unit $U(1)$-charge. Thus, for generic speeds of sound at infinity one should use $\dot\varphi_0$ as the "mean baryon mass" from \cite{Shapiro:1983du}.}
\begin{equation}
\dot{M} \equiv - 4 \pi r^2 T^{r}_{t} \, .
\label{AccretionGenRel}
\end{equation}
It follows then from the energy-momentum tensor~(\ref{cs emt}), the steady-state ansatz~(\ref{ansatz}), 
the $U(1)$-charge conservation, and the charge accretion rate \eqref{Charge_Accretion}, that
this rate is constant,
\begin{equation}
 \dot{M} =-4 \pi r^2\,\dot{\varphi}\,J^{r}=-4 \pi r^2n \dot{\varphi}\,u^{r}=\dot{\varphi}_0\,\dot{Q}
	\, .
\label{accretion_rate_M/Q}
\end{equation}
This expression reaffirms the physical interpretation of $\dot{\varphi}_0$ as the chemical potential at infinity. 
Note that the \emph{stationary} accretion rate above does not involve off-diagonal "dissipative" terms $q_{\mu}$ and $\Pi_{\mu\nu}$, but only the perfect fluid part of the EMT \eqref{cs emt}, \eqref{T decomp}, diagonal in the $u_{\mu}$-basis, contributes to it. 
It is convenient to rewrite the above accretion rate as 
\begin{equation}
 \dot{M}=4\pi r_{\scr S}^{2}\,\dot{\varphi}_{0}^{2}\,\rho_{\scr S}^{2}=4\pi r_{\scr S}^{2}\,
    \Bigl( \frac{\dot{\varphi}_{0}^{4}}{\lambda} \Bigr) \,\beta^{2}
	\, .
\label{general accretion rate}
\end{equation}
The dimensionless parameter $\beta$ is not independent, 
but is instead fixed by the steady-state assumption.
It is typically of order~$\mathcal{O}(1)$, as will be shown 
explicitly in subsequent sections. For instance, this can
already be seen in the~$P(X)$ 
limit~\cite{Frolov:2004vm,Babichev:2005py} (also cf.~\cite{Feng:2021qkj}), discussed in
Sec.~\ref{sec: EFT of the complex scalar} as the limit~$\xi^2\!\gg\!1$.
In this limit,~$\beta$ can be express using \eqref{C_s} as 
\begin{equation}
\beta^{2}
    =
    \Bigl( \frac{2c_{s}^{2}}{1 \!-\! 3c_{s}^{2}} 
        \Bigr)_{\!r_{\scr S}}
    \Bigl( \frac{1 \!-\! 3c_{s}^{2}}{1 \!-\! c_{s}^{2}} 
        \Bigr)_{\!\infty}
	\, .
\label{general_beta_C_s}
\end{equation}
Here in the first parentheses the sound speed is evaluated at 
the BH horizon, while in the second it is evaluated at spatial infinity.

The fact that the accretion rate in~(\ref{general accretion rate})
is inversely proportional to the quartic self-coupling 
constant~$\lambda$ is a consequence of the test-field 
approximation. This can be inferred from the 
action~(\ref{csAction}) already: upon an appropriate rescaling of the fields in~\eqref{dimensionless def},~$\lambda$ appears only as an inverse overall factor, similarly to the EFT 
case given by~(\ref{EFTexample}). The inverse dependence 
on~$\lambda$ in~(\ref{general accretion rate}) implies that, 
for a fixed chemical potential, one cannot take~$\lambda$ arbitrarily small  without violating the 
test-field approximation. Since~$\beta$ is dimensionless, the 
last equality in~(\ref{general accretion rate}) shows that the 
accretion rate is proportional to the characteristic energy 
density at spatial infinity,~$\dot{\varphi}_{0}^{4}/\lambda$, 
and to the area of the horizon, $4\pi r_{\scr S}^2$. While in the~$P(X)$ limit~$\beta$ depends only on~$\mu$, in the general complex-scalar case we will show that $\beta \!=\! \beta(\xi,\mu)$, so that it depends on both~$\mu$ and~$\xi$.
We find the difference between the accretion rates to be  significant for a large range of parameters.
\\
\\
Sometimes the formulation of accretion rate in terms of the $\dot{\varphi}_{0}^{2}$ may be inconvenient and one would prefer the energy density at infinity instead. Using 
\eqref{energy density} one obtains for our boundary conditions $\epsilon_{\infty}=\frac{1}{2}\rho^{2}_{\infty}\dot{\varphi}^{2}_0+V\left(\rho_{\infty}\right)$, where $\rho_{\infty}=\dot{\varphi}_{0}\sigma_{+}/\sqrt{\lambda}$, with $\sigma_{+}$ from \eqref{boundary conditions}. Solving with respect to $\dot{\varphi}_{0}$  results in 
\begin{align}
\label{connection phi/E}
\text{without spontaneous symmetry breaking:}
	\qquad & \dot{\varphi}_{0\left[+\right]}^{2}=\frac{m^{2}}{3}\left(1+2\sqrt{1+3\lambda\frac{\epsilon_{\infty\left[+\right]}}{m^{4}}}\right)	\, , 
\\ 
\text{for vanishing mass parameter:}
	\qquad & \dot{\varphi}_{0}^{2}=2\sqrt{\frac{\lambda\epsilon_{\infty}}{3}}
			\, ,
\\
\text{for spontaneous symmetry breaking:}
	\qquad & \dot{\varphi}_{0\left[-\right]}^{2}=\frac{\left|m^{2}\right|}{3}\left(\sqrt{1+12\lambda\frac{\epsilon_{\infty\left[-\right]}}{m^{4}}}-1\right)
		\, . 
\end{align}
These expressions are useful to translate the results from $\mu$ and $\dot{\varphi}_{0}$ to energy density at infinity $\epsilon_{\infty}$ and mass $m$.

\subsection{Validity of test-field approximation and classical description}
\label{sec: Validity_Test_Field}

Backreaction in accretion and the applicability of test-field approximation were studied 
in~\cite{Babichev:2012sg,Kimura:2021dsa}. Nonetheless, it is useful to provide simple 
estimates of when the test-field approximation can be assumed. For simplicity, we will 
assume that the self-interaction energy is parametrically of the same order as the energy in 
the mass term. For the hydrodynamic limit this corresponds to a sound speed of order one.

Several criteria can be considered when assessing the applicability of the test-field approximation. First, one can require that (i) the BH does not grow substantially during 
a Hubble time, $\dot{M}H^{-1} \!\lesssim\! M$. If the scalar field makes a non-negligible contribution to the current cosmic energy density, then the Hubble parameter at spatial infinity is given by $H^{2}\simeq G\dot{\varphi}_0^{4}/\lambda$ (see definition of $\sigma$ in \eqref{dimensionless def}). 
For $\beta \!=\! \mathcal{O}(1)$ the accretion rate 
in~(\ref{general accretion rate}) yields
\begin{equation}
\label{our_condition}
\lambda\gtrsim G^{3}M^{2}\dot{\varphi}_0^{4}\simeq\xi^{2}\,G\,\dot{\varphi}_0^{2}\,,
\end{equation} 
which, for\footnote{As we will discuss later, see \eqref{mu_v_c}, these two scales generically differ by $\mathcal{O}\left(1\right)$, at least for $P(X)$.} 
$m \!\simeq\! \dot{\varphi}_0$ is stronger than \eqref{eq:lambda_condition_Feng_Colpi} by a factor of $\xi^2$, which was used in~\cite{Feng:2021qkj}. Thus, 
requiring weak coupling, $\lambda\!\lesssim\!1$, this condition yields the largest 
possible $\xi$ compatible with the test field approximation: $\xi \!\lesssim\! M_{\rm \scr Pl}/\dot{\varphi}_{0}$. Neglecting the difference between the scales $m$ and $\dot{\varphi}_0$, 
and requiring weak coupling, we obtain the 
condition~$m\!\lesssim\! M_{\rm \scr Pl}\sqrt{M_{\rm \scr Pl}/M}$. 
For a solar mass BH this yields $m \!\lesssim\!1\,\text{GeV}$, 
for a supermassive BH of $10^{6}\,M_{\odot}$ one has $m\!\lesssim\!1\,\text{MeV}$, 
while for a PBH of $10^{21}\,\text{g}$, one finds $m \!\lesssim\!10^{3}\,\text{TeV}$, which can include new high-energy physics beyond the Standard Model. By requiring that 
quantum-gravitational effects of BH are negligible, $M \!\gg\! M_{\rm \scr Pl}$,
 for $m \!\simeq\! \dot{\varphi}_0$, our condition  \eqref{our_condition} also implies an absolute lower bound
\begin{equation}
\lambda\gg G^{2}m^{4}\,,
\end{equation}
which is even weaker than \eqref{eq:lambda_condition_Feng_Colpi}.

Demanding negligible BH growth on cosmological time scales might be too restrictive,
and one might instead require that (ii) accretion be negligible over the light-crossing time $\dot{M}r_{\scr S} \!<\! M$. Here we assume that the sound speed is of order unity, so that 
characteristic scales of the accretion configuration are parametrically the same as $r_{\scr S}$. 
Surprisingly, this requirement produces the same condition~(\ref{our_condition}). 

Finally, one can require that (iii) the characteristic curvature of the black hole (the square root of the Kretschmann invariant) be larger than the curvature due to the complex scalar field, which is neglected in the test-field approximation. This yields the following condition at radius $r$:
\begin{equation}
\frac{M}{r^{3}}>\frac{\dot{\varphi}_0^{4}}{\lambda}\,,
\end{equation}
where $G$ dependence cancels out, as we are comparing curvatures due to $M$ and $\Psi$. Thus, one can neglect the curvature due to $\Psi$ only for radii satisfying 
\begin{equation}
\label{MaxRad}
r\lesssim R_K
	\simeq
	\Big( \lambda\frac{M}{\dot{\varphi}_0^{4}} \Big)^{\!1/3}\,. 
\end{equation}
Requiring that $R_K$ is larger than the characteristic scale of the accreting configuration, 
which, for sound speeds of order unity, is $r_{\scr S}$, one again obtains \eqref{our_condition}. 
Thus, all three criteria (i)-(iii) reduce to the same \emph{lower bound} on the self-interaction coupling constant \eqref{our_condition}. 

\bigskip

One can also obtain an \emph{upper bound} on~$\lambda$ by requiring that the occupation numbers are sufficiently large so that the classical 
approximation of the accretion configuration is valid. Namely, 
for $\xi \!=\! \dot{\varphi}_0r_{\scr S} \!\simeq\! mr_{\scr S} \!\gg\! 1$ 
the particles of the radial field\footnote{Note that to find accretion we solve the classical equation  \eqref{dimless profile eq} for the modulus field.} with wavelengths of characteristic scale of accretion $r_{\scr S}$ are non-relativistic. Therefore, the occupation number on this scale can be estimated as 
\begin{equation}
\label{occupation}
n_{S}\simeq\frac{\dot{\varphi}_0^{4}}{\lambda}\,r_{\scr S}^{3}\,\frac{1}{m}\,.
\end{equation}
Requiring that $n_{S} \!\gg\! 1$ yields $\lambda \!\ll\! G^{3}M^{3}m^{3} \!\simeq\! {\xi^3}$, which is trivially satisfied by assumption.
In the opposite limit $\xi \!=\! \dot{\varphi}_0r_{\scr S} \!\simeq\! mr_{\scr S} \!\ll\! 1$, 
particles of the modulus field are relativistic with energy $r_{\scr S}^{-1}$. 
In that case, the occupation number can be estimated as 
\begin{equation}
n_{S}\simeq\frac{\dot{\varphi}_0^{4}}{\lambda}\,r_{\scr S}^{3}\,\frac{1}{r_{\scr S}^{-1}}\,,
\end{equation}
so that classicality $n_{S}\gg 1$ gives 
\begin{equation}
\label{upper_bound_small_xi}
\lambda\ll G^{4}M^{4}m^{4}\,.
\end{equation}
This is easily compatible with \eqref{our_condition}, since the naive condition for 
negligible quantum-gravitational effects requires $M\!\gg\! M_{\scr \rm Pl}$. The upper bound \eqref{upper_bound_small_xi} can be written as $\lambda\ll \xi^4$, providing a strong constraint on quartic self-interaction for $\xi\ll 1$. 

Cases in which $\dot{\varphi}_0$ is parametrically different from $m$ correspond to an accretion length scale parametrically different from $r_{\scr S}$ and require a more sophisticated analysis, beyond the scope of the present discussion.

\section{Accretion for $P(X)$}
\label{sec: Accretion for P(X)}

The complex scalar model~(\ref{CS action}) behaves very much like a~$P(X)$ model when 
gradients of its modulus are negligible. At the level of the profile 
equation~(\ref{dimless profile eq}), this formally corresponds to the limit~$\xi\!\to\!\infty$,
in which the gradient terms are neglected,
\begin{equation}
	\mathcal{U}' \bigl( \mu^2, \beta^2; f, \sigma \bigr)
	=
	\mu^2 \sigma + \sigma^3
	- \frac{\sigma}{f} \biggl[ 1 - \frac{ (1\!-\!f)^4 \beta^4 }{ \sigma^4 } \biggr]
	= 0
	\, .
\label{P(X) profile equation}
\end{equation}
This is the profile equation of the~$P(X)$ model, in which the dependence on the black 
hole mass appears only through a rescaling of the coordinates. Steady-state accretion 
in this model was studied previously by Frolov~\cite{Frolov:2004vm}. We find it useful 
to recast his analysis in terms of quantities that are more suitable for the discussion 
in subsequent sections.\footnote{
Making the substitutions~$M^4\!\to\! \dot{\varphi}_0^4/(2\lambda)$ 
and~$A \!\to\! \mu^2$ in Frolov's paper~\cite{Frolov:2004vm} reproduces 
Eq.~(\ref{P(X) profile equation}).
}
This reformulation facilitates direct comparison with the complex scalar model 
and clarifies the deviations between the two.

\begin{figure}[h!]
\centering
\includegraphics[width=4.9cm]{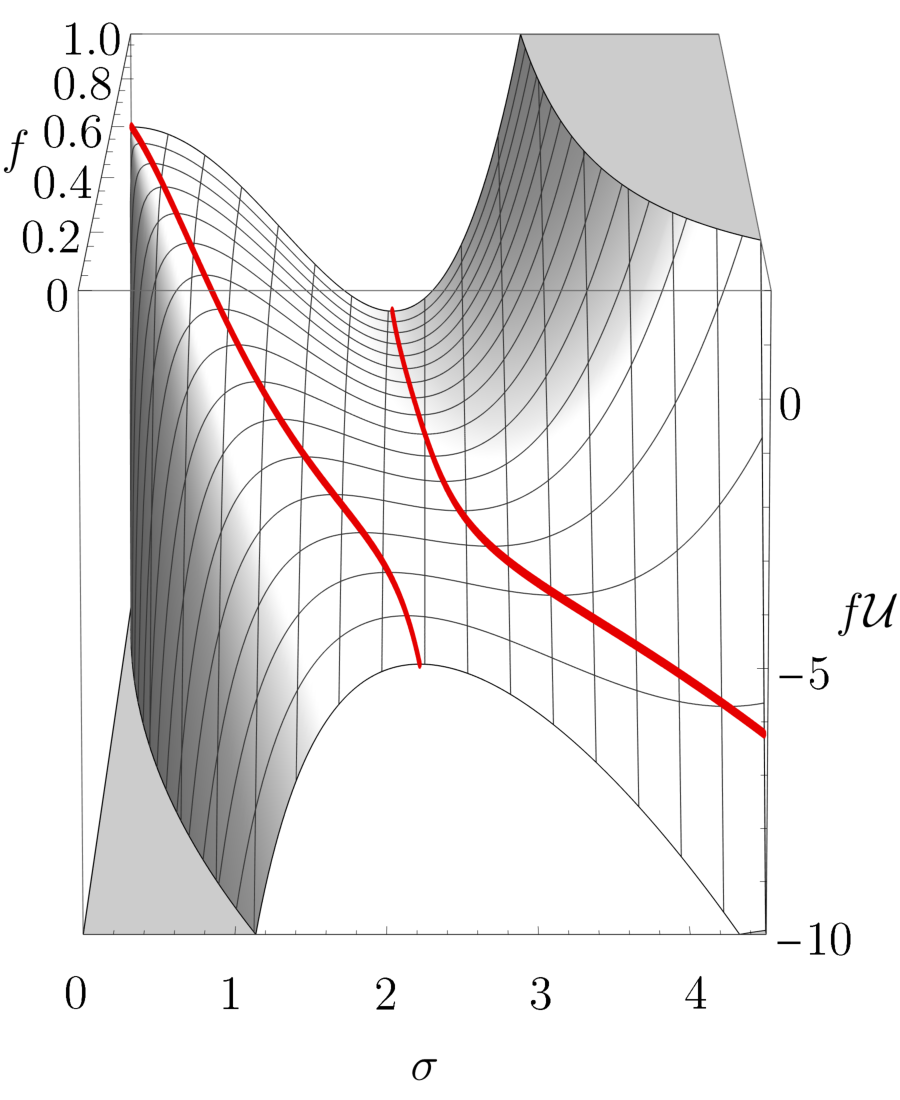}
\hfill
\includegraphics[width=4.9cm]{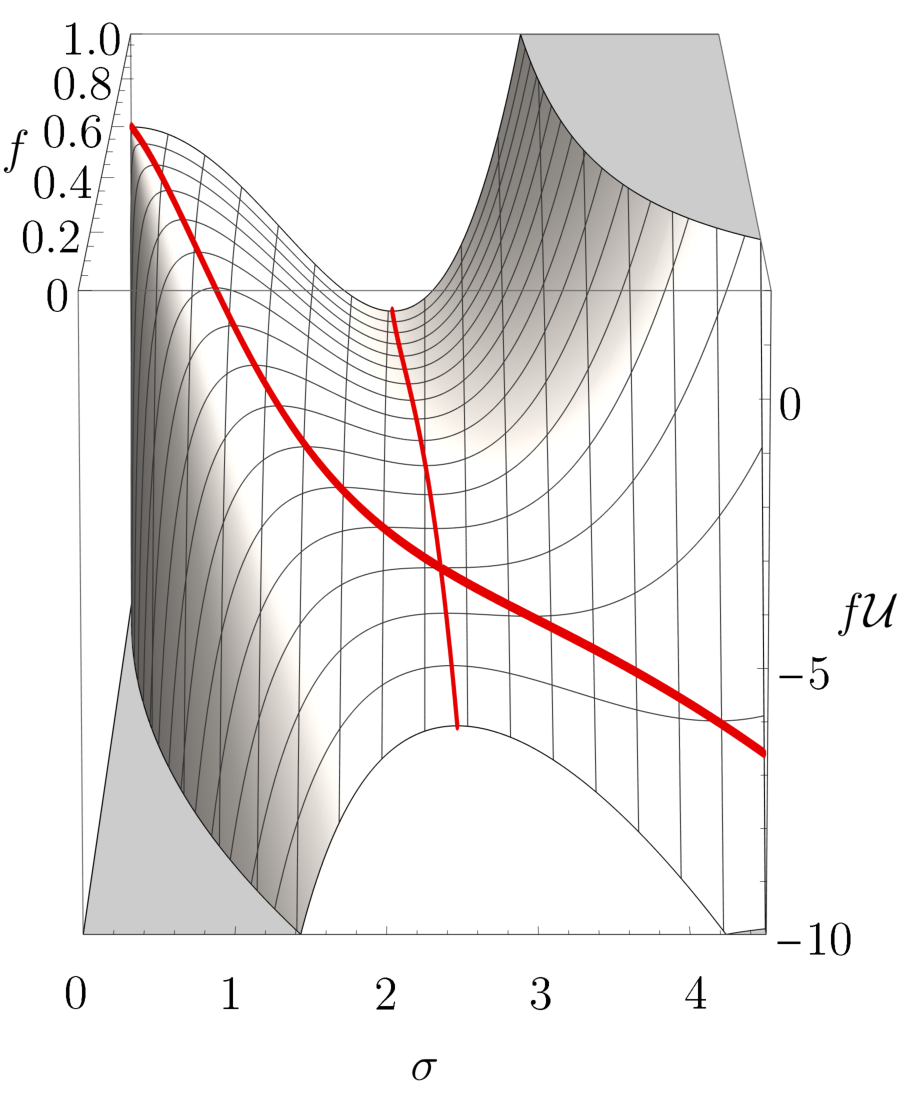}
\hfill
\includegraphics[width=4.9cm]{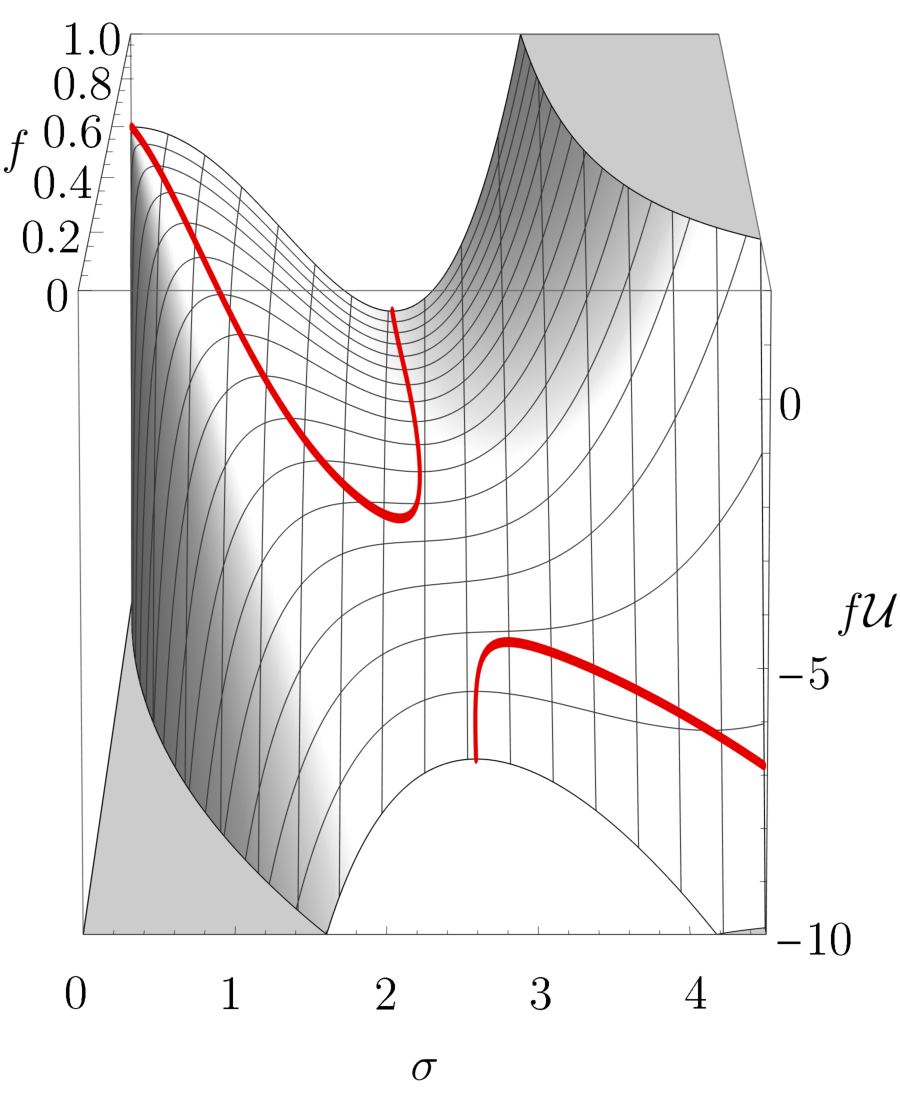}
\vskip-3mm
\caption{Effective potential landscape with red curves solving the 
algebraic~$P(X)$ profile equation~(\ref{P(X) profile equation}):
{\it left:} subcritical case, in which the curves connect the boundaries, but not the correct boundary points;
{\it middle:} critical case where two curves touch tangentially connecting
the correct boundary conditions;
{\it right:} supercritical case, in which the curves do not connect the two boundaries.
}
\label{PX landscape}
\end{figure}

The profile equation~(\ref{P(X) profile equation}) can be regarded as a cubic equation 
for~$\sigma^2$. However, following~\cite{Frolov:2004vm}, rather than applying the cubic 
formula directly, it is more instructive to first examine the corresponding phase-space diagram. 
The profile equation implicitly defines algebraic curves in the~$(\sigma,f)$ plane, 
from which considerable information can be extracted by analyzing their endpoints 
and how these curves connect. It is straightforward to show, by asymptotic analysis, 
that equation~(\ref{P(X) profile equation}) admits two positive real solutions near the horizon,
\begin{equation}
\sigma 
	\ \overset{f\to 0}{\longsim} \ \beta \, \sqrt{ 1 - \frac{1}{2} \bigl( 4 \!-\! \mu^2 \!-\! \beta^2 \bigr) f }
	\xrightarrow{f\to0} \beta
	\, ,
\qquad \text{or} \qquad
\sigma
	\ \overset{f\to 0}{\longsim} \  \sqrt{ \frac{1}{f} - \mu^2 }
	\xrightarrow{f\to 0} \infty 
	\, ,
\label{bc f0}
\end{equation}
and likewise two distinct real solutions at radial infinity,
\begin{equation}
\sigma
	\ \overset{f\to 1}{\longsim} \ \sqrt{ \frac{1}{f} - \mu^2 }
	\xrightarrow{f\to 1} \sqrt{ 1 \!-\! \mu^2 } \, ,
\qquad \quad \text{or} \qquad \quad
\sigma
	\ \overset{f\to 1}{\longsim} \ \frac{ (1 \!-\! f) \beta }{ (1 \!-\! \mu^2f)^{\frac{1}{4}} }
	\xrightarrow{f\to 1} 0 \, .
\label{bc f1}
\end{equation}
The manner in which these endpoints connect is determined by the parameter~$\beta$, 
which dictates the shape of the effective potential. The three qualitatively distinct 
cases for the potential and corresponding solutions are illustrated in Fig.~\ref{PX landscape}.
\begin{figure}[h!]
\vskip+7.mm
\centering
\includegraphics[width=14.5cm]{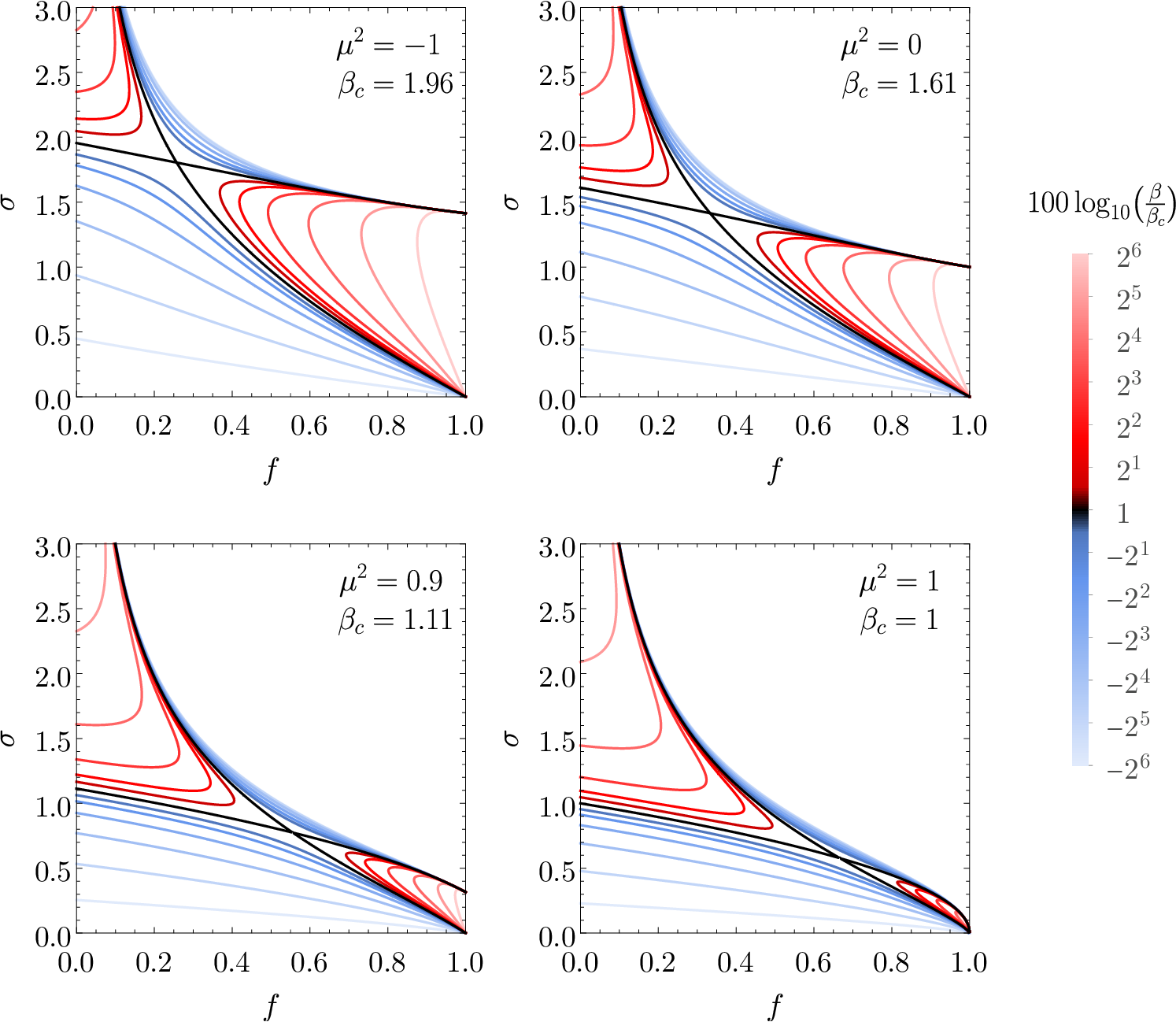}
\vskip-1mm
\caption{Phase diagrams of solutions of the~$P(X)$ profile equation~(\ref{P(X) profile equation}) 
for four different choices of~$\mu^2$. The finite boundary 
conditions~(\ref{boundary conditions}) are connected only in special cases when 
the solution curves intersect at a singular point. This happens for special values 
of~$\beta\!=\!\beta_c$, shown in black. Curves for~$\beta<\beta_{c}$ are shown
in blue and curves for~$\beta>\beta_{c}$ are shown in red. }
\label{phase diagrams}
\end{figure}

All curves satisfying~(\ref{P(X) profile equation}) terminate at some of the
asymptotic roots. The way the solution curves connect the asymptotic roots depends 
on the number of singular (or critical) points that 
Eq.~(\ref{P(X) profile equation}) admits. These are the points where the concept of a tangent 
vector ceases to be defined in the usual sense. In our case, 
critical points are characterized by
\begin{equation}
\frac{\partial}{\partial \sigma} \mathcal{U}' \bigg|_{c} = 0 \, ,
\qquad
\qquad
\frac{\partial}{\partial f} \mathcal{U}' \bigg|_{c} = 0 \, .
\label{singular point}
\end{equation}
At such points, the implicit function theorem fails, indicating that distinct curves intersect. 
Another way to understand the necessity for the critical point is by requiring finite $d\sigma/df$ throughout the configuration. From \eqref{P(X) profile equation} one obtains 
\begin{equation}
\frac{\partial\mathcal{U}'}{\partial\sigma}\,\frac{d\sigma}{df}+\frac{\partial\mathcal{U}'}{\partial f}=0\,,
\end{equation}
yet boundary conditions \eqref{boundary conditions} imply that $\partial\mathcal{U}'/\partial\sigma>0$ on the horizon and $\partial\mathcal{U}'/\partial\sigma<0$ at spatial infinity. Thus, by continuity, there should exist some $\left(f_{c},\sigma_{c}\right)$ where $\partial\mathcal{U}'/\partial\sigma=0$. Finiteness of $d\sigma/df$ then implies the second of equations \eqref{singular point}, see also Appendix \ref{app: Derivatives at the critical point}. Critical-point equations \eqref{singular point}, together with the profile 
equation~(\ref{P(X) profile equation}), provide three conditions that define the 
critical points:
\begin{align}
\mu^2 + \sigma_{c}^2 - \frac{1}{ f_{c} } 
	\biggl[ 1 - \frac{ (1\!-\! f_{c} )^4 \beta_c^4 }{ \sigma_{c}^4 } \biggr] 
	={}&
	0
	\, ,
\\
\mu^2 + 3 \sigma_{c}^2 - \frac{1}{ f_{c} } 
	\biggl[ 1 + \frac{ 3(1\!-\! f_{c} )^4 \beta_c^4 }{ \sigma_{c}^4 } \biggr] 
	={}&
	0
	\, ,
\\
1 - \frac{ (1 \!-\! f_{c} )^3 (1 \!+\! 3 f_{c} ) \beta_c^4 }{ \sigma_{c}^4 } 
	={}&
	0
	\, .
\end{align}
These equations specify the critical point location~$(f_c,\sigma_c)$ and the 
critical parameter~$\beta_c$ for which such a point exists. There is only one such 
point, located at\footnote{In the limit~$\mu^2\!=\!1$, another critical point appears 
at~$f\!=\!1$,~$\sigma\!=\!0$,~$\beta\!=\!0$; this marginal case is not considered here.}
\begin{equation}
f_{c} = \frac{1}{6 \mu^2} \Bigl[ 6 - \mu^2 - \sqrt{ \mu^4 - 36 \mu^2 + 36 } \, \Bigr] \, ,
\qquad \quad
\sigma_{c} = 
    \sqrt{\frac{2 (1\!-\!f_c) }{ f_c(1\!+\!3f_c) }}
  \, ,
\label{critical f sigma}
\end{equation}
which occurs when the dimensionless flux parameter takes the critical value
\begin{equation}
\beta^2_c = 
     \frac{2 }{\sqrt{f_{c}^2 ( 1 \!-\! f_{c} ) ( 1 \!+\! 3 f_{c} )^3 }}
    \, .
\label{critical beta}
\end{equation}

\begin{figure}[h!]
\vskip+3mm
\centering
\includegraphics[width=15cm]{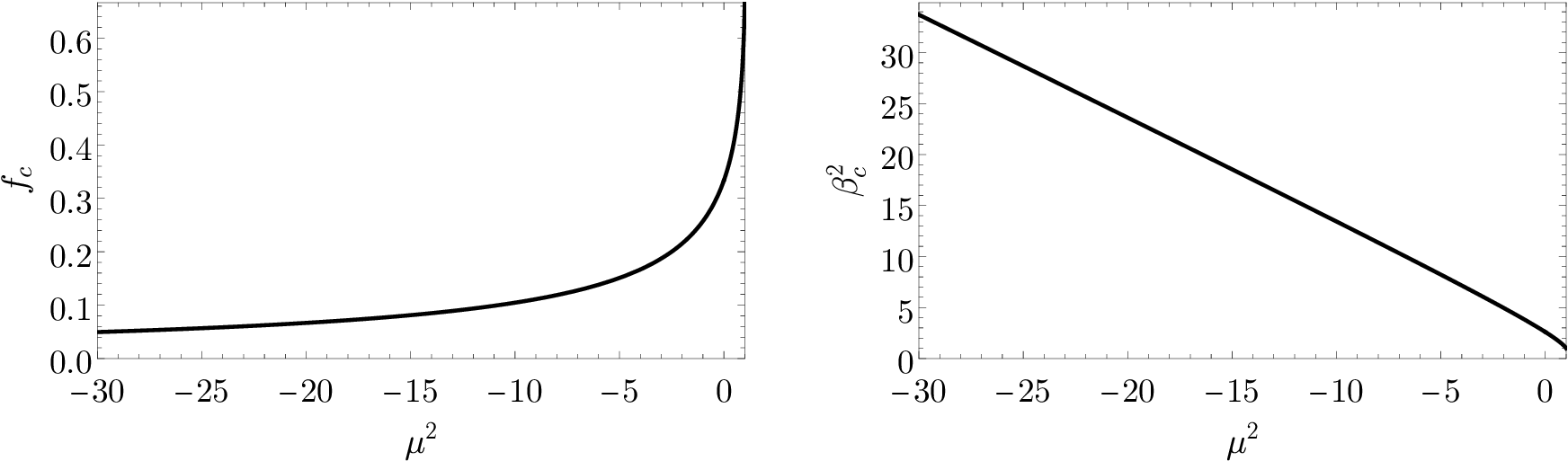}
\vskip-2mm
\caption{Dependence of $f_{c}$ and $\beta_{c}^{2}$ on the mass parameter $\mu^{2}$ in the range $-30 \leq \mu^{2} \leq 1$.}
\label{criticals}
\end{figure}

For any other value of the dimensionless flux parameter, there are no critical points and, consequently, 
no intersections of solution curves. Since each asymptotic root must be connected 
to at least one curve, it follows that two distinct curves exist for any non-critical~$\beta$, 
neither of which connects the required boundary conditions~(\ref{boundary conditions}). 
Such a connection is only possible when the curves intersect, which occurs exclusively 
for the critical value~(\ref{critical beta}). This behavior is illustrated in more detail 
in Fig.~\ref{phase diagrams}, while the dependence of the critical radius and the critical 
flux on the mass parameter~$\mu^2$ is shown in Fig.~\ref{criticals}. 

It is worth mentioning that the blue curves located below both black separatrices in Fig.~\ref{phase diagrams} start at spatial infinity with vanishing $\sigma$ thereby violating the validity of polar decomposition. Moreover, for $m^2>0$ this also corresponds to vanishing energy density at spatial infinity, which violates one of our main assumptions. On the other hand, the physical accretion setup is only applicable up to a finite distance away from the BH, see e.g. discussion around \eqref{MaxRad}. Moreover, a BH is usually not in a complete isolation, but is a part of the galaxy and is surrounded by other stars. Thus, it is not unreasonable to exclude the spatial infinity from the consideration and consider accretion configurations only up to a finite radius, as done in \cite{Hui:2019aqm}. In this setup the blue solutions below both black separatrixes in Fig.~\ref{phase diagrams} could be relevant. However, similarly to solutions in \cite{Hui:2019aqm}, it is unclear which physical processes would fix the peculiar boundary conditions needed to maintain stationary flow from this finite distance.

\begin{figure}[h!]
\vskip+3mm
\centering
\includegraphics[width=15cm]{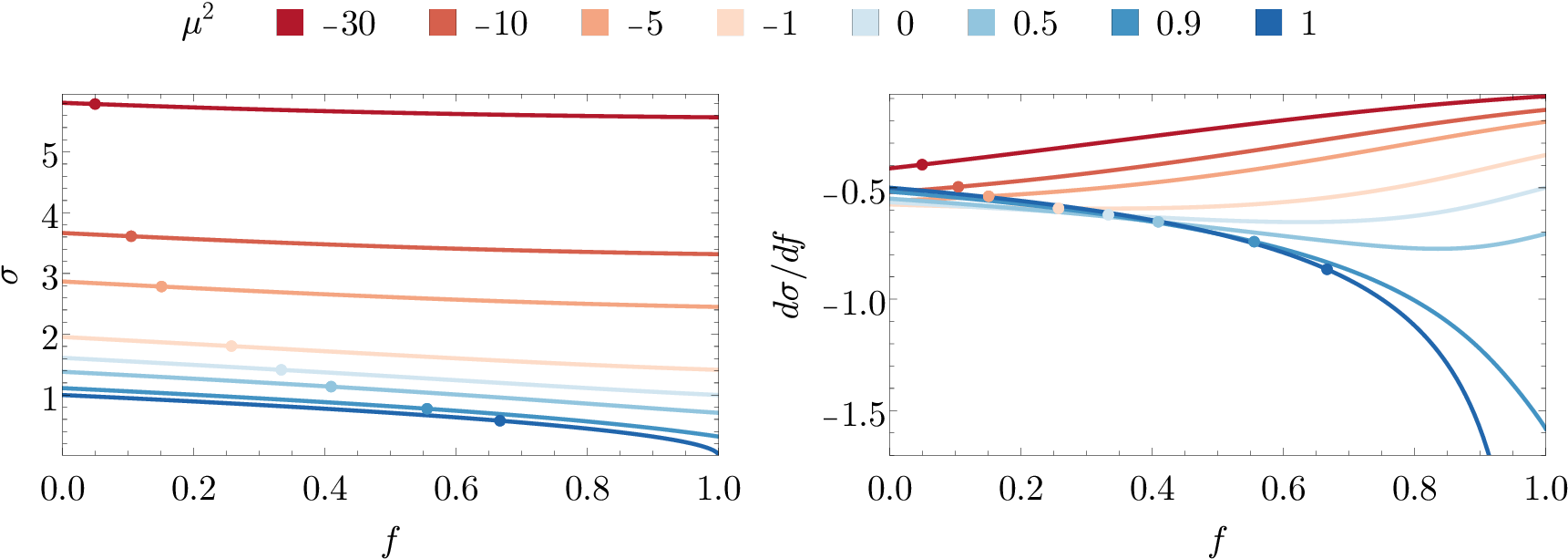}
\caption{Dimensionless modulus-field profile~$\sigma_{0}(f)$ {\it (left)} 
and its $f$-derivative {\it (right)} for the~$P(X)$ model, plotted for different values 
of the mass parameter~$\mu^2$, including tachyonic values.
Bullet points on the curves represent the location 
of the critical point, which is the acoustic or sound horizon, see e.g. \cite{Moncrief,Frolov:2004vm}.
}
\label{P(X)figureProfiles}
\end{figure}

Having determined the critical value of~$\beta$ in~(\ref{critical beta}), the scalar-modulus
profiles can now be obtained from the cubic equation~(\ref{P(X) profile equation}).
These profiles, together with their derivatives, are shown in the upper panels of Fig.~\ref{P(X)figureProfiles}
for different values of the mass parameter. The lower panels display the equation of 
state parameter and the speed of sound, again for various values of~$\mu^2$. The bullet points
on the curves denote the position of the acoustic horizon, which coincides with the 
critical point~$f_c$ in~(\ref{critical f sigma}), whose existence is guaranteed by 
the flux~$\beta$ taking its critical value. The fact that the dimensionless flux~$\beta$ 
is fixed by imposing specific boundary conditions on the solution suggests that the 
profile equation~(\ref{P(X) profile equation}) represents a non-linear eigenvalue problem. 
This concept will generalize naturally to the complex scalar model discussed in 
the following sections. Expressions~\eqref{critical beta} 
and~\eqref{critical f sigma} determine accretion rate in terms 
of~$\mu^2$, which physically is the ratio between the square of the mass parameter to the squared chemical potential at infinity,~$m^2/\dot\varphi^2_0$.

\bigskip

It is instructive to relate our description of accretion to the usual hydrodynamic picture used in~\cite{Bondi:1952ni,Michel:1972oeq,Moncrief,Shapiro:1983du,Babichev:2004yx,Babichev:2005py,Babichev:2008dy,Babichev:2008jb,Babichev:2012sg,Babichev:2013vji,Feng:2021qkj}, in order to clarify the physical meaning of the critical point.
critical point. Namely, using the expressions for the sound speed~(\ref{C_s}) and for the physical velocity~(\ref{definition_vc}), and evaluating them at the critical point defined 
by~(\ref{critical f sigma}) and~(\ref{critical beta}),
one obtains
\begin{equation}
c_{s}^{2}(f_{c}) =
    \frac{1 \!-\! f_{c}}{3 (1 \!-\! f_{c} )+\mu^{2} f_{c} (1 \!+\! 3f_{c} )} \, ,
    \qquad  \text{and} \qquad 
v^{2}(f_{c}) =
    \frac{1 \!-\! f_{c}}{1 \!+\! 3f_{c}} \,.
\label{critical C_s and v}
\end{equation}
A short algebraic manipulation then shows that the squares of both speeds take the same value,
\begin{equation}
c_{s}^{2}(f_{c})=v^{2}\left(f_{c}\right)\equiv v^2_c \, ,
\label{acoustic horizon}
\end{equation}
equal to the square of the {\it critical speed}~$v_c$.
This shows that the critical point represents 
the \emph{acoustic horizon} for the $P(X)$ EFT given by \eqref{specific P}. 

It follows from~(\ref{critical C_s and v}) that the position of the acoustic horizon
in the Schwarzschild radial coordinate can be expressed in terms of the critical speed~$v_c$ from~(\ref{acoustic horizon}),
\begin{equation}
r_{c} = \frac{r_{\scr S}}{4} \Bigl( 3 + \frac{1}{v_{c}^{2}} \Bigr) \, .
\label{acoustic horizon radius}
\end{equation}
Therefore, for potentials without symmetry breaking, we have that~$r_c \!>\! 3r_{\scr S}/2$,
while for symmetry-breaking potentials, we have~$r_c \!<\! 3r_{\scr S}/2$,
as also seen from Fig.~\ref{criticals}, where the border between the two cases corresponds to $\mu^2=0$ and $f_c=1/3$.

By inverting~(\ref{critical C_s and v}) for~$f_c$, and substituting it into 
the critical flux~(\ref{critical beta}), we can also express the accretion rate in terms of the critical speed,
\begin{equation}
\label{dot_M_phi}
\dot{M} =
    2 \pi M^2 G^2
    \Bigl( \frac{\dot{\varphi}_{0}^{4}}{\lambda} \Bigr)
    \,
    \frac{1}{|v_{c}|}\,
    \frac{ ( 1 \!+\! 3v_{c}^{2} )^3 }{ 1 \!-\! v_{c}^2 }
    \, .
\end{equation}
This result for the steady-state accretion rate~(\ref{dot_M_phi}) applies to cases both with 
and without symmetry breaking and can be considered as a differential equation for $M$. 
Naively, this accretion rate diverges when $v_{c}\!\rightarrow\! 0$ or $v_{c}\!\rightarrow \! -1$. 
However, it should be stressed that $\dot{\varphi}_{0}$ and $v_{c}$ are not independent 
parameters, and we will see below how this apparent issue is avoided.
Crucially, the accretion rate given by~\eqref{dot_M_phi} is manifestly finite in the relativistic, or massless, limit $m^2\!=\!0$, 
in which~$v^2_c\!=\!1/3$,
\begin{equation}
\dot{M} \xrightarrow{m \to 0}
    4\pi r_{\scr S}^2
    \Bigl( \frac{\dot{\varphi}_{0}^{4}}{\lambda} \Bigr)
    \,
    \Big( \frac{3\sqrt{3}}{2} \Big) \, ,
\label{AccretionRelativisticLimit}
\end{equation}
where the last factor corresponds to the value of $\beta^2$. This behavior is consistent 
with findings in~\cite{Frolov:2004vm,Babichev:2005py},
but naively contradicts~\cite{Feng:2021qkj} which reports a \emph{diverging} accretion rate for $v^2_c\!\rightarrow\!1/3$. 

To make a closer comparison between our result~(\ref{dot_M_phi}) and the literature, 
we first use~(\ref{critical f sigma}) and~(\ref{C_s}) to obtain
\begin{equation}
\label{mu_v_c}
\mu^{2} = \frac{ ( 1 \!-\! 3v_{c}^2 ) ( 1 \!+\! 3v_{c}^2 ) }{ 1 \!-\! v_{c}^2 } \, .
\end{equation}
Using this formula together with \eqref{C_s}, we can express the critical velocity 
in terms of the speed of sound at the spatial infinity
\begin{equation}
\label{critical_vs_c_S}
v_{c}^{2}=\frac{1-3c_{s}^{2} (\infty) + \sqrt{1+66c_{s}^{2} (\infty) 
	- 63c_{s}^{4} (\infty) } }{18 \big[ 1-c_{s}^{2} (\infty ) \big] }\, .
\end{equation}
The same relation was obtained in~\cite{Feng:2021qkj}, where it was realized that even 
for $c_{s}^{2}( \infty ) \!\rightarrow\! 0$ the critical velocity does not vanish, but is still relativistic: $v_{c}^{2} \!\rightarrow\! 1/9$. In fact, this is the lowest value of the critical 
velocity for the system, and the limit $v_{c}\!\rightarrow\! 0$ cannot be taken, thereby avoiding the naive divergent behavior of~(\ref{dot_M_phi}) mentioned above. Moreover, this established $v_{c}>1/3$
implies that $r_c<3 r_{\scr S}$ on top of the limits mentioned below \eqref{acoustic horizon radius}. 

On the other hand, one can use \eqref{mu_v_c} to translate
the dependence on the boundary data~$\dot{\varphi}_0$ in~(\ref{dot_M_phi}) 
to the dependence on the parameter of the model, $m$:  
\begin{equation}
\dot{M}=2\pi G^{2}M^{2}\,
	\Big( \frac{m^{4}}{\lambda} \Big)
	\,\frac{1-v_{c}^{2}}{|v_{c}|}\,
	\frac{ (1+3v_{c}^{2} )}{ ( 1-3v_{c}^{2} )^{2}}\,,
\label{dot_M_m}
\end{equation}
which is again a differential equation for $M$. Here it is clearly seen that the accretion rate 
does not diverge for~$v_c \!\rightarrow\! -1$, which via \eqref{critical_vs_c_S} corresponds to $c_{s}^{2}(\infty) \!\rightarrow\! 1$. On the contrary, the accretion rate vanishes in 
this limit. This happens because, in order to achieve $v_c \!\rightarrow\! -1$, one has to assume a tachyonic mass $m^2\!<\!0$, so that $\dot{\varphi}_{0}\!\rightarrow \!0$ and the field approaches the spontaneously broken vacuum which does not accrete.  
Moreover, \eqref{dot_M_m} agrees with expressions reported 
previously~\cite{Frolov:2004vm,Babichev:2005py}, 
but it deviates\footnote{
In our notation, Eq.~(2.19) from~\cite{Feng:2021qkj} reads
\begin{equation*}
\dot{M}=2\pi G^{2}M^{2}
	\Big( \frac{m^{4}}{\lambda} \Big)
	\sqrt{\frac{1-v_{c}^{2}}{v_{c}^{2}}} 
	\Big( \frac{1+3v_{c}^{2}}{1-3v_{c}^{2}} \Big)^{\!3/2}\,,
\end{equation*}
which differs from our result~(\ref{dot_M_m}) by a factor of~$\mu$,
i.e. $\dot{M}_{[87]}={(m/\dot{\varphi}_{0})}\dot{M}_{\text{here}}$, see footnote \ref{footnote_rate}, given as a square in~(\ref{mu_v_c}). Using (\ref{mu_v_c}),
the dependence on~$m$
in this formula can be exchanged for the dependence on~$\dot{\varphi}_{0}$,
\begin{equation*}
\dot{M}
	=
	2\pi G^{2}M^{2}
	\Big( \frac{\dot{\varphi}_{0}^{4}}{\lambda} \Big)
	\frac{\left(1-3v_{c}^{2}\right)^{1/2}
	(1+3v_{c}^{2} )^{7/2}}{ (1-v_{c}^{2})^{3/2} |v_{c}|} \, .
\end{equation*}
This implies that in the radiation limit~$m\!\to\!0$ when~$v_{c}^{2} \!\rightarrow\! 1/3$, the accretion rate \emph{vanishes}, which is clearly unphysical. 
} from Eq.~(2.19) obtained in~\cite{Feng:2021qkj}. 
Even though, in the limit $c_{s}^{2}(\infty) \!\rightarrow\! 0$,
our result \eqref{dot_M_m} 
agrees with~\cite{Feng:2021qkj}, there is substantial disagreement for relativistic sound speeds at spatial infinity. As we have already mentioned in footnote \ref{footnote_rate}, we believe that this disagreement can be attributed 
to~\cite{Feng:2021qkj} assuming that each conserved charge falling into the BH  contributes a particle rest mass $m$ defined in potential \eqref{eq:Potential}. However, at spatial infinity it is the chemical potential $\upmu=\dot{\varphi}_{0}$ which corresponds to energy per charge, see \eqref{UPmu}. This is directly confirmed by generally-covariant
definition~\eqref{AccretionGenRel} of accretion rate. 

Written in the form~(\ref{dot_M_m}), the accretion rate seems to diverge in the 
limit~$v_c^2 \!\to\! 1/3$. However, this is only true when keeping $m$ constant 
and climbing high up the potential by indefinitely increasing the modulus so 
that~$\lambda \rho^4 \!\gg\! 2 |m^2|\rho^2$. Thus, this divergence is due to the 
divergence of $\dot \varphi_0$, and hence to the divergence of the energy density at infinity, 
and not to any singularity in the accretion process for $v_c^2\!\to\!1/3$. Of course, parametrically 
large energy density at infinity violates the assumption of the test-field approximation;
see the discussion in Sec.~\ref{sec: Validity_Test_Field} and the 
condition~\eqref{our_condition}.

\section{Accretion for the complex scalar field}
\label{sec: Accretion for complex scalar field}

In this section, we solve the profile equation~(\ref{dimless profile eq}) for the complex
scalar model,
\begin{equation}
\frac{(1\!-\!f)^4}{\xi^2} \biggl[ f \frac{d^2\sigma}{df^2} + \frac{d\sigma}{df} \biggr]
	- \bigl( \mu^2 \sigma + \sigma^3 \bigr)
	+ \frac{\sigma}{f} \biggl[ 1 - \frac{ (1\!-\!f)^4 \beta^4 }{ \sigma^4 } \biggr]
	=
	0
	\, .
\label{CS profile equation}
\end{equation}
Even though this equation  qualitatively differs from
the~$P(X)$ equation~(\ref{P(X) profile equation}), being a differential equation rather than an algebraic one,
the two share some important structural features. 

The singular points of the profile 
equation~(\ref{CS profile equation}) are located at the endpoints, and they enforce the 
same form of the boundary conditions~(\ref{boundary conditions}) at radial infinity and at the 
horizon, that can be satisfied only for a particular value of the dimensionless
flux~$\beta$, mirroring the same feature observed for the~$P(X)$ theory. However, this critical 
value for the flux will differ from the~$P(X)$ case and will depend on~$\xi$. In this way, the sensitivity
to gradients shows up in the boundary condition at the horizon. We will see that this universally 
translates into the lowering of the field profile at the horizon.

After briefly discussing the qualitative properties of the profile equation~(\ref{CS profile equation}),
we first find approximate analytic solutions of the profile equation~(\ref{CS profile equation}) 
in two limits: (i)~$\xi^2\!\gg\!1$, which corresponds to the limit of small gradient corrections 
close to the~$P(X)$ case, and (ii)~$\xi^2 \!\ll\! 1$, which
corresponds to the limit where gradients 
play the dominant role over the large interval. Then we turn to the central part of the paper 
devoted to solving the profile equation for finite~$\xi^2$, which must be done numerically.

\subsection{Qualitative considerations for the profile equation\label{qualitative}}

The profile equation~\eqref{CS profile equation} can be rewritten in a less singular and more intuitive form:
\begin{equation}
\frac{1}{\xi^{2}}\,\frac{d^{2}\sigma}{dx_{*}^{2}}+\frac{2}{\xi^{2}}\,f_{*}(1\!-\!f_{*})\,\frac{d\sigma}{dx_{*}}+\mathscr{V}'=0\,,\label{mechanica}
\end{equation}
in terms of the non-compact tortoise coordinate
\begin{equation}
x_* = \frac{1}{1\!-\!f} + \ln \Bigl( \frac{f}{1\!-\!f} \Bigr) \, ,
\label{TortoiseDefinition}
\end{equation}
with~$f_* \!=\! f(x_*)$ denoting the inverse of Eq.~(\ref{TortoiseDefinition}). The effective potential in Eq.~(\ref{mechanica}) is obtained by switching the sign and 
making an $x_*$-dependent rescaling of the potential in~(\ref{eff potential}),
\begin{equation}
\label{V_nice}
\mathscr{V}(\sigma,x_*)
	=
	- f_*\frac{(\sigma^{2} \!-\! \sigma_{+}^{2} )^{2}}{4}
	+
	(1 \!-\! f_* ) \frac{\sigma^{2}}{2}
	+
	\frac{(1\!-\!f_*)^{4} \beta^{4} }{2\sigma^{2}}\,,
\end{equation} 
with $\sigma_{+}$ given in~(\ref{boundary conditions}). This profile equation, 
written as in~(\ref{mechanica}), does not exhibit singular points either at the horizon or at spatial infinity, as does 
Eq.~(\ref{CS profile equation}) that is written in terms of the compactified radial coordinate.

The form of the profile equation~(\ref{mechanica}) lends itself to the following simple 
mechanical analogy. This equation describes a particle of mass~$\xi^{-2}$ evolving in 
``time'' $x_*$ from $-\infty$ to $+\infty$, in a time-dependent 
potential~$\mathscr{V}(\sigma,x_*)$ under the friction force with the time-dependent, 
but positive friction coefficient~$2\xi^{-2}f_{*}(1 \!-\! f_{*})$. The friction vanishes in the 
infinite past ($f_*\!\rightarrow\! 0$) and in the infinite future ($f_*\!\rightarrow\! 1$). 
Moreover, 
in the infinite past, the potential $\mathscr{V}(\sigma,-\infty)$ has a minimum 
at~$\sigma\!=\!\beta$ with value $\beta^2$, while in the infinite 
future,~$\mathscr{V}(\sigma,+\infty)$ is an inverted ``Mexican hat'' potential that vanishes 
at the maximum, located at $\sigma_{+}$. This picture is 
insensitive to the sign of~$\mu^2$. Thus, in the infinite past the particle starts with 
zero velocity at the minimum of the potential,~$\sigma\!=\!\beta$,
and with (non-conserved) 
energy~$\beta^2$, and ends its motion in the infinite future with the vanishing velocity and 
vanishing energy at the maximum of the potential $\sigma_{+}$.\footnote{Such finely tuned 
dynamics resembles situations considered in the semiclassical theory of false vacuum 
decay~\cite{Coleman:1977py}. }
Clearly, by considering overshooting and undershooting, one can find the amount of energy in 
the infinite past $\beta^2$ needed to arrive at a given $\sigma_{+}$ in the infinite future. 
Thus, our boundary problem should have a unique solution. 

In this mechanical analogy, the limit~$\xi\!\rightarrow \!\infty$ corresponds to a massless particle and vanishing friction. It is clear that the lighter the particle is, the larger the 
distance it can traverse. That is why the largest distance traveled, $\beta \!-\! \sigma_{+}$, 
will be observed  in this limit. Therefore, one can expect that the accretion rate is maximal 
in the~$P(X)$ limit. This is indeed confirmed numerically (see Fig.~\ref{accretion ratio} in the 
concluding section).

The opposite limit, $\xi\!\rightarrow \!0$, corresponds to an infinitely massive particle 
with infinitely large friction. Due to enormous inertia the particle will stay at virtually the same 
position~$\beta$ for a very long time, even if the potential has already substantially changed 
and $\beta$ is not its minimum anymore. Thus, one can expect the appearance of a very 
long plateau in $\sigma$ in this case, and a very small traversed distance $\beta-\sigma_{+}$. 
This explains why the accretion rate is suppressed for $\xi\!\rightarrow \!0$. 

In general, the 
particle will follow the time-dependent minimum of the potential \eqref{V_nice} until the 
adiabatic condition is violated. In particular, this happens slightly before the ``effective 
frequency''  $\mathscr{V}''$ vanishes. Note that this necessarily happens during the evolution, 
as in the infinite past $\mathscr{V}''\!>\!0$, while in the infinite future $\mathscr{V}''\!<\!0$. 

Thus, for any finite $\xi$ the value of the dimensionless radial field on horizon $\beta(\mu,\xi)$ is always in the interval  $1-\mu^2<\beta^2(\mu,\xi)<\beta^2(\mu,\infty)$, providing lower and upper bounds for the accretion.

\subsection{Solving the profile equation for large~$\xi^2$}
\label{subsec: Solving profile equation for large xi}

Given that the profile equation~(\ref{CS profile equation}) formally reduces to the~$P(X)$ model
in the limit~$\xi^2\!\to\!\infty$, it is natural to seek solutions as power series in~$1/\xi^2$
in the regime where~$\xi^2\!\gg\!1$. This expansion is assumed for both the modulus field and the flux:
\begin{equation}
\sigma = \sigma_0 + \frac{\sigma_1}{\xi^2} + \frac{\sigma_2}{\xi^4} 
    + \dots \, ,
\qquad \qquad
\beta = \beta_0 + \frac{\beta_1 }{\xi^2} + \frac{\beta_2 }{\xi^4} 
    + \dots \, ,
\label{pert expansion}
\end{equation}
where~$\sigma_0$ and~$\beta_0$ are the field profile and dimensionless flux 
found in the exact~$P(X)$ case in Sec.~\ref{sec: Accretion for P(X)}.\footnote{Any 
non-analytic dependence on~$\xi$ should be at least exponentially suppressed 
compared to any power-law correction in the limit~$\xi^2\!\gg\!1$.}
The second of the two expansions above is crucial, as it guarantees the 
existence of a solution and can be viewed as correcting the critical point 
conditions from Sec.~\ref{sec: Accretion for P(X)}. The dimensionless flux retains the 
property that it is determined by the profile equation  itself, and the boundary 
conditions it dictates. 

The first subleading correction to the scalar profile follows directly by substituting 
the expansions~(\ref{pert expansion}) into the profile equation~(\ref{dimless profile eq}) 
and organizing the terms in powers of~$1/\xi^2$,
\begin{equation}
\sigma_1 = \frac{ \mathcal{D}^2 \sigma_0 - \frac{4 (1-f)^4 \beta_0^3}{f \sigma_0^3} \beta_1 }{ \mathcal{U}''(\mu, \beta_0; f, \sigma_0) } \, .
\label{sigma1 solution}
\end{equation}
However, there is seemingly no expression that determines the correction~$\beta_1$ to the flux. 
To obtain it, we must recognize that the denominator in the solution above vanishes at the~$P(X)$ 
critical point~$f_c$ given in~(\ref{critical f sigma}). 
Thus, the solution becomes ill-defined unless the numerator also vanishes at the same point,
and this regularity condition fixes~$\beta_1$,
\begin{equation}
\beta_1 = \biggl[ \frac{f \sigma_0^3 \times \mathcal{D}^2 \sigma_0 }{4 (1\!-\!f)^4 \beta_0^3} 
	\biggr]_{f = f_{\scr c} } \ .
	\label{beta_1}
\end{equation}
This completes the first subleading correction. The solution~(\ref{sigma1 solution})
follows from an algebraic equation, leaving no freedom to choose boundary conditions; rather,
they emerge directly from the solution itself,
\begin{equation}
\sigma_1\xrightarrow{f\to0}
	\beta_1 \, ,
\qquad \qquad
\sigma_1 \xrightarrow{f\to1} 0 \, .
\end{equation}
It is remarkable that these conditions
automatically reproduce the correct boundary conditions corresponding 
to the first subleading order in the~$1/\xi^2$ expansion of the exact boundary 
conditions~(\ref{boundary conditions}). 
Thus, treating corrections to the~$P(X)$ solution as 
an expansion in~$1/\xi^2$,
and requiring that they remain well defined everywhere, automatically produces corrections with 
the correct boundary behavior.
This property persists at the second subleading order and is expected to hold at all orders. 

\begin{figure}[h!]
\vskip+3mm
\centering
\includegraphics[width=15cm]{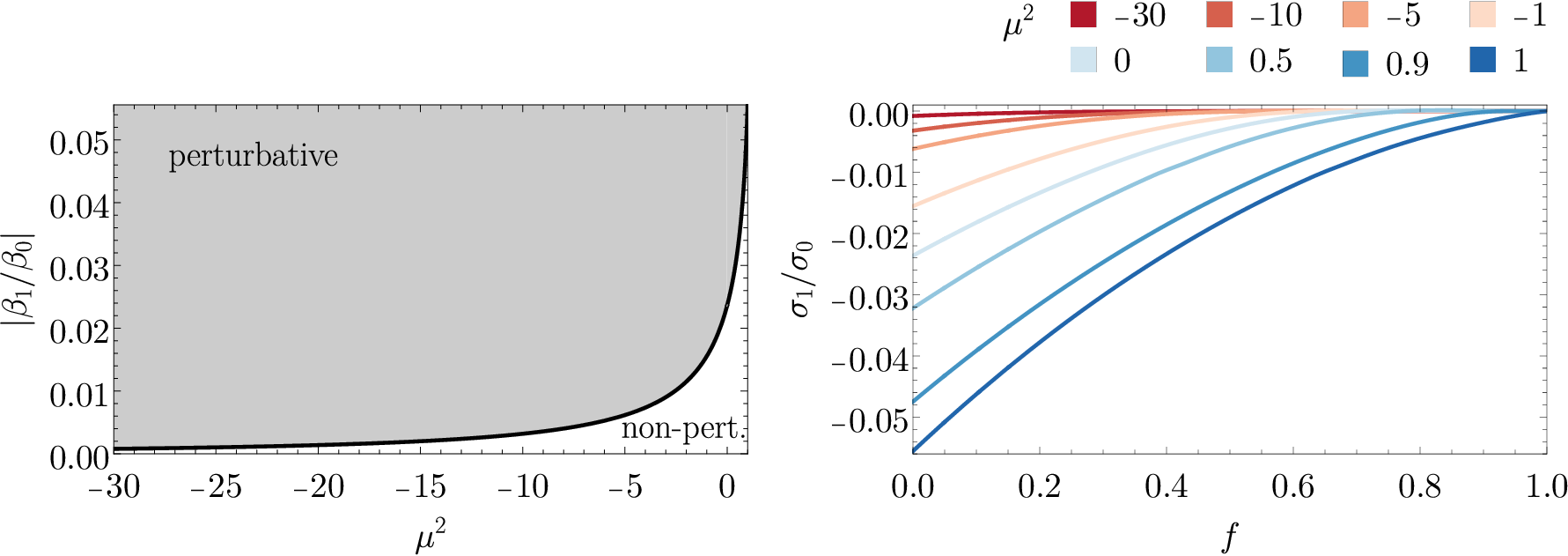}
\caption{Bounds on the applicability of the perturbative approach. 
{\it Left:} The solid curve depicts the relative correction to the dimensionless flux~(\ref{beta_1})
in units of~$\xi^2$, as a function of~$\mu^2$. For values
of~$\xi^2$ greater than
this relative correction (shaded region) the perturbative expansion in~(\ref{pert expansion}) 
is justified, while for values of~$\xi^2$ lower than this relative correction
the perturbative expansion is not justified (white region).
{\it Right:} Relative corrections to the modulus field profile to leading order in $1/\xi^{2}$, 
in units of~$1/\xi^2$. The values of the curves at~$f\!=\!0$
correspond to~$\beta_1/\beta_0$. 
}
\label{perturbative_bound_plot}
\end{figure}

The correction to the dimensionless flux~(\ref{critical beta}) can be used 
to place a bound on the applicability of the perturbative approach,
which breaks down when the first correction to~$\beta$ in~(\ref{pert expansion}) becomes 
comparable to the leading-order term, yielding the condition
\begin{equation}
\xi^2 \gg \beta_1 / \beta_0 \, .
\end{equation}
This bound depends on~$\mu^2$ and is shown in the left panel of Fig.~\ref{perturbative_bound_plot}.

\begin{figure}[h!]
\centering
\vskip+3mm
\includegraphics[width=10cm]{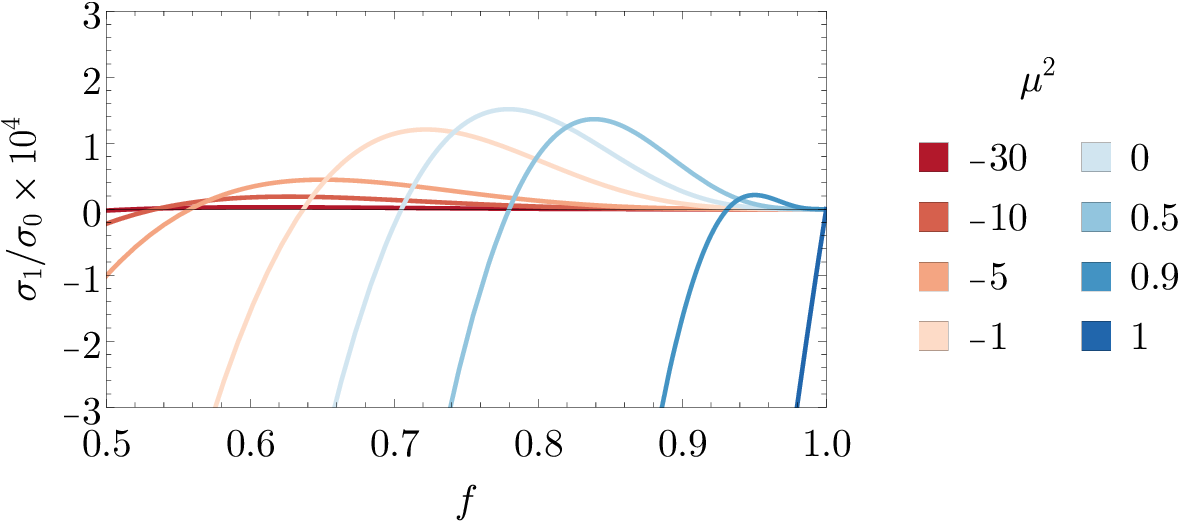}
\vskip-2mm
\caption{Relative correction to the radial field profile (in units of~$1/\xi^2$)
in the regime~$\xi^2\!\gg\!1$, far from the horizon.}
\label{overshootingP(X)}
\end{figure}

The right panel of Fig.~\ref{perturbative_bound_plot} shows the first relative correction
to the modulus profile. While this figure may appear to suggest 
that the complex scalar modulus profile always lies below the corresponding~$P(X)$ profile, 
this is not the case far from the horizon. To see this, we examine the behavior of the 
profile correction~(\ref{sigma1 solution}) for~$f\!\to\!1$,
\begin{equation}
\frac{\sigma_1}{\sigma_0}
    \ \overset{f \to 1}{\longsim} \
    \frac{(1\!-\!f)^4}{8 \sigma_0^6}
    \Bigl[
    1 - 2\mu^2 - 16 \beta_0^3 \beta_1
    \Bigr]
    \, ,
\qquad\qquad
\sigma_0 \ \overset{f \to 1}{\longsim} \
    \sqrt{ \frac{1}{f} - \mu^2 }
    \, .
\label{Inequality1}
\end{equation}
The coefficient in the brackets determines the behavior far from the horizon.
As shown in Appendix~\ref{app: Derivatives at the critical point}, it is almost always
positive, except for a narrow mass interval where~$\mu^2$ is close to unity. 
This implies that, for most values of~$\mu^2$ considered here, the complex scalar profile lies above the
corresponding~$P(X)$ profile far from the horizon. 
This behavior is confirmed visually in Fig.~\ref{overshootingP(X)}, which zooms in on the relevant region. 
By comparison, the positive correction far from the horizon is about two orders of magnitude smaller 
than the negative correction near the horizon shown in the
right panel of Fig.~\ref{perturbative_bound_plot}.

The large-$\xi^2$ regime considered in this subsection effectively corresponds to the situation 
in which gradients of the scalar field modulus are small. Therefore, the expansion 
in powers of~$1/\xi^2$ outlined here can also be derived within the EFT framework presented in 
Sec.~\ref{sec: EFT of the complex scalar}, and the first subleading results should follow
from the first gradient correction in~(\ref{EFTexample}).

\subsection{Solving the profile equation for small~$\xi^2$}
\label{subsec: Solving profile equation for small xi}

In the limit of large~$\xi^2$, the analytic~$P(X)$ solution provided
a good departure point for perturbatively corrected solutions in the
preceding section. The opposite limit of small~$\xi^2$ can also be 
described analytically, as we do in this subsection, but finding a
uniform approximation requires more finesse.

Inspecting the profile equation~(\ref{CS profile equation}) in the
limit of~$\xi^2 \!\ll\! 1$ reveals two qualitatively different regions of 
the equation: (i) a relatively wide interval where the prefactor of the derivative terms
is large,~$(1\!-\!f)^4/\xi^2\!\gg\!1$, such that all potential terms can 
be discarded at leading order; and (ii) a relatively narrow interval that
requires special treatment, where the prefactor of derivative terms is of order 
one or smaller,~$(1\!-\!f)^4/\xi^2 \!\sim\! 1$.
Note that the latter narrow interval, close to~$f\!=\!1$, where the behavior of the equation changes abruptly, becomes smaller as~$\xi^2$ decreases. Such 
behavior is typical of problems for which the technique 
known as {\it boundary layer theory}
(see e.g. chapter 9 in~\cite{Bender:1999box}) was developed. This is a 
systematic approximation method tailored to approximate solutions to differential
equations where behavior changes abruptly in a small interval controlled
by a small parameter, such that the smaller the parameter is, the narrower 
the layer of abrupt change, and the more drastic the change. 

We apply this method to determine the field profiles in the small~$\xi^2$ limit. 
Following the terminology of boundary layer theory, we refer 
to the wide interval as the {\it outer layer}, and to the narrow interval 
as the {\it inner layer}. First we derive reliable approximations in the two layers,
followed by the matching procedure that constructs a uniform approximation across
the entire interval.

\paragraph{Outer layer.}
In between the horizon and a narrow layer close to~$f\!\sim\!1$ we can apply
naive perturbation theory as a power series in small~$\xi$. This region is referred to as the {\it outer layer} in the terminology of boundary layer theory. There we assume that 
both the profile and the flux are well approximated by the power series in $\xi$,
\begin{equation}
\sigma(f) =
\sigma_{\rm out}(f) = \sigma_{\rm out}^{0)}(f)  + \xi \sigma_{\rm out}^{1)}(f)
    + \xi^2 \sigma_{\rm out}^{2)}(f) + \dots \, ,
\qquad
\beta = \beta_0 + \xi \beta_1 + \xi^2 \beta_2 + \dots \, .
\label{OuterExpansion}
\end{equation}
Consequently, the boundary condition~(\ref{boundary conditions}) at the horizon
is expanded in the same manner,
\begin{equation}
\sigma_{\rm out}^{0)}(0) = \beta_0 \, ,
\qquad \quad
\sigma_{\rm out}^{1)}(0) = \beta_1 \, .
\label{OuterBoundary}
\end{equation}
Organizing the profile equation~(\ref{CS profile equation}) 
in powers of~$\xi$ then produces the same equation for the leading 
and subleading profiles,
\begin{equation}
\biggl[ f \frac{d^2 }{df^2} + \frac{d }{df} \biggr] \sigma_{\rm out}^{0)} = 0 \, ,
\qquad \quad
\biggl[ f \frac{d^2 }{df^2} + \frac{d }{df} \biggr] \sigma_{\rm out}^{1)} = 0 \, ,
\label{OuterEqs}
\end{equation}
They are both solved by a linear combination of a constant and~$\ln(f)$,
but the latter is forbidden by the boundary conditions~(\ref{OuterBoundary})
leaving just constants as solutions,
\begin{equation}
\sigma_{\rm out}^{0)}(f) = \beta_0 \, ,
\qquad \quad
\sigma_{\rm out}^{1)}(f) = \beta_1 \, .
\end{equation}
The value of the flux is not fixed by the boundary conditions. Rather,   it will be fixed 
by the procedure of matching the solution in the outer layer with the one in the inner 
layer that we work out next.

\paragraph{Inner layer.}
In the thin region close to~$f\!\sim\!1$ the behavior of the
modulus field profile changes abruptly, and this feature
only becomes more pronounced and more localized with smaller~$\xi$. 
Upon closer examination we uncover that our inner layer has a typical width~$\sim\xi$, which suggest that perturbative solution should be 
sought for only after zooming in on the layer by rescaling the coordinate,
\begin{equation}
F\equiv \frac{1\!-\!f}{\xi} \, .
\end{equation}
This rescaling transforms the profile equation~(\ref{CS profile equation})
to
\begin{equation}
F^4 \biggl[
    (1 \!-\! \xi F) \frac{d^2 \sigma}{dF^2}  
    - 
    \xi \frac{d\sigma}{dF}
    \biggr]
    -
    \mu^2 \sigma
    -
    \sigma^3
    +
    \frac{\sigma}{1 \!-\! \xi F}
    \biggl[ 1 \!-\! \frac{ \xi^4 F^4 \beta^4 }{ \sigma^4 } \biggr]
    =
    0
    \, .
\label{InnerLayerEq}
\end{equation}
The remaining~$\xi$-dependence in this equation that is not contained 
in the coordinate~$F$, can now be treated perturbatively.
This is accomplished by assuming a power-series form of the solution,
\begin{equation}
\sigma(f) =
    \sigma_{\rm in}(F)
    =
    \sigma_{\rm in}^{0)}(F) 
    + 
    \xi \sigma_{\rm in}^{1)}(F)
    + 
    \xi^2 \sigma_{\rm in}^{2)}(F)
    + \dots \, ,
\end{equation}
together with the same expansion for the flux, introduced in~(\ref{OuterExpansion}),
and the boundary conditions~(\ref{boundary conditions}),
\begin{equation}
\sigma_{\rm in}^{0)}(F\!=\!0) = \sqrt{1 \!-\! \mu^2}  \, ,
\qquad \quad
\sigma_{\rm in}^{1)}(F\!=\!0) = 0 \, .
\label{InnerBoundary}
\end{equation}
Likewise, the profile equation~(\ref{InnerLayerEq}) is also organized in powers
of~$\xi$. The leading order equation,
\begin{equation}
F^4 \frac{d^2 \sigma_{\rm in}^{0)}}{d F^2} + (1\!-\!\mu^2) \sigma_{\rm in}^{0)}
    - \bigl( \sigma_{\rm in}^{0)} \bigr)^3
    = 0
    \, ,
\end{equation}
is solved by a constant, after boundary conditions in~(\ref{InnerBoundary})
are imposed,
\begin{equation}
\sigma_{\rm in}^{0)}(F) = \sqrt{ 1 \!-\! \mu^2 } \, .
\end{equation}
The subleading equation is consequently inhomogeneous,
\begin{equation}
F^4 \frac{d^2 \sigma_{\rm in}^{1)}}{dF^2}
    - 2 (1 \!-\! \mu^2) \sigma_{\rm in}^{1)}
    =
    -
    F \sqrt{ 1 \!-\! \mu^2 }
    \, ,
\end{equation}
and, upon enforcing boundary conditions~(\ref{InnerBoundary}),
it is solved by,
\begin{equation}
\sigma_{\rm in}^{1)}(F) = \frac{F}{ 2\sqrt{1\!-\!\mu^2} }
    \biggl[ 1 + C
    \exp\biggl( - \frac{ \sqrt{2(1\!-\!\mu^2)} }{ F } \biggr) \biggr]
    \, ,
\end{equation}
where~$C$ is a constant of integration.

\paragraph{Matching.} Following boundary layer theory, as the last step,
the two approximations for the inner and outer layers now need to 
be asymptotically matched in the region of overlap as~$\xi\!\to\!0$. 
This region formally corresponds to~$f\!\to\!1$ limit of the 
outer layer, and~$F\!\to\!\infty$ limit of the inner layer,
where the two asymptotic forms must match,
\begin{equation}
\sigma_{\rm out}(f)
    \ \overset{f \to 1}{\longsim} \
    \sigma_{\rm match}(f)
    \ \overset{F \to \infty}{\longsim} \
    \sigma_{\rm in}(F)
    \, ,
    \qquad
    {\rm as} \quad \xi \to 0 \, .
\end{equation}
The leading order matching condition is trivial,
\begin{equation}
\beta_0 = \sqrt{ 1 \!-\! \mu^2 } \, ,
\end{equation}
but it determines the leading flux. The subleading condition
reads
\begin{equation}
\beta_1 \ \overset{F \to \infty}{\longsim} \
    \frac{1}{2}  \bigg(\frac{F(1 \!+\! C)}{\sqrt{1\!-\!\mu^2}}
    - C \sqrt{2} \bigg)\, .
\end{equation}
The only way for the two sides of this relation to be asymptotic to each other 
is for the $F$ dependence on the right-hand side to vanish, which 
fixes the integration constant~$C\!=\!-1$, and consequently 
yields the flux correction~$\beta_1\!=\! 1/\sqrt{2}$. Thus, for small $\xi$, the accretion rate has a lower bound via 
\begin{equation}
    \beta^{2}=1-\mu^{2}+\xi\sqrt{2\left(1-\mu^{2}\right)}+...>1-\mu^{2}\,.
    \label{lower_bound}
\end{equation}
The uniform approximation for the solution is now given by
(see ch.~9 in~\cite{Bender:1999box}),
\begin{equation}
\sigma(f) \ \overset{ \xi^{2} \ll 1}{\longsim}  \
    \sigma_{\rm out}(f) + \sigma_{\rm in}(F) - \sigma_{\rm match}
    \, ,
\end{equation}
which in our case evaluates to
\begin{equation}
\sigma(f) \ \overset{\xi^2 \ll 1}{\longsim} \
    \sqrt{1 \!-\! \mu^2} 
    +
    \frac{1 \!-\! f}{2 \sqrt{1 \!-\! \mu^2} }
    \biggl[
    1 - 
    \exp\biggl( - \frac{\xi \sqrt{2(1\!-\!\mu^2)} }{ 1\!-\! f } 
    \, \biggr)
    \biggr]
    \, .
\label{SmallXiApprox}
\end{equation}

\begin{figure}
\centering
\includegraphics[width=15cm]{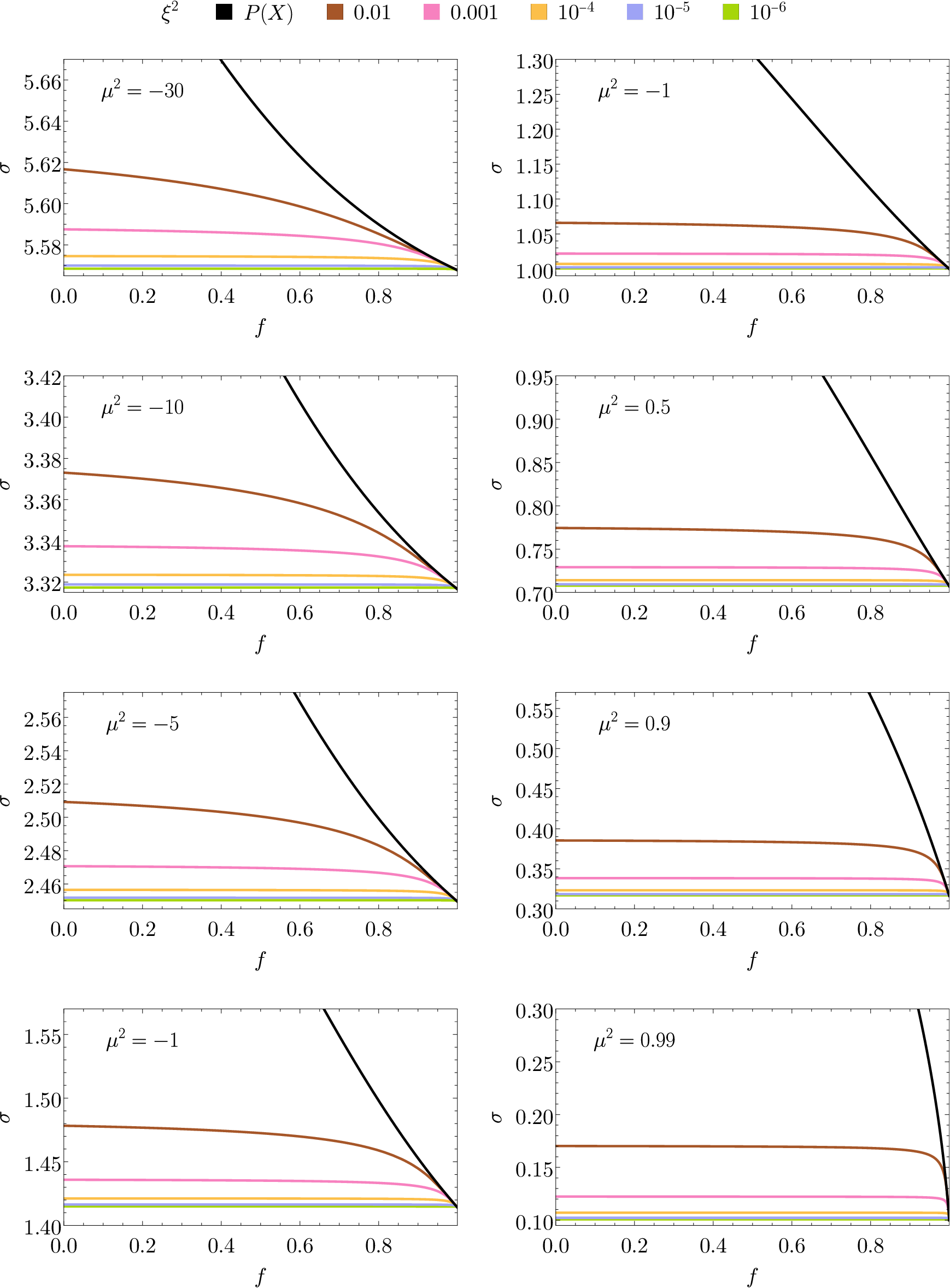}
\caption{Field profiles in the limit~$\xi^2 \!\ll\! 1$ (in colours) 
compared to their~$P(X)$ counterparts, shown in black. The coloured profiles are
plotted using the approximation in~(\ref{SmallXiApprox}). Note that we did
not include the exact~$\mu^2\!=\!1$ case, for which the approximation 
obtained by boundary layer theory breaks down.}
\label{small xi profiles}
\end{figure}

The expression in~(\ref{SmallXiApprox}) represents an
approximation for the field 
profile that is reliable over the entire interval. Plots of field
profiles for different values of parameters in this regime are given in 
Fig.~\ref{small xi profiles}. While it is clear from these plots that, in this regime,
field profiles lie well below their~$P(X)$ counterparts, it is worth noting that 
far away from the horizon the complex scalar profiles still go very slightly above the
corresponding~$P(X)$ ones. This point is discussed by the end of the following subsection.

\subsection{Solving the profile equation for finite~$\xi^2$}
\label{subsec: Solving profile equation for finite xi}

The profile equation~(\ref{CS profile equation}) for finite~$\xi$, subject to 
the boundary conditions~(\ref{boundary conditions}), constitutes a variant of 
a non-linear eigenvalue problem. Much like in the~$P(X)$ case, it admits only 
a single eigenvalue for the flux~$\beta$ that allows a solution to exist. 
Furthermore, the behavior of the profile equation near the boundaries at~$f\!=\!0$ 
and~$f\!=\!1$ is sufficiently singular to fully determine the boundary conditions, 
implying that the integration constants must be contained in the non-analytic 
asymptotic behavior close to the boundaries. We first examine these asymptotic 
behaviors to gain insight that will better inform our choice of numerical methods 
for solving the profile equation.

\paragraph{Radial infinity limit.}
The profile equation~(\ref{CS profile equation}) has an irregular singular 
point at~$f\!=\!1$, which is manifested by no free constants appearing in the 
Taylor series,
\begin{equation}
\sigma(f) \ \overset{ f\to 1 }{\longsim} \
    \sigma_+ + (f\!-\!1) \sigma_+^{1)} + \frac{1}{2}(f\!-\!1)^2 \sigma_+^{2)}
    + \frac{1}{3!}(f\!-\!1)^{3} \sigma_+^{3)} +
     \frac{1}{4!}(f\!-\!1)^4 \sigma_+^{4)} 
    + \mathcal{O}\bigl[ (f\!-\!1)^5 \bigr]
    \, ,
\label{TaylorPlus}
\end{equation}
where the first several coefficients are
\begin{subequations}
\begin{align}
\sigma_+^{1)} ={}& 
    - \frac{1}{2\sigma_+}\,,
\qquad \quad
\sigma_+^{2)} =
    \frac{1}{\sigma_+}
    -
    \frac{1}{4\sigma_+^3}
    \, ,
\qquad \quad
\sigma_+^{3)} =
    -
    \frac{3}{\sigma_+}
    +
    \frac{3}{2\sigma_+^3}
    -
    \frac{3}{8\sigma_+^5}
    \, ,
\\
\sigma_+^{4)} ={}& 
    \frac{12}{\sigma_+}
    -
    \frac{9}{\sigma_+^3}
    +
    \frac{9}{2 \sigma_+^5}
    -
    \frac{15}{16 \sigma_+^7}
    +
    \frac{3}{\sigma_+^5} \biggl[ \frac{2\sigma_+^2 \!-\! 1}{\xi^2} - 4\beta^4 \biggr]
    \, .
\label{SigmaPlus4}
\end{align}
\end{subequations}
First, we see that the difference with respect to the~$P(X)$ case appears
only at quartic order, via the dependence on~$\xi$ and~$\beta$. Furthermore,
no free constants of integration appear at any order in this Taylor expansion.

This signals the presence of non-analytic behavior of the field profile
at radial infinity, that must harbor the constants of integration. Revealing this 
contribution is not straightforward in terms of the compactified radial coordinate.
However, the tortoise coordinate, $x_*$, defined in \eqref{TortoiseDefinition}, is well adapted for this task, as it removes the singularity from the derivative terms
in the profile equation, and puts it in the form~(\ref{mechanica}).
We can examine the behavior of the solutions to this equation 
at radial infinity,~$x_* \!\to\! \infty$,
by shifting the field profile by its asymptotic value,
\begin{equation}
\sigma = \sigma_+ + \delta\sigma \, ,
\end{equation}
and specializing the linearized profile equation to this regime where~$(1\!-\!f_*) \!\sim\! 1/x_* \!\ll\! 1$,
\begin{equation}
\biggl[ \frac{1}{\xi^2} \Bigl( \frac{d^2}{dx_*^2} 
        + \frac{2}{x_*} \frac{d}{d x_*} \Bigr)
    - 2 \sigma_+^2 \biggr] \delta\sigma
    \ \, \overset{ x_* \to \infty }{\longsim} \ - \frac{\sigma_+}{x_*} \, .
\label{TortoiseEq}
\end{equation}
It is straightforward to check that the general solution to this sourced equation is
\begin{equation}
\delta\sigma \ \overset{x_*\to\infty}{\longsim} \
    \frac{1}{2\sigma_+x_*}
    +
    \frac{A_-}{x_*} e^{-\sqrt{2}\xi \sigma_+ x_*}
    +
    \frac{A_+}{x_*} e^{\sqrt{2}\xi \sigma_+ x_*}
    \, ,
\label{TortoiseSolution}
\end{equation}
where~$A_+$ and~$A_-$ are free constants of integration. 

In terms of the compactified radial variable, the asymptotic 
solution~(\ref{TortoiseSolution}) reads
\begin{equation}
\delta\sigma \ \overset{f\to1}{\longsim} \
    (1\!-\!f) \sigma_+^{1)}
    +
    A_- (1 \!-\! f)^{1 + \sqrt{2} \xi \sigma_+}
    \exp \Bigl[ - \frac{\sqrt{2}\xi \sigma_+}{1 \!-\! f} \Bigr]
    +
    A_+ (1\!-\!f)^{1-\sqrt{2}\xi \sigma_+}
    \Bigl[ \frac{ \sqrt{2}\xi \sigma_+ }{1\!-\!f} \Bigr]
    \, ,
\label{DeltaSigmaInfinitySolution}
\end{equation}
The first term in the solution, which descends from the source in Eq.~(\ref{TortoiseEq}),
reproduces the first term in the Taylor series~(\ref{TaylorPlus}).
The remaining two terms with constants of integration are the non-analytic 
contributions that we were after.
Requiring the finiteness of the profile at radial infinity immediately 
requires~$A_+ \!=\! 0$, given that the last term diverges exponentially.
This also explains why the equation is numerically very unstable close to~$f\!=\!1$.

\paragraph{Horizon limit.}
Close to the horizon the regular power series,
\begin{equation}
\sigma(f) \ \overset{ f\to 0 }{\longsim} \
    \sigma_- + f \sigma_-^{1)} + \frac{1}{2}f^2 \sigma_-^{2)}
    + \mathcal{O}( f^3 )
    \, ,
\label{TaylorMinus}
\end{equation}
is also not able to capture the free constants of integration of the differential 
equation. Rather, the coefficients of the expansion are all fixed in terms of the 
parameters of the equation,
\begin{equation}
\frac{\sigma_-^{1)}}{\sigma_-} =
    - \frac{ 4 - \mu^2 - \beta^2 }{ 4 \!+\! 1/\xi^2 }
    \, ,
\qquad \quad
\frac{\sigma_-^{2)}}{\sigma_-} = \frac{6 + {\sigma_-^{1)}}/{\sigma_-}\,\,  
        \bigl( 6\,\sigma_-^{1)}/\sigma_- + 12 + \mu^2 + 3\beta^2 + 4/\xi^2 \bigr)}{2 \bigl( 1 \!+\! 1/\xi^2\bigr)}
\, .
\end{equation}
The difference compared with the radial infinity is that already the first term
differentiates between the complex scalar and its~$P(X)$ limit.

The reason why the Taylor expansion   misses the free constants of integration is that
the profile equation~(\ref{CS profile equation}) 
also has a singular point at the horizon,
but this time a regular one. This point is best elucidated for the small perturbation
of the profile close to the horizon,
\begin{equation}
\sigma \ \overset{f\to 0}{\longsim} \ \sigma_- + \delta \sigma \, ,
\end{equation}
for which the linearized equation of motion close to the horizon reads,
\begin{equation}
\biggl[
\frac{1}{\xi^2} \Bigl( f \frac{d^2}{df^2} + \frac{d}{df} \Bigr)
    + \frac{4}{f} \biggr]\delta\sigma
	\ \, \overset{ f \to 0 }{\longsim} \
	- \sigma_- (4 \!-\! \mu^2 \!-\! \sigma_-^2)
	\, .
\end{equation}
The particular solution to this equation captures the first term in the 
Taylor series~(\ref{TaylorMinus}), while the homogeneous part exhibits non-analytic
power-law behavior ubiquitous to differential equations with regular singular
points for which the Frobenius method applies\footnote{Here the Frobenius method yields imaginary powers.},
\begin{equation}
\delta \sigma \ \overset{f \to 0}{\longsim} \
    f \sigma_-^{1)} + A_c \cos\bigl[ 2\xi \ln(f) \bigr]
    + A_s \sin\bigl[ 2\xi \ln(f) \bigr] \, .
\label{DeltaSigmaHorizonSolution}
\end{equation}
The homogeneous solutions do not have a well-defined limit as they approach the horizon
and keep oscillating faster and faster. That is why both constants of integration
need to be fixed to vanish,~$A_c\!=\!A_s\!=\!0$. Because the homogeneous solutions
here are bounded this makes the singular point at the horizon much milder
numerically than the singular point at radial infinity.

\paragraph{Numerics.}
For finite~$\xi^2$ the modulus-field profiles and the flux can only really be determined numerically.
The examination of the asymptotic behavior at the endpoints indicates that one should 
not expect significant numerical complications at the horizon at~$f \!=\! 0$, which represents 
a regular singular point of the profile equation. In the vicinity of the horizon, the 
homogeneous solutions in~(\ref{DeltaSigmaHorizonSolution}) oscillate rapidly, though with a 
bounded amplitude. In contrast, near radial infinity at~$f \!=\! 1 $, which harbors an irregular 
singular point, the behavior is expected to be numerically exponentially unstable due to 
the presence of a runaway homogeneous mode in~(\ref{DeltaSigmaInfinitySolution}).

\begin{figure}[h!]
\centering
\includegraphics[width=15cm]{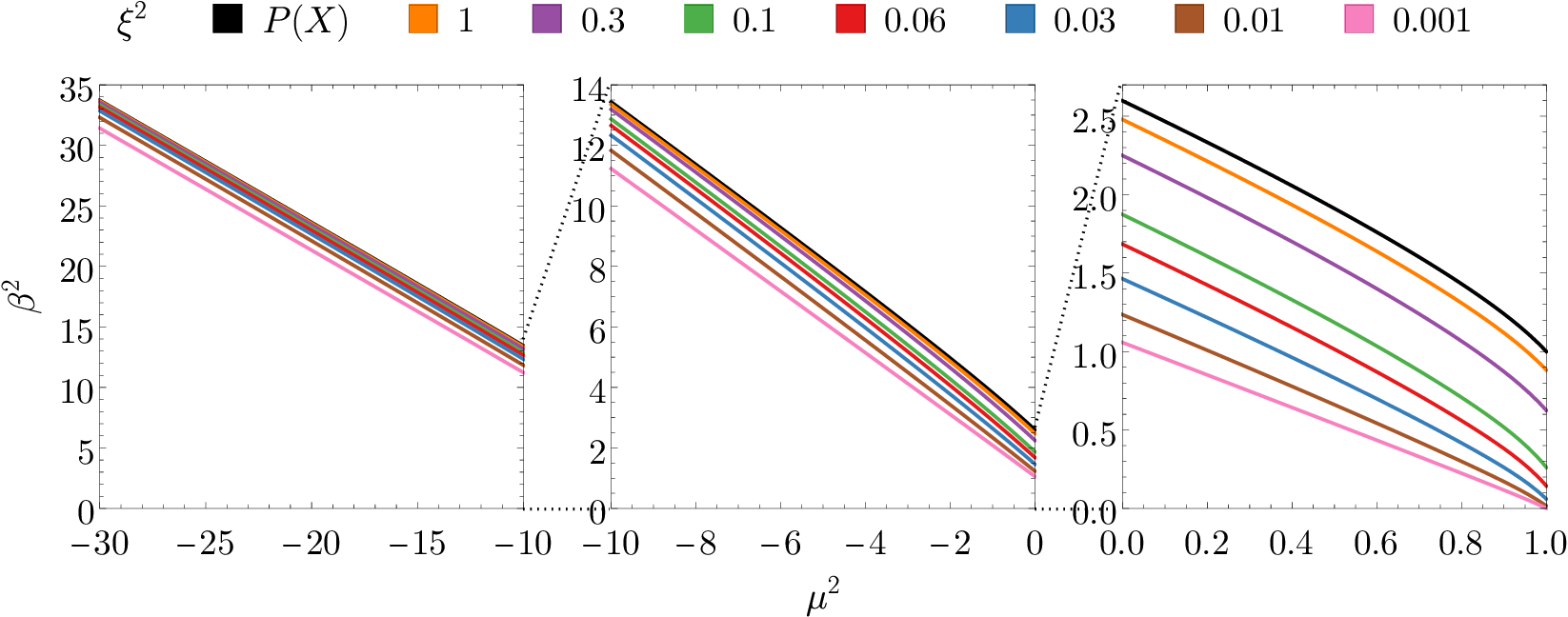}
\\
\vspace{-0.2cm}
\caption{Dimensionless flux $\beta^2\!=\! \lambda \mathcal{F}/\xi^2$ 
as a function of~$\mu^2$ for different values of~$\xi^2$.}
\label{beta CS}
\end{figure}

These asymptotic behaviors inform our choice of numerical methods to determine the critical 
value~$\beta$, as well as the modulus-field profile and its derivative. Although the 
shooting method cannot recover the profile near~$f \!=\! 1$ because of the exponential instability, 
it is well suited for determining the dimensionless flux~$\beta$ with relatively high accuracy. 
We set the initial conditions encoded in~(\ref{TaylorMinus}) at~$f \!=\! 0$. The profile is then 
extended numerically as far as possible toward~$f \!=\! 1$, while avoiding the onset of runaway 
behavior. The resulting numerically determined steady-state accretion flux is shown 
in Fig.~\ref{beta CS}. Notably, it now depends on both~$\mu$ and~$\xi$.

\begin{figure}
\centering
\includegraphics[width=15cm]{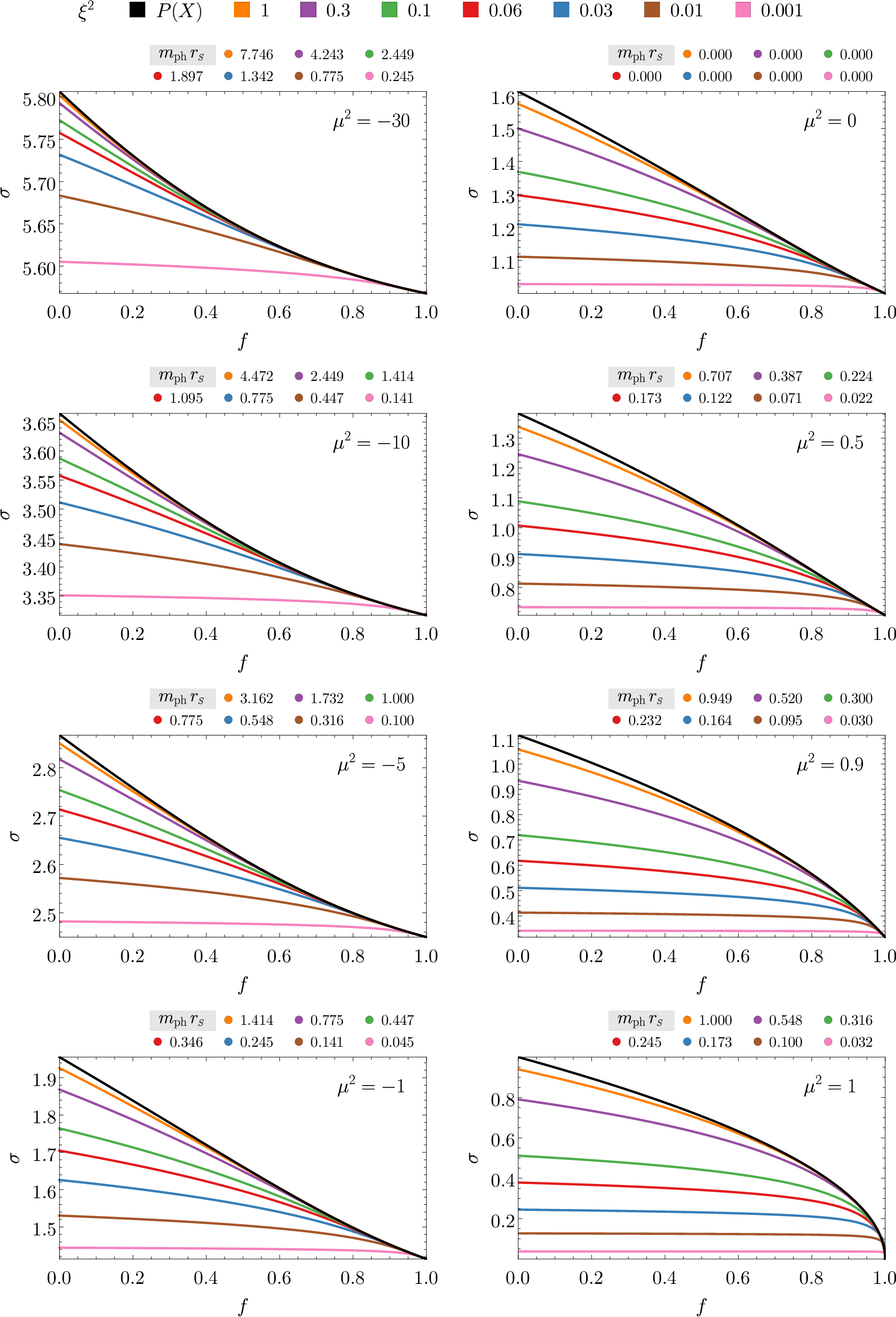}
\caption{Numerically obtained modulus field profiles for different choices 
of parameters~$\xi^2$ and~$\mu^2$. The numbers above the panels indicate the field mass corresponding to the true potential minimum in flat space, 
expressed in units of the inverse Schwarzschild radius. }
\label{sigma profiles}
\end{figure}

\begin{figure}
\centering
\includegraphics[width=15cm]{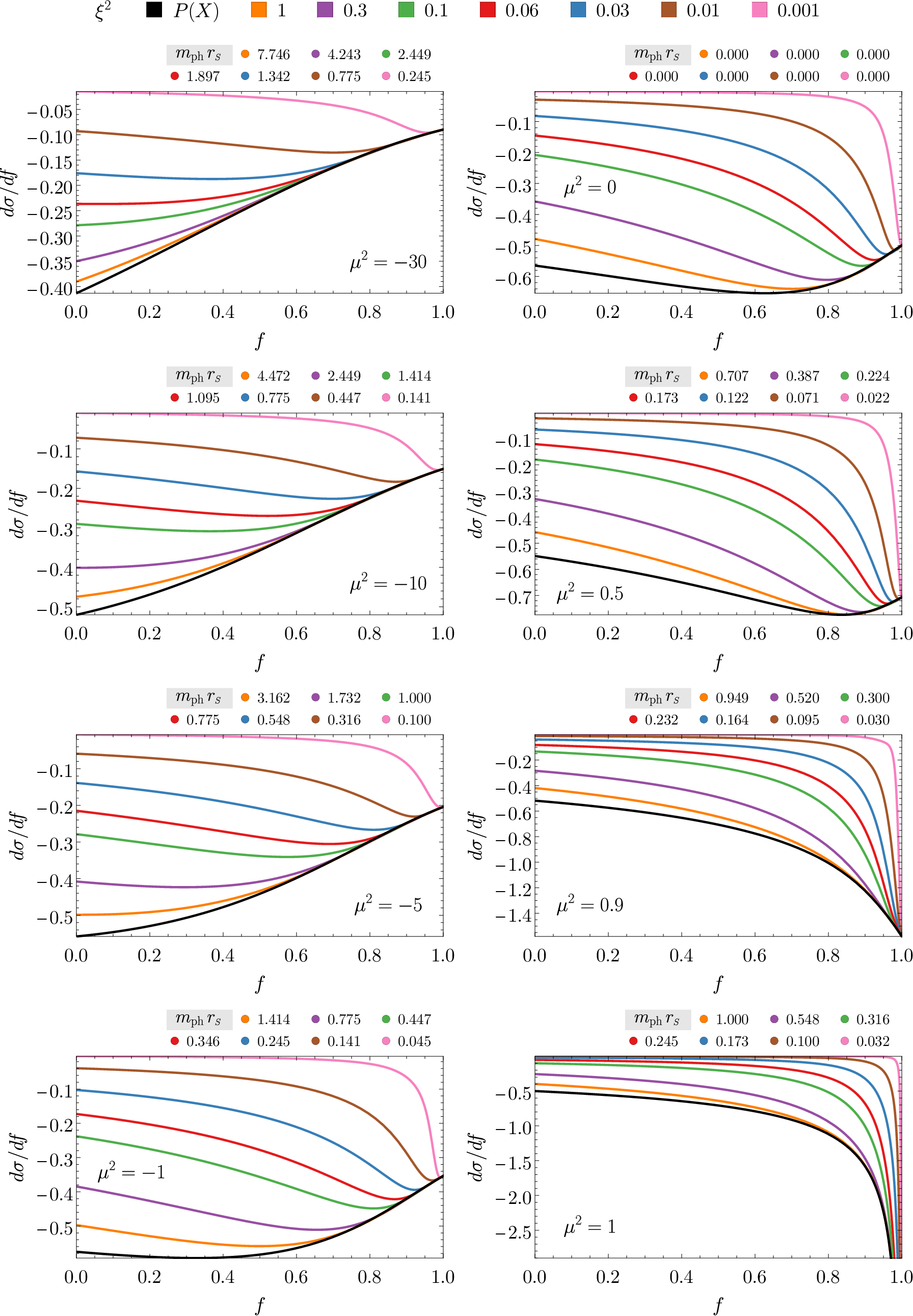}
\caption{Numerically obtained derivative of modulus field profiles for different choices 
of parameters~$\xi^2$ and~$\mu^2$. The numbers above the panels indicate the field mass corresponding to the true potential minimum in flat space, 
expressed in units of the inverse Schwarzschild radius. }
\label{dsigma}
\end{figure}

After determining the flux, we compute the modulus field profile using a variant of the 
variational method. We adopt a higher-order rational function ansatz, constrained to reproduce 
the boundary conditions in~(\ref{TaylorPlus}) and~(\ref{TaylorMinus}), and determine the free 
coefficients by minimizing the residual of the profile equation~(\ref{CS profile equation}). 
The profiles obtained from this fitting procedure are shown in Fig.~\ref{sigma profiles}, and 
their first derivatives in Fig.~\ref{dsigma}.

The modulus field profile~$\sigma(f)$ generally differs from the~$P(X)$ solution and may deviate 
significantly depending on the values of~$\xi$ and~$\mu^{2}$. As expected, the limit~$\xi^{2}\!\ll\!1$ 
exhibits the largest deviations from the~$P(X)$ behavior. In this limit, the closed-form 
approximation~(\ref{SmallXiApprox}) reproduces the numerical solutions: the brown 
and pink curves in Figs.~\ref{small xi profiles} and~\ref{sigma profiles} correspond to the same 
case and show excellent agreement.

\begin{figure}
\centering
\includegraphics[width=15cm]{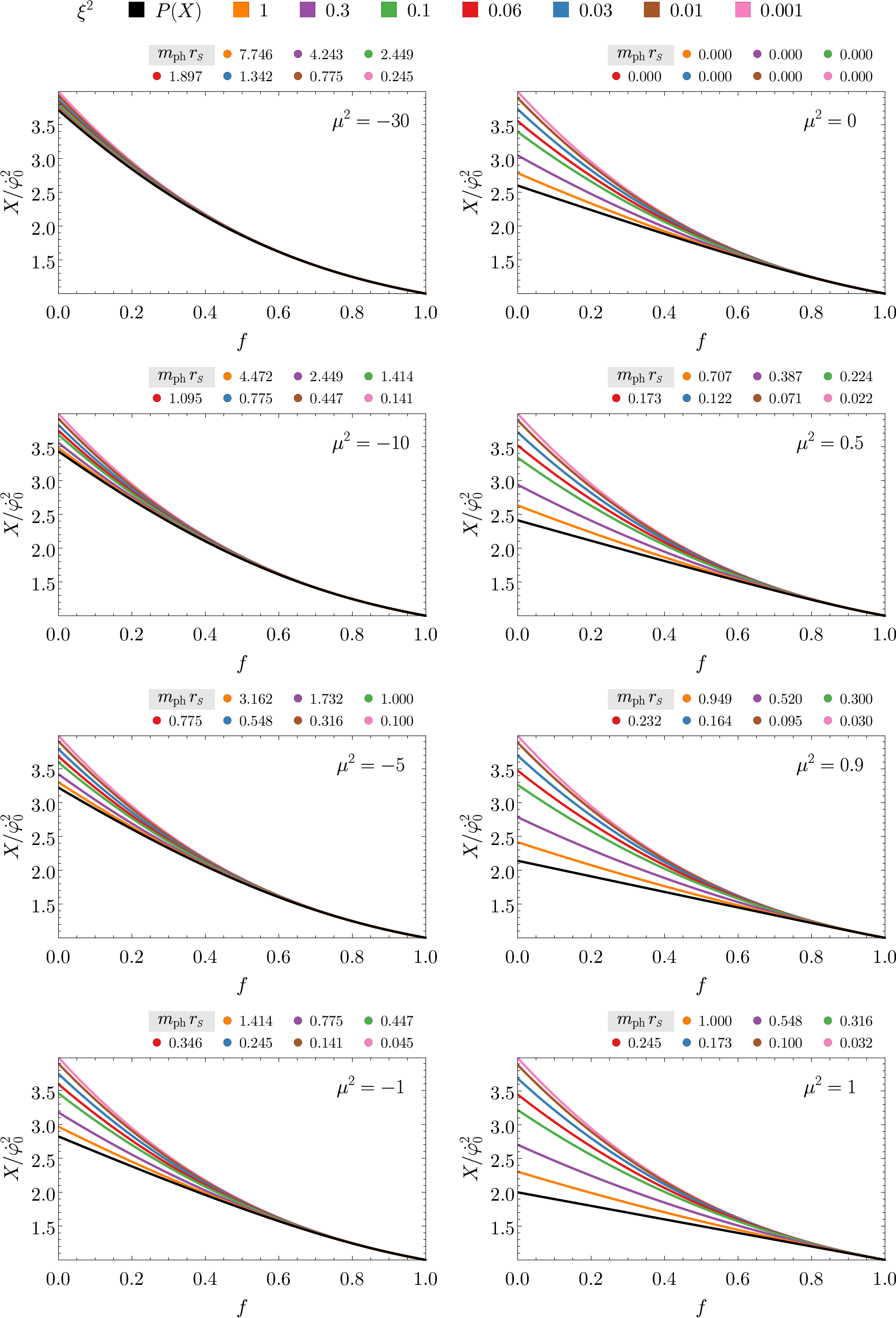}
\caption{Radial dependence of the phase-field kinetic term~(\ref{WandX}) for 
different choices of parameters~$\xi^2$ and~$\mu^2$. 
There is a general tendency for~$X$ at finite~$\xi^{2}$ to exceed its limiting 
$P(X)$ value across all~$\mu^{2}$, consistent with enhanced gradient contributions 
and stronger deviations from perfect-fluid behavior at smaller~$\xi^{2}$.}
\label{X plots}
\end{figure}

\bigskip

We find that, in general, the modulus field profiles for finite~$\xi^2$ always lie 
below the corresponding~$P(X)$ profiles near the horizon, and that the difference 
increases as~$\xi^2$ decreases. Far from the horizon, Fig.~\ref{sigma profiles} 
might suggest that the profiles asymptote to the corresponding~$P(X)$ profiles from 
below. This is, however, not the case, just as in the limit~$\xi^{2}\!\gg\!1$ discussed 
in Sec.~\ref{subsec: Solving profile equation for large xi}. To zoom in on the
behavior as~$f\!\to\!1$, we derive the asymptotic behavior directly from 
Eq.~(\ref{CS profile equation}). It is safe to neglect the nonanalytic contributions
in this limit, as they are certainly smaller than any power-law corrections.
Far away from the horizon we can treat the terms explicitly proportional to~$(1\!-\!f)^4$ 
as perturbations. Then the leading-order behavior at radial infinity is
\begin{equation}
\sigma(f) \ \overset{f \to 1}{\longsim} \ \sigma_0(f) = \sqrt{ \frac{1}{f} - \mu^2 } \, ,
\end{equation}
the same as for the~$P(X)$ model (cf.~the left expression in Eq.~(\ref{bc f1})). Considering
the small perturbation of this behavior,~$\sigma\!=\! \sigma_0 \!+\! \delta \sigma$,
as being sourced by the terms neglected in the equation, it is straightforward to
derive that
\begin{equation}
\delta\sigma
    \ \overset{f \to 1}{\longsim} \
    \frac{(1\!-\!f)^4}{2 \sigma_0^5}
    \biggl[ \frac{1 - 2\mu^2}{4\xi^2} - \beta^4 \biggr]
    \, .
\end{equation}
Note that this expression captures the leading behavior for any allowed choice of
the mass parameter, including~$\mu^2\!=\!1$. The corresponding expression for the~$P(X)$
profile is obtained by taking~$\xi^2\!\to\!\infty$ in the first term in the brackets, 
and taking~$\beta\!\to\! \beta_c$ in the second term. Therefore, the difference of profiles 
is
\begin{equation}
\Delta\sigma = \delta \sigma - \delta \sigma\bigr|_{P(X)}
    \ \overset{f \to 1}{\longsim} \
    \frac{(1\!-\!f)^4}{2\sigma_0^5}
    \biggl[ \frac{1 - 2\mu^2}{4\xi^2} - \beta^4 + \beta_c^4 \biggr]
    \, .
\end{equation}
From the plot of the flux parameter in Fig.~\ref{beta CS}, it follows that this
quantity can be positive, and that the complex scalar profile can indeed rise above
the corresponding~$P(X)$ one far away from the horizon. However, this difference is
much smaller than the deviation near the horizon, where the complex scalar profiles 
always lie below them, and the difference is too small to be visible in Fig.~\ref{sigma profiles}.

Having determined the modulus-field profiles, we can infer the kinetic term
of the phase field from~(\ref{WandX}),
\begin{equation}
Y = \frac{X}{\dot{\varphi}_0^2} 
    = \frac{1}{f} \biggl[ 1 - \frac{ (1\!-\!f)^4 \beta^4 }{\sigma^4} \biggr] \, ,\label{Ydef}
\end{equation}
which is depicted in Fig.~\ref{X plots} for different values of~$\mu^2$
and~$\xi^2$. From these results, it is evident that the complex scalar gradient 
corrections lift the phase-field kinetic term above the corresponding~$P(X)$ value.
This behavior persists even far from the horizon, where the value of~$X$
approaches the~$P(X)$ value from above, as seen from the asymptotic form,
\begin{equation}
\Delta Y
    \ \overset{f\to1}{\longsim} \
    \frac{(1\!-\!f)^4}{\sigma_0^4}
    \bigl( \beta_c^4 \!-\! \beta^4 \bigr)
    > 0
    \, .
\end{equation}
%

\section{Equation of state}
\label{sec: Equation of state}

The energy-momentum tensor governing steady-state accretion in both the~$P(X)$ 
and complex scalar field models can be constructed from the solutions for the 
modulus field profile discussed in Sections~\ref{sec: Accretion for P(X)}
and~\ref{sec: Accretion for complex scalar field}. It is most conveniently 
expressed in terms of the four fluid quantities defined in 
Eqs.~(\ref{energy density})--(\ref{heat transfer}), 
whose physical behavior we analyze in this section.

\subsection{$P(X)$ model}
\label{subsec: P(X) model}

\begin{figure}[h!]
\vskip+3mm
\centering
\includegraphics[width=15cm]{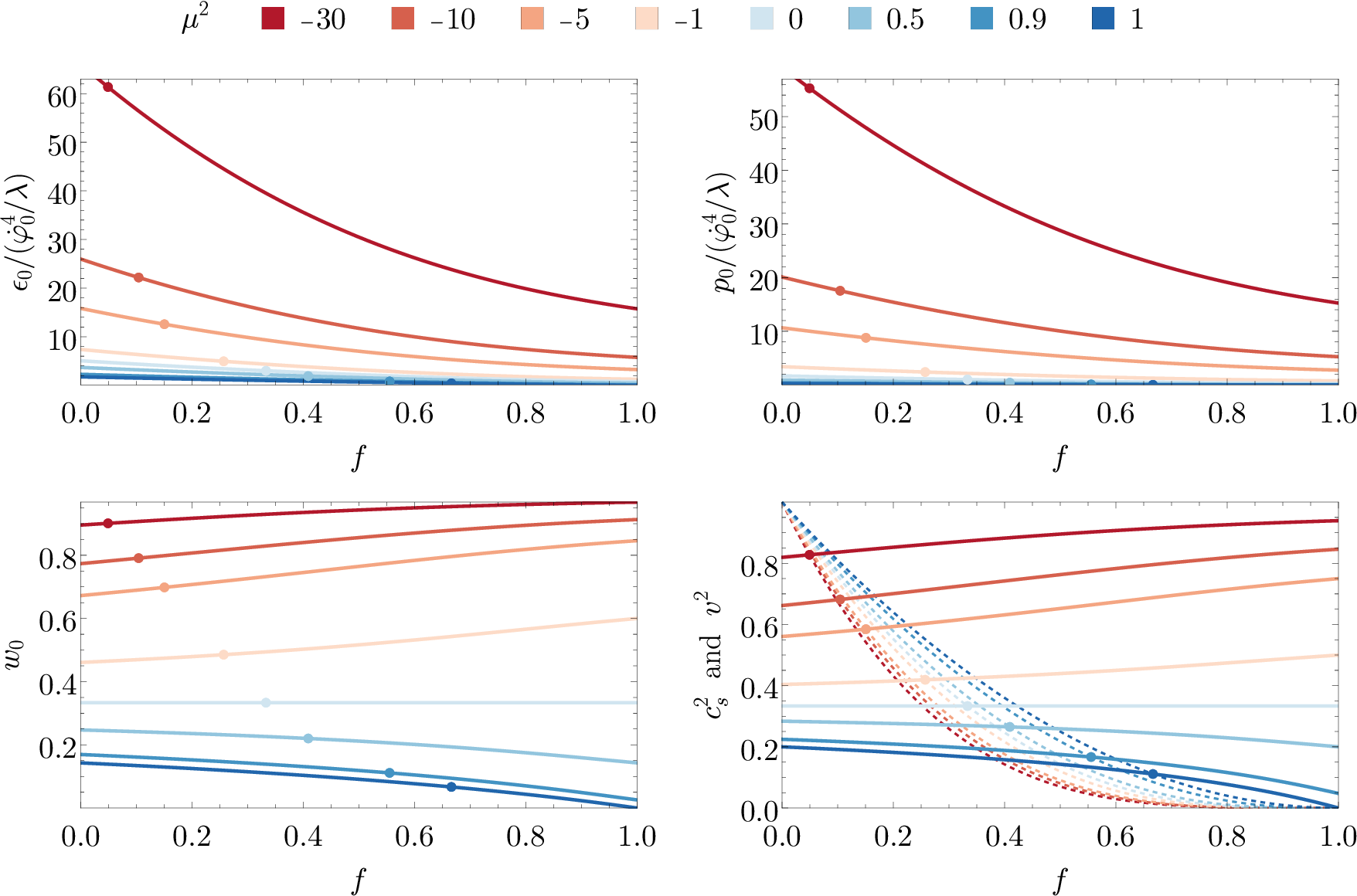}
\caption{Radial dependence of the energy density ({\it top left}), 
pressure ({\it top right}), 
equation-of-state parameter~$w\!=\!p/\epsilon$ ({\it bottom left}), 
and sound speed~$c_s^2 \!=\! dp/d\epsilon$ ({\it bottom right})
in the exact~$P(X)$ model, for different 
values of the mass parameter~$\mu^2$. 
Bullet points on the curves mark the location of the critical point, which 
coincides with the acoustic horizon \eqref{acoustic horizon}. 
In the lower-right panel, dashed curves show the squared shift charge velocity 
given in~(\ref{definition vc2}), while solid curves denote the sound 
speed~$c_s^2$. The crossing of these two sets of curves identifies the 
acoustic horizon, where the shift charge velocity first
exceeds the local sound speed.}
\label{P(X)EpsilonPressureFig}
\end{figure}

The~$P(X)$ model introduced in Sec.~\ref{subsec: Application to quartic self-coupling} describes a perfect fluid whose accretion solutions are presented in Sec.~\ref{sec: Accretion for P(X)}. 
The energy density and pressure depend solely on the modulus profile, $\sigma$, which in turn is a function of the phase kinetic term $X$, due to \eqref{eq:Rho_X_quartic}. For steady-state accretion 
in the specific~$P(X)$ model of~(\ref{specific P}) they are given by
\begin{equation}
\epsilon_0 
	=
	\frac{ \dot{\varphi}_0^4 }{ \lambda }
	\biggl[ \frac{3 \sigma^4}{4}
		+ \mu^2 \sigma^2
		+ \mathcal{V}_0
		\biggr]
	\, ,
\qquad \qquad
p_0 =	
	\frac{\dot{\varphi}_0^4 }{ \lambda } 
	\biggl[ \frac{\sigma^4}{4}
		- \mathcal{V}_0
		\biggr]
	\, .
	\label{px_en_pres}
\end{equation}
The radial dependence of the energy density and pressure is thus inherited from the 
modulus profile shown in Fig.~\ref{P(X)figureProfiles}, and 
is presented in Fig.~\ref{P(X)EpsilonPressureFig} for different choices 
of the mass parameter, including tachyonic values.
On the other hand, solving for the relativistic Bondi accretion of the perfect fluid with the equation of state \eqref{eq:eospx} yields the same energy density and 
pressure profiles as those shown in Fig.~\ref{P(X)EpsilonPressureFig}, obtained directly 
from the field-theoretic model.

Another feature characteristic of~$P(X)$ theories is illustrated by the bullet points in all
panels of Fig.~\ref{P(X)EpsilonPressureFig}, which indicate the locations of 
the critical point~(\ref{critical f sigma}). At this point, the sound speed of perturbations 
in the~$P(X)$ model, shown in the bottom-right panel of 
Fig.~\ref{P(X)EpsilonPressureFig}, becomes smaller
than the fluid velocity. Consequently, perturbations can no longer propagate outward, 
signaling the presence of an \emph{acoustic horizon}
--- a causal boundary beyond which 
acoustic perturbations are trapped, analogous to an event horizon in general 
relativity~\cite{Frolov:2004vm}. It is not clear whether any such structure appears 
in the complex scalar model, or whether it emerges only in the strict~$P(X)$ limit.

\subsection{Complex scalar model}
\label{subsec: Complex scalar model}

For the complex scalar model, a general ideal-fluid description does not exist, 
as the equation of state depends on the chosen frame. It is most natural to adopt 
the Eckart frame, in which the shift charge current defines the normalized 
four-velocity~(\ref{velocity Eckart}) of the fluid. The quantities characterizing
the energy–momentum tensor --- energy density, pressure,
heat flux, and anisotropic stress --- are, according to~(\ref{cs emt})
and~(\ref{T decomp}), given by:
\begin{align}
\epsilon ={}&
	\frac{ \dot{\varphi}_0^4 }{\lambda} \biggl[
		\frac{3\sigma^4}{4} 
		+ \mu^2\sigma^2
		+ \mathcal{V}_0
		- \frac{\sigma \mathcal{D}^2 \sigma}{2\xi^2}
		+ \frac{(1\!-\!f)^4}{\xi^2} \frac{(2 \!-\! fY)}{2Y} \Bigl( \frac{d\sigma}{df} \Bigr)^{\!2}
		\biggr]
		\, ,
\label{eck_en}
\\
p ={}&
	\frac{ \dot{\varphi}_0^4 }{\lambda} \biggl[
		\frac{ \sigma^4}{4} 
		- \mathcal{V}_0
		- \frac{\sigma \mathcal{D}^2 \sigma}{2\xi^2}
		+ \frac{(1\!-\!f)^4}{\xi^2} \frac{(2 \!-\! 3fY)}{6Y} \Bigl( \frac{d\sigma}{df} \Bigr)^{\!2}
		\biggr] \, ,
\label{eck_pres}
\\
\label{heat_transfer_spec}
q \equiv\sqrt{-q_\mu q^\mu} ={}&
	\frac{ \dot{\varphi}_0^4 }{\lambda} \biggl[
		\frac{(1\!-\!f)^4}{\xi^2} \frac{\sqrt{ 1 \!-\! fY }}{Y} \Bigl( \frac{d\sigma}{df} \Bigr)^{\!2} 
		\biggr]
		\, ,
\\
\label{anisotropic_stress_spec}
\Pi \equiv \sqrt{\Pi_{\mu\nu} \Pi^{\mu\nu}} ={}&
	\frac{ \dot{\varphi}_0^4 }{\lambda} \biggl[
		\frac{(1\!-\!f)^4}{\xi^2} \sqrt{\frac{2}{3}} \frac{1}{Y} 
		\Bigl( \frac{d\sigma}{df} \Bigr)^{\!2} 
		\biggr]
		\, ,
\end{align}
where the dimensionless kinetic term~$Y$ of the phase field
is given in~(\ref{Ydef}).

The quantities $q$ and $\Pi$ describe deviations from perfect-fluid behavior, 
representing respectively the heat flux and anisotropic stress generated by spatial 
gradients of the modulus field. This model therefore generalizes the~$P(X)$ case 
discussed previously, recovering the perfect-fluid limit when these quantities vanish.

We first examine the above quantities in the regime of small gradients, 
that is, in the limit~$\xi^2 \!\gg\!1$, where simple perturbation theory provides 
accurate analytic results. We then turn to the case of finite~$\xi^2$, where gradients 
are not negligible, and use the numerically determined profiles from 
Sec.~\ref{subsec: Solving profile equation for finite xi} to infer the corresponding 
fluid quantities.

\subsubsection{Large $\xi^2$ regime}
\label{largexisquaredenergypressure}

\begin{figure}[h!]
\centering
\includegraphics[width=15cm]{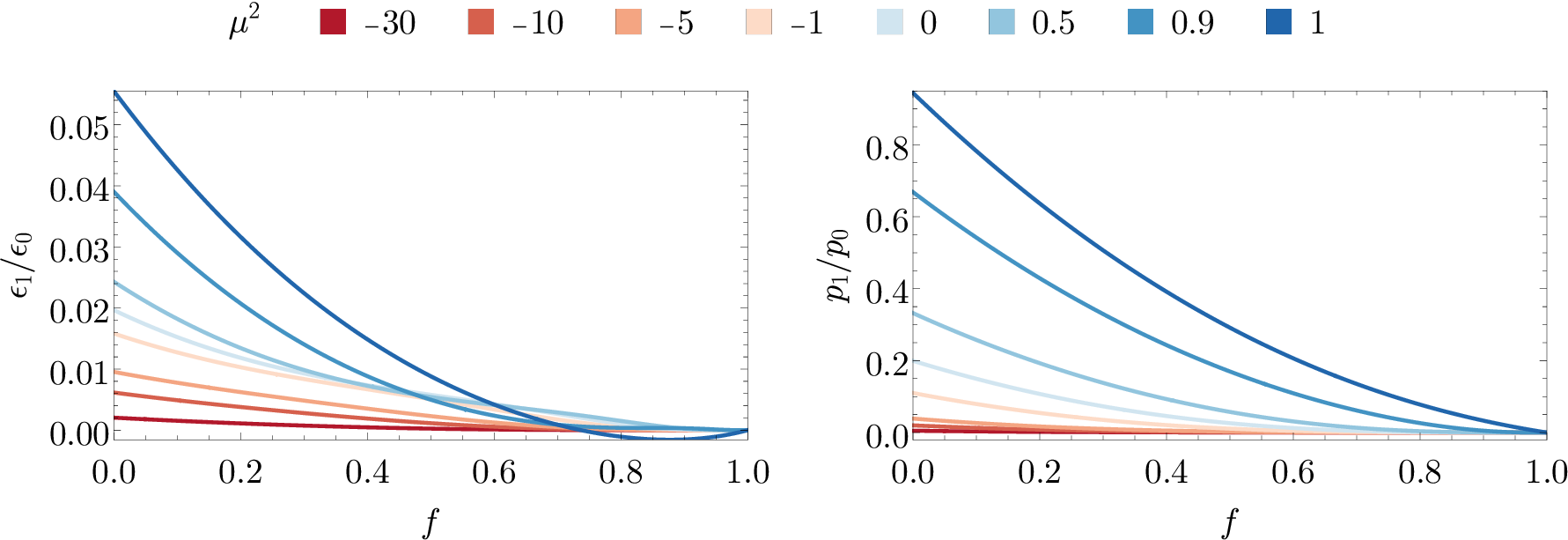}
\vskip-2mm
\caption{Radial dependence of relative corrections to the energy density
and pressure in the limit of large~$\xi^2$, for different values of 
the parameter~$\mu^2$.}
\vskip-2mm
\label{LargeXiEpsilonPressureFig}
\end{figure}

In the limit of large~$\xi^2$, the corrections to the~$P(X)$ results 
are expected to be small and can be captured by the perturbative 
expansion~(\ref{pert expansion}), which also applies to the energy 
density and pressure,
\begin{equation}
\epsilon = \epsilon_0 + \frac{\epsilon_1}{\xi^2}
    + \frac{\epsilon_2}{\xi^4} + \dots
    \, ,
\qquad\quad
p = p_0 + \frac{p_1}{\xi^2}
    + \frac{p_2}{\xi^4} + \dots
    \, .
\end{equation}
The leading corrections follow by substituting the expansion~(\ref{pert expansion})
into~(\ref{eck_en}) and~(\ref{eck_pres}),
\begin{align}
\epsilon_1 ={}&
	\frac{ \dot{\varphi}_0^4 }{\lambda} \biggl[
		3\sigma_0^3 \sigma_1
		+ 2 \mu^2\sigma_0 \sigma_1
		- \frac{1}{2} \sigma_0 \mathcal{D}^2 \sigma_0
		+ (1\!-\!f)^4 \frac{(2 \!-\! fY)}{2Y} \Bigl( \frac{d\sigma_0}{df} \Bigr)^{\!2}
		\biggr]
		\, ,
\\
p_1 ={}&
	\frac{ \dot{\varphi}_0^4 }{\lambda} \biggl[
		\sigma_0^3 \sigma_1
		- \frac{1}{2} \sigma_0 \mathcal{D}^2 \sigma_0
		+ (1\!-\!f)^4 \frac{(2 \!-\! 3fY)}{6Y} \Bigl( \frac{d\sigma_0}{df} \Bigr)^{\!2}
		\biggr] \, .
\end{align}
The radial variation of the relative corrections to the energy density and pressure
is shown in Fig.~\ref{LargeXiEpsilonPressureFig}.

\begin{figure}[h!]
\centering
\includegraphics[width=15cm]{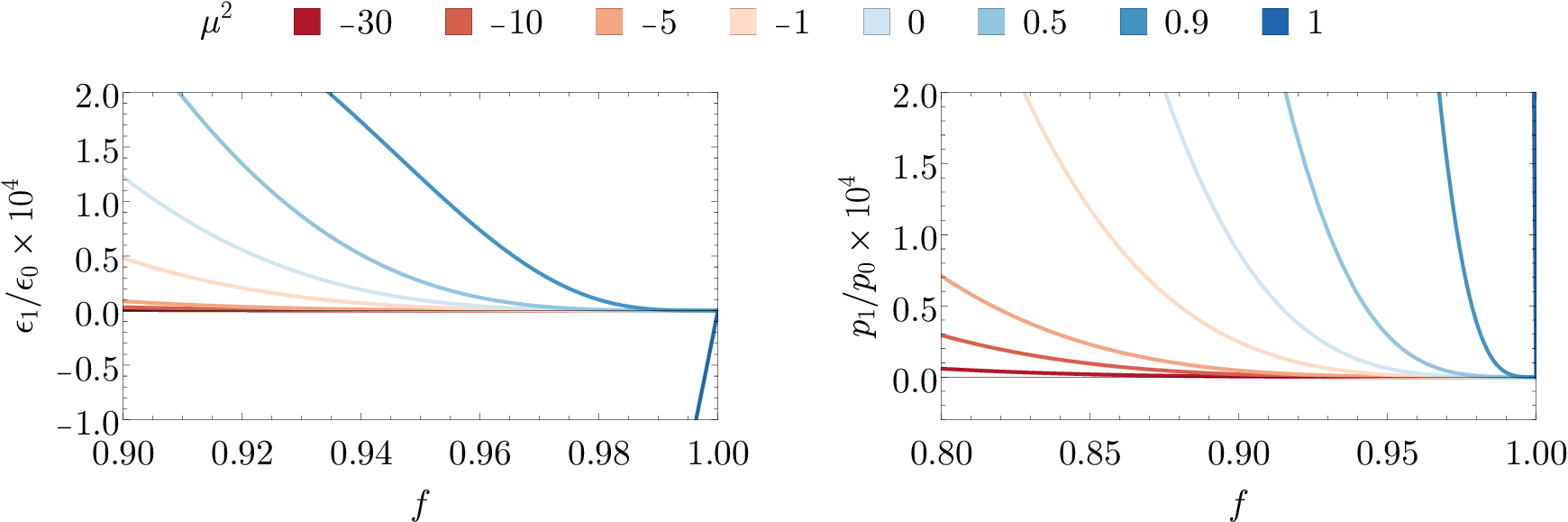}
\caption{Radial dependence far away from the horizon of 
the relative corrections to the energy density
and pressure in the limit of large~$\xi^2$, for different values of 
the parameter~$\mu^2$.}
\vskip-2mm
\label{LargeXiEpsilonPressureFigZoomed}
\end{figure}

At first glance, the plots in Fig.~\ref{LargeXiEpsilonPressureFig} 
may suggest that both the energy density and pressure remain above the corresponding 
$P(X)$ values, that is, that the relative corrections are always positive. 
However, this is not the case for the energy density. 
Far away from the horizon, the energy-density correction can become negative, 
while the pressure correction remains positive, as illustrated in 
Fig.~\ref{LargeXiEpsilonPressureFigZoomed}.

To clarify this behavior, we derive the asymptotic form of the energy density 
and pressure far from the horizon,
\begin{equation}
\frac{\epsilon_1}{\epsilon_0}
	\ \overset{f \to 1}{\longsim}
    \left\{
    \begin{matrix}
    \displaystyle
	\frac{ (1\!-\!f )^4 }{\sigma_0^6}
		\frac{ \bigl[ 3 \!-\! 5 \mu^2 \!-\! 16( 3 \!-\! \mu^2 ) \beta_0^3 \beta_1 \bigr] }{2(3 \!+\!\mu^2)}
		\, ,
    \\
    \displaystyle
	(1\!-\!f )^4
		\frac{ \bigl[ 3 \!-\! 5 \mu^2 \!-\! 16( 3 \!-\! \mu^2 ) \beta_0^3 \beta_1 \bigr] }
        {2 (1 \!-\! \mu^2)^2 (3\!-\!2\mu^2)}
        \, ,
    \end{matrix}
    \right.
\qquad
\frac{p_1}{p_0}
	\ \overset{f \to 1}{\longsim}
    \left\{
    \begin{matrix}
    \displaystyle
	\frac{(1\!-\!f)^4}{\sigma_0^6}
		\frac{ \bigl[ - 1 \!-\! 48\beta_0^3 \beta_1 \bigr]}{6}
	\, ,
    &
    \ \ \mu^2 \ge 0 \, ,
    \\
    \displaystyle
	(1\!-\!f)^4
		\frac{ \bigl[ - 1 \!-\! 48\beta_0^3 \beta_1 \bigr] }{6 (1\!-\!\mu^2) (1\!-\!2\mu^2)}
		\, ,
    &
    \ \ \mu^2 < 0
    \, .
    \end{matrix}
    \right.
\label{LargeXiFoneEpsilonPasymp}
\end{equation}
Here the asymptotic behavior of~$\sigma_0$ is given by the right-hand expression
in~(\ref{Inequality1}). The dependence of the bracketed coefficients in these two 
expressions on the mass parameter~$\mu^2$ is plotted in 
Appendix~\ref{app: Derivatives at the critical point}. 
It is found that the first coefficient is always negative, except for values of~$\mu^2$
close to unity, where it becomes positive, while the latter coefficient is always positive.

\medskip

\begin{figure}[h!]
\vskip+3mm
\centering
\includegraphics[width=15cm]{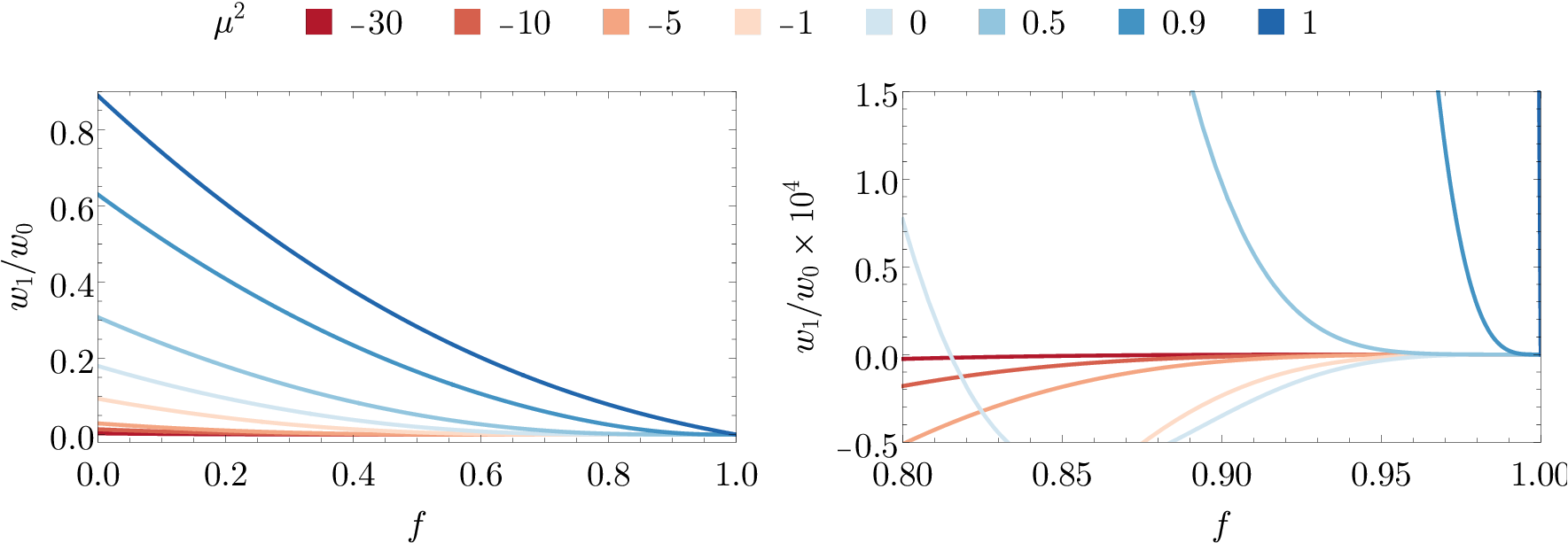}
\vskip-2mm
\caption{Radial dependence of the relative correction to the equation-of-state parameter in the limit of large~$\xi^2$, for different values of 
the parameter~$\mu^2$. The right panel shows a zoomed-in region 
near~$f\!=\!1$.}
\label{LargeXiwFig}
\end{figure}

The first correction~$w_1$ to the equation-of-state parameter,
\begin{equation}
w = w_0 + \frac{w_1}{\xi^2} + \frac{w_2}{\xi^4} + \dots \, ,
\end{equation}
can be expressed in terms of the relative energy-density and pressure corrections,
and the~$P(X)$ equation of state parameter as
\begin{equation}
w_1 = w_0 \Bigl( \frac{ p_1 }{p_0} - \frac{\epsilon_1 }{\epsilon_0} \Bigr)
\, .
\end{equation}
The relative correction to the equation-of-state parameter is shown in the left
panel of Fig.~\ref{LargeXiwFig}.

We observe that the equation-of-state parameter approaches the~$P(X)$ value 
far from the horizon, but not always from above, as is evident from the right panel
in Fig.~\ref{LargeXiwFig}. This can be understood analytically by examining
the asymptotic form of this correction,
\begin{equation}
\frac{w_1}{w_0} \ \overset{f \to 1}{\longsim} \
    \left\{
    \begin{matrix}
    \displaystyle
    \frac{ (1\!-\!f)^4}{ \sigma_0^6}
	\frac{ \bigl[ - 6 \!+\! 7 \mu^2 \!-\! 48\mu^2 \beta_0^3 \beta_1 \bigr]}
        {3 (3 \!+\! \mu^2) }
	\, ,
    &
    \ \ \mu^2 \ge 0 \, ,
    \\
    \displaystyle
    (1\!-\!f )^4
		\frac{ \bigl[ - 6 \!+\! 19\mu^2 \!-\! 16\mu^4 \!-\! 48 \mu^2  \beta_0^3 \beta_1 \bigr] }
        {3 ( 3 \!-\! 2\mu^2 ) (1\!-\!2\mu^2) (1 \!-\! \mu^2)^2}
        \, ,
    &
    \ \ \mu^2<0 \, .
    \end{matrix}
    \right.
\label{dw2}
\end{equation}
As shown in Appendix~\ref{app: Derivatives at the critical point},
the coefficients of these expressions are generally negative, except 
for positive values of~$\mu^2$, for which the coefficient 
can become positive.

\medskip

\begin{figure}[h!]
\centering
\includegraphics[width=15cm]{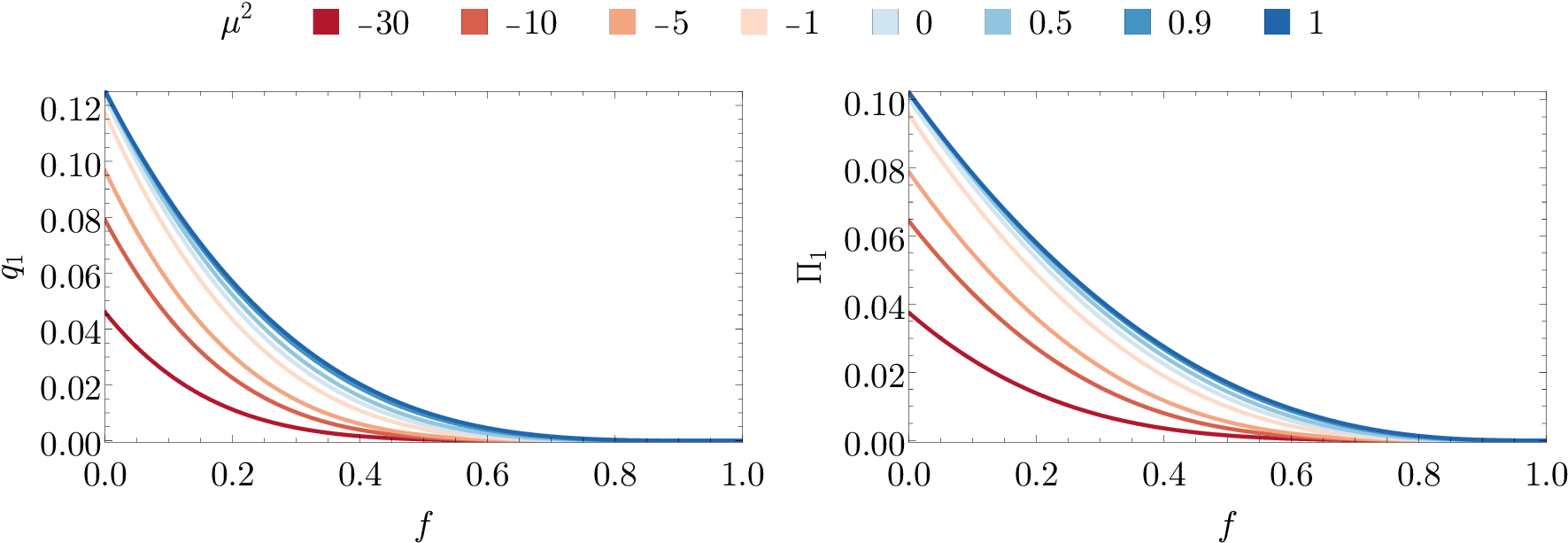}
\vskip-2mm
\caption{Radial dependence of the leading corrections to the heat flux
and anisotropic stress in the limit of large~$\xi^2$, for different values of 
the parameter~$\mu^2$.}
\vskip-2mm
\label{LargeXiqPiFig}
\end{figure}

In the~$P(X)$ limit, the magnitudes of the heat flux and anisotropic stress vanish, 
but they are generated by complex-scalar gradient corrections. 
Their leading Eckart-frame contributions in the limit of large~$\xi^2$ are
\begin{equation}
q_1 =
	\frac{ \dot{\varphi}_0^4 }{\lambda} \biggl[
		(1\!-\!f)^4 \frac{\sqrt{ 1 \!-\! fY }}{Y} \Bigl( \frac{d\sigma_0}{df} \Bigr)^{\!2} 
		\biggr]
		\, ,
\qquad\quad
\Pi_1 =
	\frac{ \dot{\varphi}_0^4 }{\lambda} \biggl[
		(1\!-\!f)^4 \sqrt{\frac{2}{3}} \frac{1}{Y} 
		\Bigl( \frac{d\sigma_0}{df} \Bigr)^{\!2} 
		\biggr]
		\, .
\end{equation}
These quantities measure the first deviations from ideal-fluid behavior,
showing that the largest departures from the perfect-fluid limit in the 
small-gradient regime occur close to the horizon, as evident from 
Fig.~\ref{LargeXiqPiFig}.

\subsubsection{Finite $\xi^2$ regime}

\begin{figure}
\centering
\includegraphics[width=15cm]{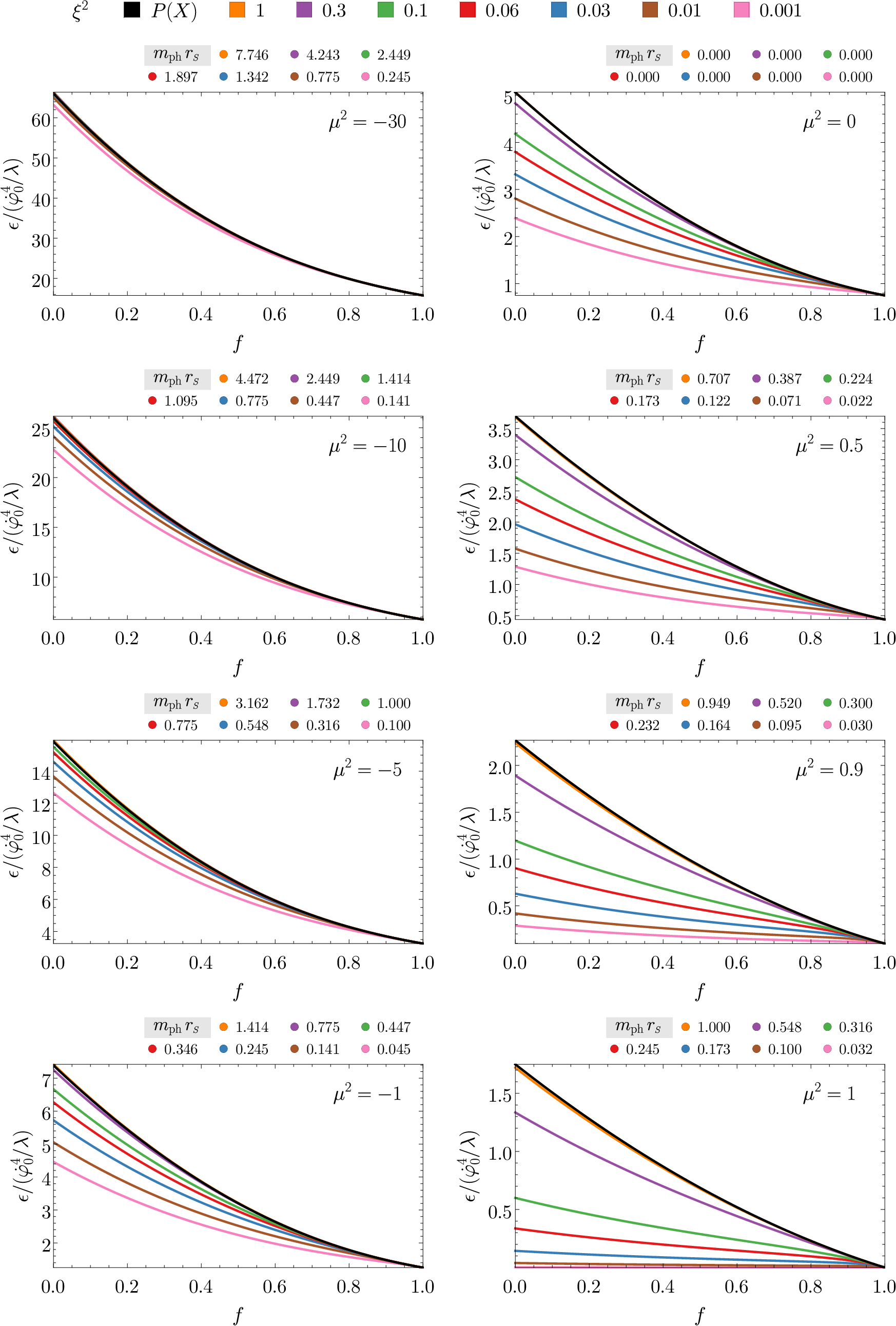}
\caption{Radial dependence of the energy density for different choices of 
parameters~$\xi^2$ and~$\mu^2$. The numbers above the panels indicate the 
field mass corresponding to the true potential minimum in flat space, 
expressed in units of the inverse Schwarzschild radius.}
\label{epsilon figure}
\end{figure}

\begin{figure}
\centering
\includegraphics[width=15cm]{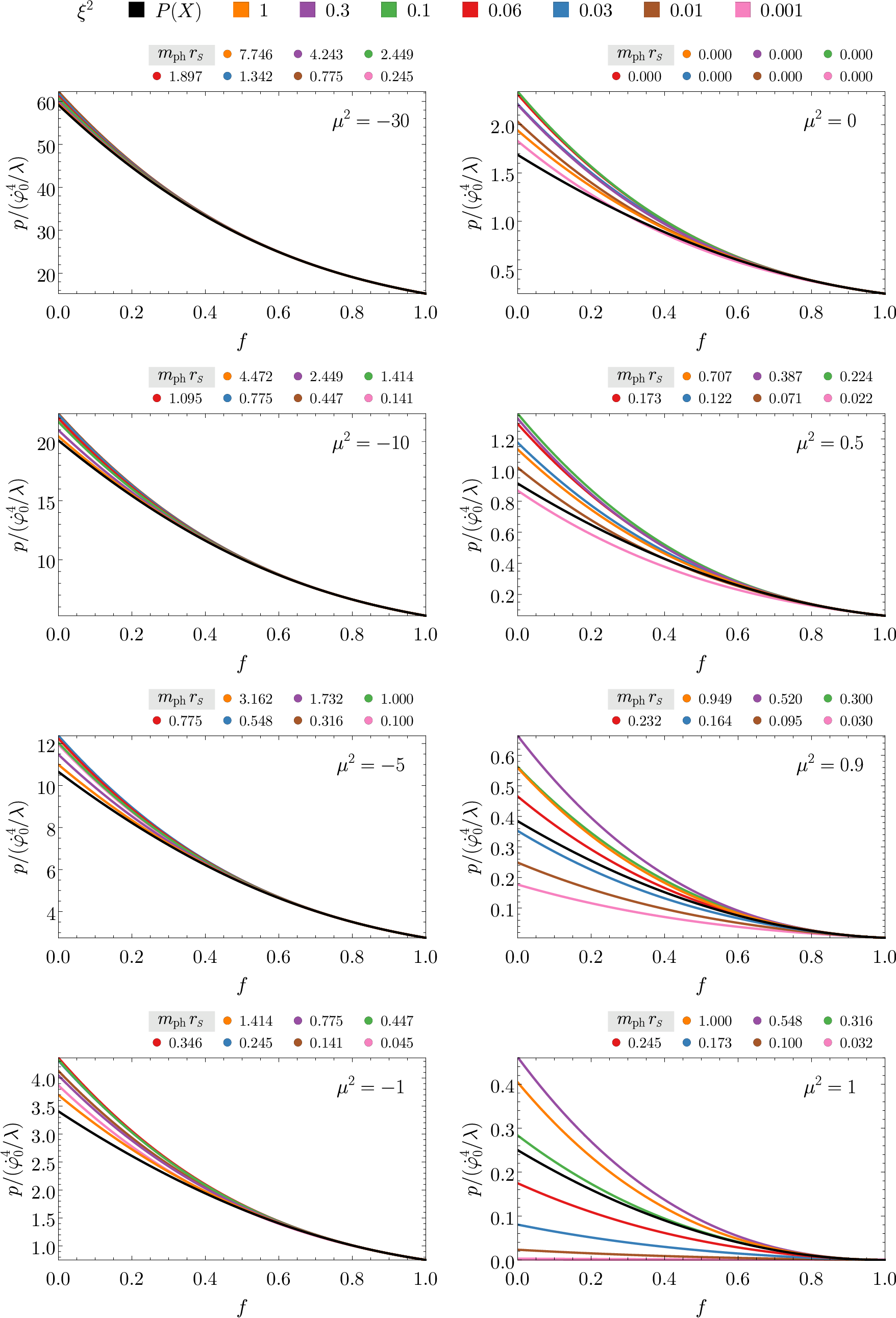}
\caption{Radial dependence of the pressure for different choices of 
parameters~$\xi^2$ and~$\mu^2$. The numbers above the panels indicate the 
field mass corresponding to the true potential minimum in flat space, 
expressed in units of the inverse Schwarzschild radius. }
\label{p figure}
\end{figure}

\begin{figure}
\centering
\includegraphics[width=15cm]{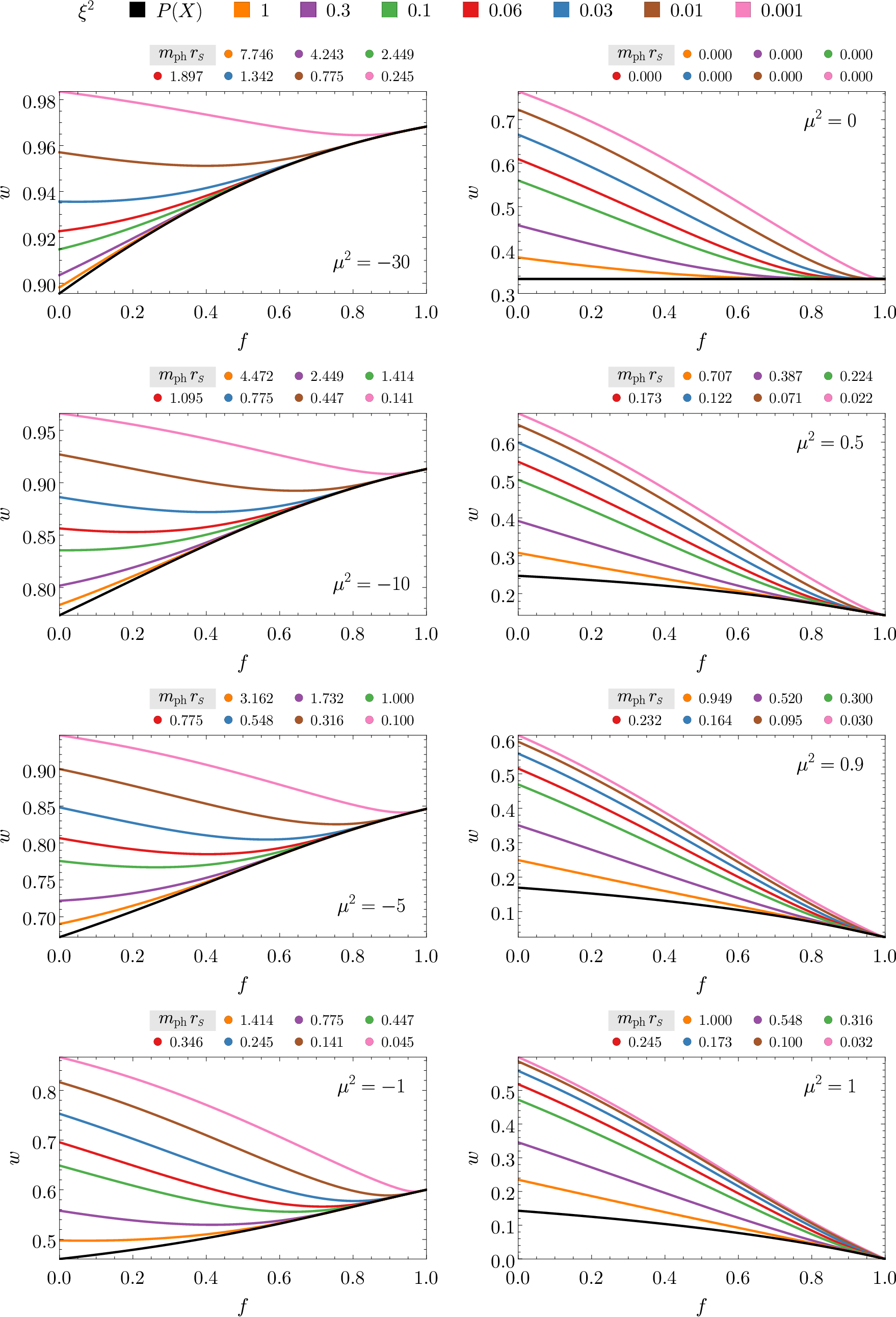}
\caption{Radial dependence of the equation-of-state parameter for different 
choices of parameters~$\xi^2$ and~$\mu^2$. For finite~$\xi^2$,
the equation of 
state is always harder close to the horizon, compared to the~$P(X)$ value.
The numbers above the panels indicate the field mass corresponding to the true 
potential minimum in flat space, expressed in units of the inverse Schwarzschild radius. 
}
\label{w plots}
\end{figure}

\begin{figure}
\centering
\includegraphics[width=15cm]{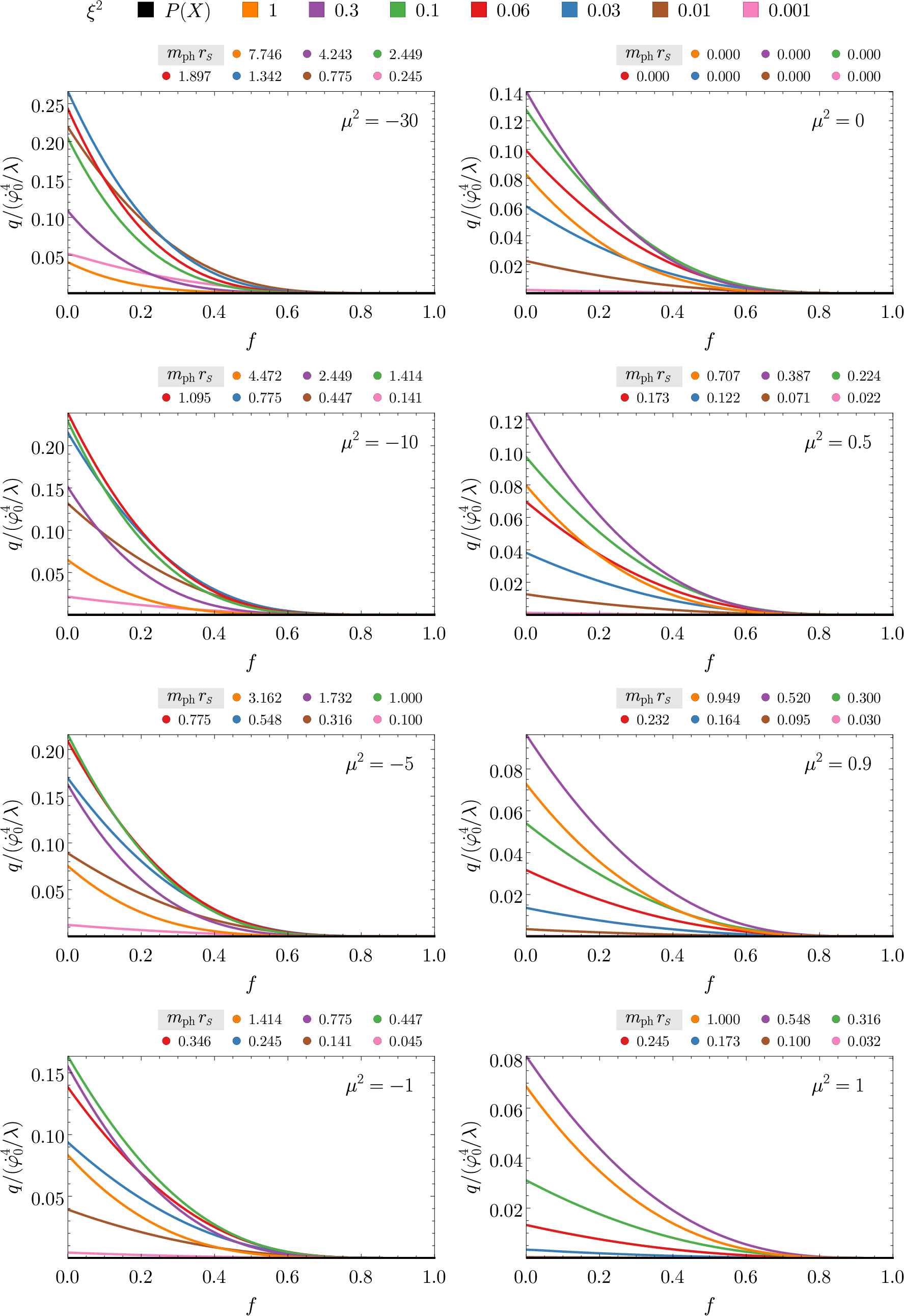}
\caption{Radial dependence of the magnitude of the heat flux~(\ref{heat_transfer_spec}) 
for different choices of parameters~$\xi^2$ and~$\mu^2$. 
The numbers above the panels indicate the field mass corresponding to the true 
potential minimum in flat space, expressed in units of the inverse Schwarzschild radius. }
\label{q plots}
\end{figure}

\begin{figure}
\centering
\includegraphics[width=15cm]{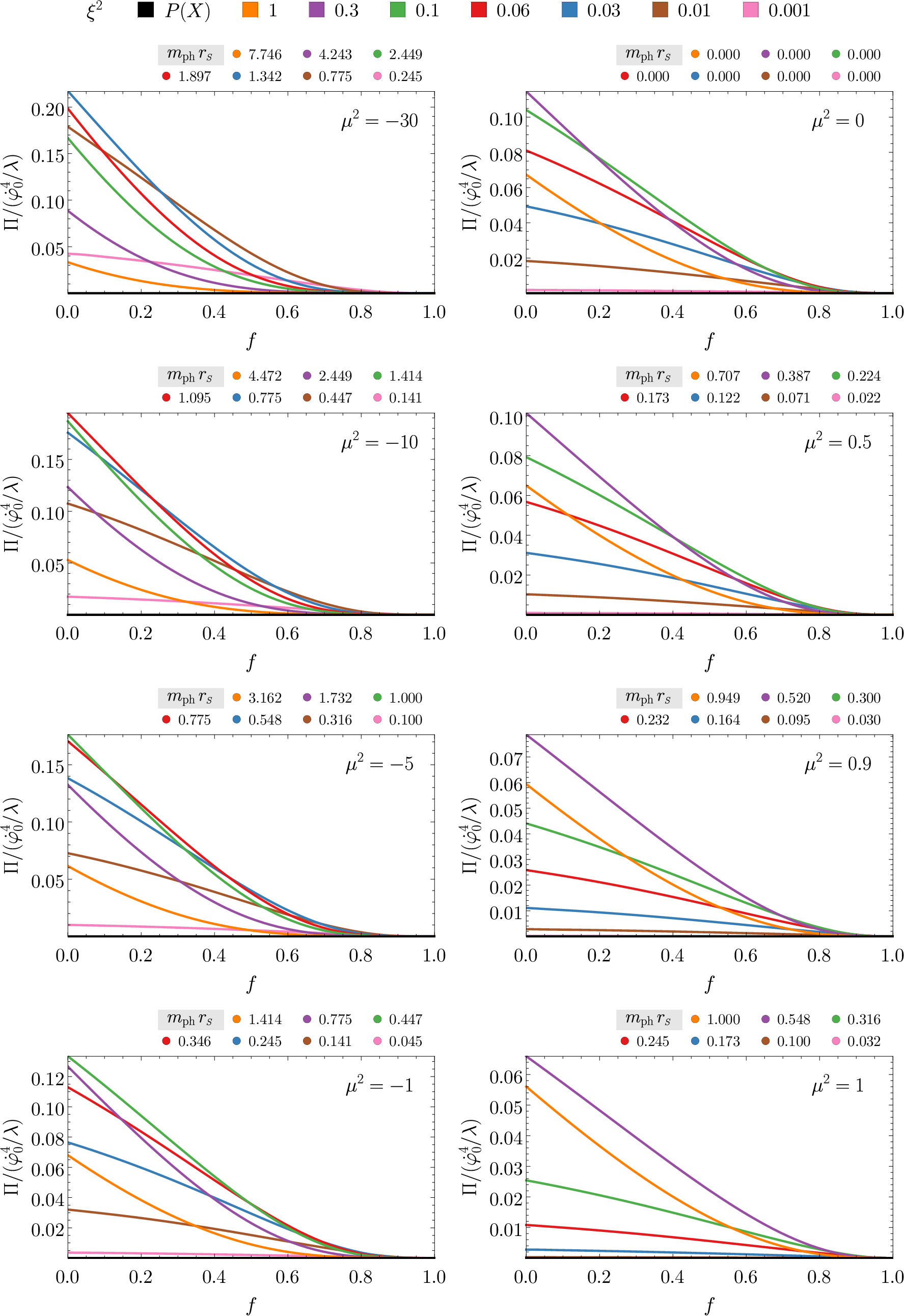}
\caption{Radial dependence of the magnitude of the anisotropic stress
(\ref{anisotropic_stress_spec}) for different choices of 
parameters~$\xi^2$ and~$\mu^2$. 
The numbers above the panels indicate the field mass corresponding to the true 
potential minimum in flat space, expressed in units of the inverse Schwarzschild radius. }
\label{Pi plots}
\end{figure}

\begin{figure}
\centering
\includegraphics[width=15cm]{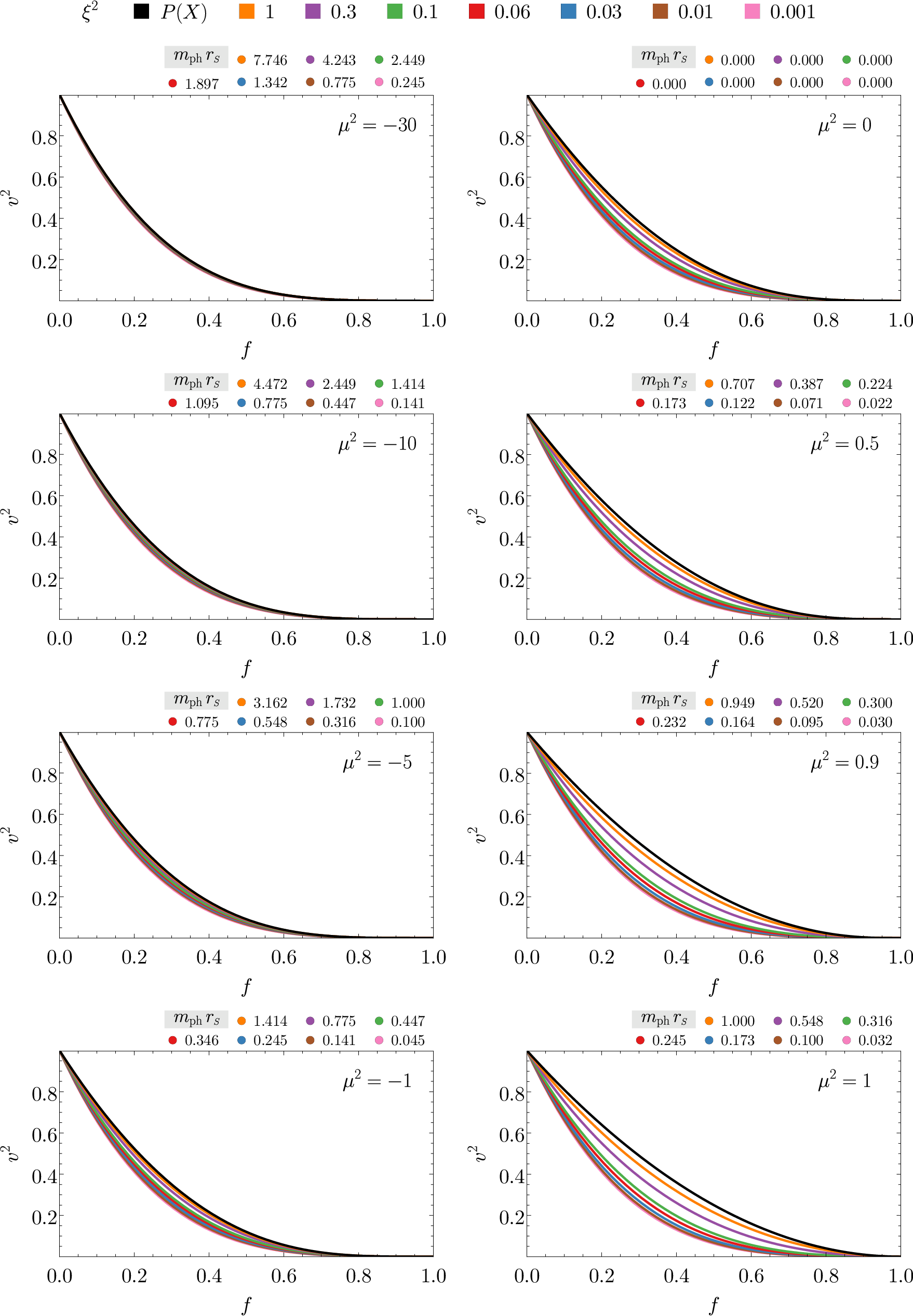}
\caption{Radial dependence of the $U(1)$ shift-charge velocity~(\ref{definition vc2})
for different choices of parameters~$\xi^2$ and~$\mu^2$. 
For finite~$\xi^{2}$ the velocity remains below the corresponding~$P(X)$ value, 
approaching it from below both near and far from the horizon. The numbers above the panels indicate the field mass corresponding to the true 
potential minimum in flat space, expressed in units of the inverse Schwarzschild radius. }
\label{vc2 plots}
\end{figure}

In the~$P(X)$ limit, $\xi^{2} \!\rightarrow\! \infty$, when the gradients are negligible, 
the definitions of the energy density and 
pressure in~(\ref{eck_en}) and~(\ref{eck_pres}) reduce to those of~(\ref{px_en_pres}), 
while the heat flux and anisotropic stress vanish. For finite~$\xi^{2}$, however, 
$q$ and~$\Pi$ acquire nonzero values due to gradient contributions that no longer 
cancel in the energy–momentum tensor, and the ideal-fluid description no longer holds. 
The following figures illustrate the radial 
variation of the energy density (Fig.~\ref{epsilon figure}), pressure 
(Fig.~\ref{p figure}), equation-of-state parameter
(Fig.~\ref{w plots}), heat flux (Fig.~\ref{q plots}), anisotropic stress 
(Fig.~\ref{Pi plots}), and shift charge velocity (Fig.~\ref{vc2 plots}) for 
different values of $(\mu^{2},\xi^{2})$, including the $P(X)$ limit.

We see that for large~$\xi^{2}$, all quantities approach their $P(X)$ values, 
as expected from the perturbative analysis in Sec.~\ref{largexisquaredenergypressure}. 
As $\xi^{2}$ decreases, gradient corrections become significant near the horizon, 
where the field varies most rapidly. Far away from the horizon, regardless of
the value of~$\xi^2$, all quantities asymptote to the corresponding~$P(X)$ ones.

We see in Fig.~\ref{epsilon figure}
that, for finite~$\xi^2$, the energy density of the steady-state 
accreting flow in the complex scalar model is generally smaller than in the 
corresponding~$P(X)$ model close to the horizon. This reverses the trend 
observed in the large-$\xi^2$ limit (see Fig.~\ref{LargeXiEpsilonPressureFig}). 
Far away from the horizon, the energy density asymptotes to that of the~$P(X)$
case, as also seen in Fig.~\ref{epsilon figure}. However, whether it asymptotes to
this limit from above or from below depends
sensitively on the parameters~$\xi^2$ and~$\mu^2$, as indicated by the 
asymptotic behavior of the energy density:
\begin{equation}
\frac{\Delta\epsilon}{\epsilon_0} 
	\ \overset{f \to 1}{\longsim} \ 
    \left\{
    \begin{matrix}
    \displaystyle
	\frac{ 2(1\!-\!f )^4 }{ ( 3 \!+\! \mu^2 ) \sigma_0^6}
	\biggl[
		\frac{ 3 \!-\! 5 \mu^2 }{4\xi^2}
		-
		( 3 \!-\! \mu^2 ) \bigl( \beta^4 \!-\! \beta_c^4 \bigr)
		\biggr]
		\, ,
    &
    \mu^2 \ge 0 \, ,
    \\
    \displaystyle
	\frac{ 2(1\!-\!f )^4 }{ (3 \!-\! 2\mu^2)(1\!-\!\mu^2)^2}
	\biggl[
		\frac{ 3 \!-\! 5 \mu^2 }{4\xi^2}
		-
		( 3 \!-\! \mu^2 ) \bigl( \beta^4 \!-\! \beta_c^4 \bigr)
		\biggr]
		\, ,
        &
        \mu^2<0
    \end{matrix}
    \right.
\end{equation}
The competition between the two terms in brackets
determines the sign of the deviation far away from the horizon, 
explaining why the relative 
behavior of~$\epsilon$ depends on~$\xi^2$ and~$\mu^2$. These details are not 
visible in Fig.~\ref{epsilon figure}, as the magnitude by which the coloured 
curves can rise above the black~$P(X)$ curve is several orders of magnitude 
smaller than the deviations close to the horizon.

The radial dependence of the pressure shows an intricate 
behavior in 
Fig.~\ref{p figure}. For a given~$\mu^2$, as we lower~$\xi^2$ 
the pressure close to the horizon first exceeds the corresponding~$P(X)$ value,
but the trend reverses at some point as~$\xi^2$ gets smaller, and the pressure 
even drops below the~$P(X)$ value. Likewise, the behavior far away from the horizon 
also depends crucially on the two parameters, as can be observed from the 
asymptotic form of the deviation from the~$P(X)$ curve,
\begin{equation}
\frac{\Delta p}{p_0} 
    \ \overset{f \to 1}{\longsim} \
    \left\{
    \begin{matrix}
    \displaystyle
	\frac{2(1\!-\!f)^4}{\sigma_0^6}
	\biggl[
		\frac{-1}{12\xi^2}
		- \bigl( \beta^4 \!-\! \beta_c^4 \bigr)
		\biggr]
		\, ,
        &
        \mu^2>0
        \, ,
    \\
    \displaystyle
	\frac{2(1\!-\!f)^4}{ (1\!-\!2\mu^2) (1\!-\!\mu^2)}
	\biggl[
		\frac{-1}{12\xi^2}
		- \bigl( \beta^4 \!-\! \beta_c^4 \bigr)
		\biggr]
		\, ,
        &
        \mu^2<0
        \, .
    \end{matrix}
    \right.
\end{equation}

Figure~\ref{w plots} shows that the equation-of-state parameter $w$ can 
deviate significantly from the $P(X)$ case, especially near the horizon.
There, $w$, of the scalar field is larger/harder than that in the $P(X)$ case, with smaller~$\xi^2$ 
producing larger deviations and, thus a stronger departure from the perfect-fluid 
limit as gradient corrections grow. 
Far from the horizon, $w$ tends to its $P(X)$ value, but not always from above: 
for certain parameter choices it can dip below the $P(X)$ value, as revealed by 
the asymptotic behavior far away from the horizon:
\begin{equation}
\frac{\Delta w}{w_0} \ \overset{f \to 1}{\longsim} \
    \frac{\Delta p}{p_0} - \frac{\Delta\epsilon}{\epsilon_0}
    \, .
\end{equation}
This relation shows that the asymptotic correction to~$w$ depends on the 
difference between the relative pressure and energy-density corrections. 
The non-uniform approach of~$w$ to its~$P(X)$ value therefore arises from the 
competition between these two quantities. However, these deviations from the~$P(X)$
curve are orders of magnitude smaller than the deviations close to the horizon,
and are not visible in Fig.~\ref{w plots}. Nevertheless, this reflects the 
increasing role of gradient terms in the complex scalar field dynamics at 
finite~$\xi^2$.

For finite~$\xi^{2}$, the emergence of nonzero heat flux and anisotropic stress 
(Figs.~\ref{q plots} and~\ref{Pi plots}) signals departures from perfect-fluid 
behavior. These quantities originate from the spatial gradients of the modulus 
field and vanish in the homogeneous limit. 
Their profiles are typically peaked near the horizon, where the gradients are 
largest, and decrease rapidly with increasing~$f$. 
For a fixed~$\mu^{2}$, the dependence on~$\xi^{2}$ is non-monotonic: in some radial 
ranges, a larger~$\xi^{2}$ yields a smaller~$q$ or~$\Pi$, whereas in others the trend 
reverses, reflecting the intricate interplay between the model parameters and the 
field’s spatial variation.

For the smallest values of~$\xi^2$ considered, magnitudes of the anisotropic stress and heat flux approach zero close to the horizon, compared to the curves with larger~$\xi^2$. However, we should not conclude that this signals irrelevance of non-ideal fluid quantities and approach to the ideal-fluid limit. This is because the energy density 
and pressure also exhibit similar behavior. A more careful investigation
of the properties of this limit would be necessary to draw conclusions.

Finally, Fig.~\ref{vc2 plots} shows the squared shift-charge velocity~(\ref{definition vc2}) 
of the accreting complex scalar with gradient corrections accounted for. We see that 
for finite~$\xi^2$ this velocity is generally below the 
corresponding~$P(X)$ limit, but that far away from the horizon and at the
horizon it approaches the~$P(X)$ value. We can also see that both of these
approaches occur from below by examining the asymptotic behavior,
\begin{equation}
\Delta v^2
    \ \overset{f \to 1}{\longsim} \
    \frac{( 1\!-\!f )^4 }{ \sigma_0^4 } \bigl( \beta^4 \!-\! \beta_c^4 \bigr)
    <
    0
    \, ,
\qquad
\Delta v^2
    \ \overset{f \to 0}{\longsim} \
    -4 f  \biggl[
    \frac{ 4 - \mu^2 - \beta_c^2 }{ 4  }
    -
    \frac{ 4 - \mu^2 - \beta^2 }{ 4 \!+\! \frac{1}{\xi^2} }
    \biggr]
    < 0 \, .
\end{equation}
It is tempting to conclude that this behavior might bring the 
acoustic horizon closer to the event horizon. However, this conclusion is not
warranted without a detailed study of the properties of perturbations, which
is beyond the scope of this work. It is not even clear whether any acoustic
horizon exists in the complex scalar model, where two modes are active,
compared to the single one in the~$P(X)$ model.

\section{Summary and discussion}
\label{sec: Discussion}

We have performed a systematic analysis of the relativistic Bondi accretion of a classical 
canonical complex scalar field~$\Psi \!=\! \rho \, e^{i \varphi}$ onto a Schwarzschild 
BH at rest. The scalar is assumed to self-interact via the $U(1)$-symmetric renormalizable 
potential~\eqref{eq:Potential}, which can either preserve the symmetry or admit spontaneous 
symmetry breaking. One of our main goals was to assess whether Bondi accretion can 
distinguish~$\Psi$ from its EFT description in the form of the perfect-(super)fluid $P(X)$ 
model~\eqref{specific P}. The latter is a valid description at leading order in a gradient 
expansion in derivatives of the modulus~$\rho$, as recalled in Sec.~\ref{sec: EFT of the complex scalar}. 

Crucially, the value of $\rho$ at spatial infinity is uniquely determined by the local 
phase velocity $\dot{\varphi}_{0}$. This value also minimizes the effective potential 
of~$\rho$ for a given asymptotically homogeneous charge density. Hence, at spatial infinity, 
where we assume an asymptotically Minkowski spacetime or a parametrically low-curvature 
cosmology, the complex scalar field $\Psi$ is in a state that admits a faithful $P(X)$ 
description. Bondi accretion therefore always interpolates from a $P(X)$-like state 
far away from the BH to a potentially very different configuration close to the BH.

There are four relevant physical parameters in our investigation: the phase field velocity $\dot{\varphi}_0$ at spatial infinity, the Schwarzschild radius $r_{\scr S}$, the quartic coupling $\lambda$, and the scalar mass squared $m^2$. We remain agnostic about their specific values, requiring only that $\lambda>0$ and assuming that it is small enough to avoid quantum strong-coupling issues. The problem then reduces to a boundary-value problem for the ordinary differential equation~\eqref{CS profile equation}, which governs the radial evolution of the dimensionless modulus $\sigma \!=\! \sqrt{\lambda}\rho/\dot{\varphi}_0$, with boundaries at the horizon and at spatial infinity. 

This \emph{master equation}, written in terms of the compactified and dimensionless
radial variable~$f\!=\!1\!-\!r_{\scr S}/r$, depends only on three dimensionless parameters: 
the mass parameter $\mu^2 \!=\! m^2/\dot{\varphi}_{0}^2$, the gradient 
parameter $\xi \!=\! r_{\scr S}\dot{\varphi}_{0}$, and, crucially, an 
a priori unknown value $\beta$ of the modulus at the horizon,~$\beta \!=\! \sigma(r_{\scr S})$. 
The latter uniquely determines the flux of the~$U(1)$ charge, see~\eqref{Jr} 
and~\eqref{dimensionless def}. Thus, the master equation contains an unknown boundary 
parameter which must be fixed by requiring the existence of a smooth solution. 
Technically, the procedure resembles finding the Coleman bounce solution for false vacuum
decay~\cite{Coleman:1977py}, see Eq.~\eqref{mechanica}. For a given BH mass and a given asymptotic phase-field velocity, 
the steady-state, spherically-symmetric flow then admits only a single value of~$\beta$.

The gradient parameter~$\xi^2$ plays a central role, and its value can vary 
by many orders of magnitude depending on the physical setup. For instance, 
in a neutron-star context, one may have $\dot{\varphi}_0 \!\sim\! 100 \, \text{MeV}$, 
which yields $\xi \!\sim\! 10^{7}$ for a PBH of mass $10^{22}\, \text{g}$. On the 
other hand, for ultralight DM, one could envision $\dot{\varphi}_0 \!\sim\! 10^{-20} \, \text{eV}$, 
which for a solar-mass BH gives $\xi \!\sim\! 10^{-10}$. In the formal limit~$\xi^{2}\!\to\!\infty$,
gradient terms are negligible in the master equation~\eqref{CS profile equation} and the 
complex scalar model reduces to the perfect-fluid $P(X)$ model, whose Bondi accretion is reviewed in Section~\ref{sec: Accretion for P(X)}. The corresponding solutions $\sigma_{0}(f)$ are plotted in Fig.~\ref{P(X)figureProfiles}, and the corresponding $\beta$ is given in Eqs.~\eqref{critical beta} and~\eqref{critical f sigma}.

Section~\ref{sec: Accretion for complex scalar field} 
examined finite-$\xi^{2}$ 
effects on the modulus profile in detail. For large but finite~$\xi^{2}$, the 
modulus field profiles $\sigma(f)$ deviate perturbatively from the $P(X)$ 
solutions $\sigma_{0}(f)$: $\sigma(f)$ generally drops below $\sigma_{0}(f)$ 
close to the horizon, but rises slightly above $\sigma_{0}(f)$ at large radii, as shown in Fig.~\ref{overshootingP(X)}. In the opposite limit of small~$\xi^{2}$, gradients 
dominate close to the horizon, and we identified the emergence of a thin boundary 
layer in the compactified coordinate near~$f\!\sim\!1$, where the modulus field varies 
sharply as gradient terms give way to the potential terms, see Fig.~\ref{small xi profiles}.

For arbitrary finite~$\xi^{2}$, we solved the profile equation numerically using the 
methods outlined in Section~\ref{subsec: Solving profile equation for finite xi}, 
obtaining the critical flux~$\beta$ and the corresponding field profiles that interpolate 
between the two analytically tractable regimes. The numerical solutions are shown in 
Fig.~\ref{sigma profiles} for different parameter choices. In all cases the value of 
the modulus at the horizon,~$\beta$, is smaller than in the $P(X)$ case for the same 
conditions at spatial infinity. Thus the accretion rate of $\Psi$ is always smaller than 
that of the corresponding $P(X)$ theory. As expected, the solutions $\sigma(f)$ are monotonic 
functions that increase towards the horizon, but they can be either convex or concave, as illustrated 
in Fig.~\ref{dsigma}. Although the numerical profiles in Fig.~\ref{sigma profiles} appear 
to satisfy~$\sigma(f)\!<\!\sigma_{0}(f)$ everywhere, in reality this inequality holds only 
up to some distance from the BH; far away from the horizon the complex scalar profile eventually 
approaches and can slightly overshoot the $P(X)$ solution, in agreement with the analytic 
asymptotics.

In Section~\ref{sec: Equation of state}, we analyzed the 
energy-momentum tensor of the accreting flow. 
The limit~$\xi^{2}\!\to\!\infty$ reproduces a 
perfect-fluid~$P(X)$ form with energy density and pressure given by~(\ref{px_en_pres}). 
At finite~$\xi^{2}$, gradient corrections induce heat flux and anisotropic stress contributions, 
defined in~(\ref{heat_transfer_spec}) and~(\ref{anisotropic_stress_spec}). 
These terms signal the breakdown of the ideal-fluid description. 
The corresponding radial dependence of these quantities, together with the energy density, pressure, 
and equation-of-state parameter~$w=p/\epsilon$, is displayed in Figs.~\ref{epsilon figure}–\ref{Pi plots}. 
These results demonstrate that gradient corrections are most pronounced near the horizon, 
where spatial derivatives of the field are largest, and that all quantities asymptotically approach 
their $P(X)$ limits far from the BH.

Another interesting feature concerns the effective equation-of-state parameter~$w$, 
see Fig.~\ref{w plots}. The following trend is apparent: for potentials preserving the~$U(1)$ 
symmetry, the equation of state always becomes substantially harder close to the horizon. 
Even when the cosmological equation of state is close to dust,~$w$ grows to~$\mathcal{O}(1)$ 
near the horizon. By contrast, for potentials that realize spontaneous symmetry breaking and 
for not too small~$\xi^2$, the equation of state can instead become softer in the 
near-horizon region. This behavior is somewhat counterintuitive and may be relevant for 
accretion of a nuclear condensate in a neutron-star core onto a central PBH. 
Note that these kinds of potentials lead to rather hard equations of state
at spatial infinity.
Accretion of real scalar fields describing nuclear matter inside neutron stars has recently 
been investigated in~\cite{Richards:2021upu,Richards:2021zbr}; 
see also~\cite{Petrich:1988zz,Babichev:2008dy,Babichev:2008jb}. In a similar spirit, it would be 
interesting to apply our results to accretion onto PBHs of the so-called quantum liquid in dwarf stars, see~\cite{Gabadadze:2008mx,Gabadadze:2009jb}.

It is also interesting to note that, even though we only imposed that $\partial_{\mu}\varphi$ 
is timelike at spatial infinity, the solutions preserve this property everywhere, see 
Fig.~\ref{X plots}. This is somewhat counterintuitive, as one might expect $\partial_{\mu}\varphi$ 
to tend toward a lightlike configuration on the horizon, cf.~\cite{Frolov:2002va}. 
Moreover, $(X \!-\! m^2)$ plays the role of an effective cutoff in the EFT, see 
Eq.~\eqref{EFTexample} and the discussion of perturbations in~\cite{Babichev:2018twg}. 
Our numerical solutions in Fig.~\ref{X plots} show that this background-dependent cutoff 
increases monotonically toward the BH. It is an intriguing open question whether other 
accreting systems beyond perfect fluids exhibit the same behavior: maintaining timelike 
gradients while the effective EFT cutoff grows closer to the BH horizon.

\bigskip

\begin{figure}[h]
\vskip+3mm
\centering
\includegraphics[width=15cm]{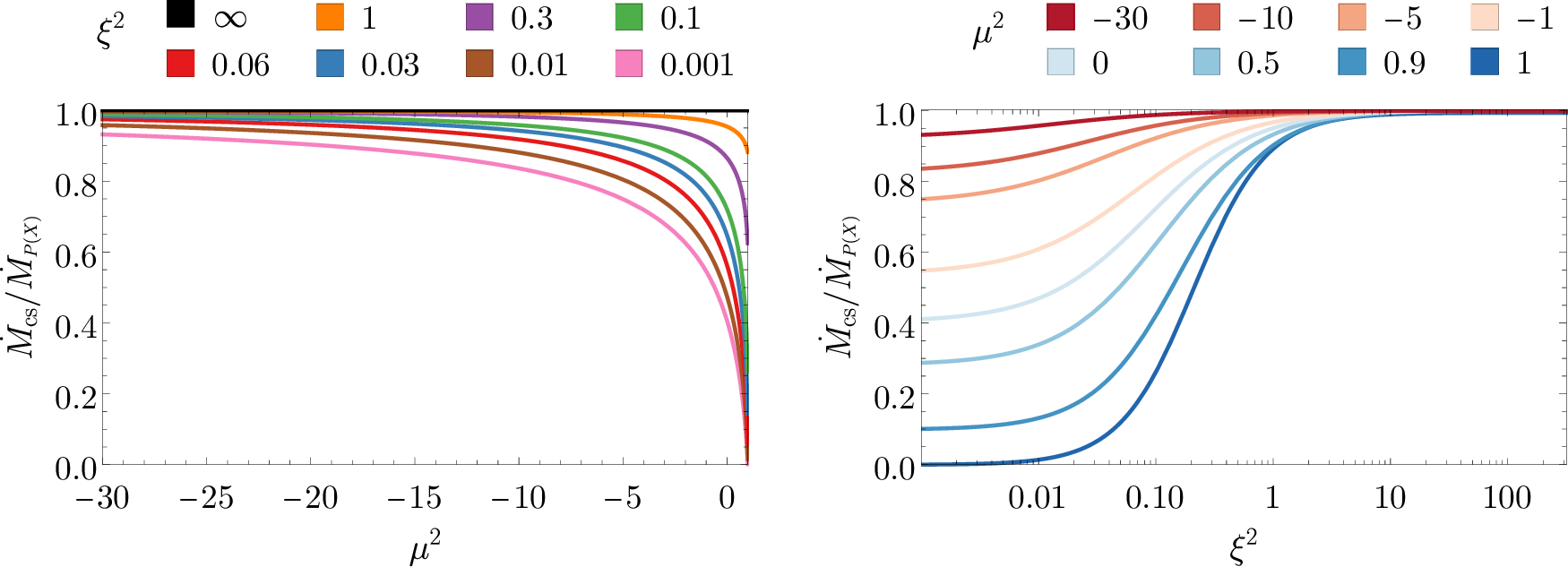}
\vskip-2mm
\caption{Ratio of the accretion rate for the complex scalar model to that of the corresponding~$P(X)$
model that it UV-completes, for different values of~$\mu^2$ and~$\xi^2$. 
Finite-gradient corrections ($\xi^2$ finite) lower the accretion rate relative to the~$P(X)$ limit, 
with the suppression increasing for smaller~$\xi^2$ and increasingly tachyonic masses.}
\label{accretion ratio}
\end{figure}

A key observable associated with this system is the mass accretion rate of the black hole, 
which provides a direct measure of the impact of gradient corrections. 
In this paper, we 
find\footnote{It is worth noting that one could estimate the rate~(\ref{nice_rate}) 
using dimensional analysis, as typically~$\beta\!=\!\mathcal{O}(1)$.}
\begin{equation}
\dot{M} 
	= 4\pi r_{\scr S}^{2} \Bigl( \frac{\dot{\varphi}_{0}^{4}}{\lambda} \Bigr) \beta^2 \, ,
\label{nice_rate}
\end{equation}
for the complex scalar model, with the corresponding rate in the $P(X)$ limit 
obtained by taking~$\xi^2\!\to\!\infty$ in the flux parameter~$\beta$. 
Thus, the ratio of mass accretion rates between the~$P(X)$ model and its ultraviolet completion 
in the form of a complex scalar is
\begin{equation}
\frac{ \dot{M}_{\rm cs} }{ \dot{M}_{\scr P(X)} }
	=
	\frac{ \beta^2(\xi^2,\mu^2) }{ \beta^2(\infty,\mu^2) }
	\, ,
\end{equation}
which is shown in Fig.~\ref{accretion ratio}. 
This figure summarizes the main quantitative result of our work: 
\textit{finite-gradient corrections systematically lower the accretion rate compared to the perfect-fluid~$P(X)$ case.} 
The suppression increases as~$\xi^{2}$ decreases, and 
decreases as the mass becomes more tachyonic, 
with the ratio approaching unity as~$\xi^{2}\!\to\!\infty$ 
or as~$\mu^{2}\!\to\!-\infty$. 
This demonstrates that the perfect-fluid limit provides an upper bound on the accretion efficiency 
of the complex scalar model. It is also worth noting that the dependence of $\beta$ on $\xi$ implies that the usual perfect-fluid differential relation $\dot M \propto M^2$ is modified for the accretion of the complex scalar field.

Moreover, as it follows from the qualitative arguments of subsection \ref{qualitative}, from  \eqref{lower_bound} and from numerics, for arbitrary finite $\xi$ the Bondi accretion rate also has a lower bound, hence, is always in the interval 
\begin{equation}
    \left(1-\mu^{2}\right)\frac{\dot{\varphi}_{0}^{4}}{\lambda}<\frac{\dot{M}}{4\pi r_{\scr S}^{2}}<\frac{\dot{\varphi}_{0}^{4}}{\lambda}\,\beta_{c}^{2}\left(\mu^{2}\right)\,,
\end{equation}
where as usual $\mu^{2}=m^{2}/\dot{\varphi}_{0}^{2}$ and $\beta_{c}^{2}\left(\mu^{2}\right)=\beta^2(\infty,\mu^2)$ is given by  \eqref{critical beta} and \eqref{critical f sigma}. Note that, for $\mu^2\leq 0$, one has $1<\beta_{c}^{2}/(1-\mu^{2})<3\sqrt{3}/2\simeq2.6$, while for $\mu^{2}\!\to\!-\infty$ one obtains  $\beta_{c}^{2}/(1-\mu^{2})\!\to\!1$. Note that the bounds can be reformulated in terms of the energy density at infinity using \eqref{connection phi/E}. 

A self-interacting complex scalar remains a viable DM candidate. 
However, if it constitutes the dominant component of DM, 
constraints on galactic halo formation~\cite{Arbey:2001qi,Arbey:2001jj} require $\xi^{2}$ 
to be extremely large for accretion onto stellar or supermassive BHs, 
placing the system deep in the perfect-fluid regime. 
Indeed, this is consistent with the assumptions of Ref.~\cite{Feng:2021qkj}, 
which analyzed Bondi accretion of a self-interacting complex scalar onto a supermassive 
BH (e.g. Sagittarius~A*) precisely in this limit by considering the 
effective~$P(X)$ description. Therefore, observationally distinguishing between the 
two models in this regime appears unlikely.

If instead the complex scalar constitutes only a subcomponent of the DM, 
the constraint on~$\xi^{2}$ is relaxed. 
In this case, finite-gradient effects could be relevant for accretion onto small black holes, 
such as PBHs, where~$r_{\scr S}$ is much smaller \cite{Carr:2020gox} and $\xi$ may take moderate values. 
Accretion of scalar-field DM onto PBHs could then 
influence their mass growth and evaporation thresholds.

\bigskip

A natural extension of the present work is to study perturbations around the steady-state 
profiles obtained here. In~\cite{Moncrief}, for accretion in the $P(X)$ model, it was found that 
``(i) no unstable normal modes exist which extend outside the sound horizon of the background 
flow; and (ii) there are no unstable modes which represent a standing shock at the sound
horizon'', see also~\cite{RivasplataPaz:2014gng}. It would be very interesting to determine 
whether these conclusions remain valid beyond the EFT description, i.e.~in the UV-completed 
complex scalar field setup. In particular, the presence of an acoustic horizon for the gapless 
mode may induce Cherenkov radiation from the fast gapped modes that appear in the UV completion, as 
discussed in~\cite{Babichev:2024uro,Sawicki:2024ryt}. Even if both the limiting $P(X)$ models 
and the complex scalar models are stable, the dynamics of perturbations could still reveal 
new observational signatures distinguishing the two.

Other important open issues for future investigation include whether our solution remains an attractor when starting from physically-motivated initial conditions, whether the test-field approximation holds, and whether backreaction (especially in the case of PBHs) can have a substantial effect. Addressing these questions will require more sophisticated numerical simulations, including numerical general relativity. Furthermore, it would be important to see whether one can perturbatively incorporate slow rotation of the BH, or slow motion of the BH through the condensate described by $\Psi$. One can also relax the assumption that the scalar field is isolated and consider couplings to other fields. In particular, an axion-like coupling of the phase to the electromagnetic field, $\varphi F_{\mu\nu}\tilde{F}^{\mu\nu}$, is promising from the point of view of phenomenological applications. Finally, one could consider other, more exotic types of BHs in the hope that accretion can differentiate them from the standard Schwarzschild BH. We believe that this paper is a first step toward a better understanding of these important open problems.

\subsection*{Acknowledgments}

We are grateful to Ignacy Sawicki and Constantinos Skordis for discussions 
about the numerical aspects of the project and to Eugeny Babichev for useful discussions about standard hydrodynamical picture of accretion. DG acknowledges the use of HPC 
cluster Phoebe of the CEICO at the FZU where some of the computations were performed. AV is grateful to Wei-Xiang Feng for useful discussions and comments on the e-preprint version of this paper. 
DG and AV were supported by project 24-13079S of the Czech Science Foundation 
(GA\v{C}R).

\appendix
\section{Derivatives at the critical point}
\label{app: Derivatives at the critical point}

The first correction to the~$P(X)$ profile in the large~$\xi^2$ limit, given in~(\ref{sigma1 solution}),
requires determining the first correction to the dimensionless flux, given in~(\ref{beta_1}). 
To obtain this, we need to compute the first two derivatives of the modulus field profile 
at the critical point. This, in turn, requires the evaluation of the second and third derivatives 
of the profile equation~(\ref{P(X) profile equation}) at the critical point. 
These derivatives are conveniently captured by expanding the modulus field to third order,
\begin{equation}
\sigma \ \overset{f \to f_c}{\longsim} \
    \sigma_c + \sigma'_c (f \!-\! f_c)
    + \frac{1}{2} \sigma''_c (f \!-\! f_c)^2 
    + \frac{1}{6} \sigma'''_c (f \!-\! f_c)^3 
    + \dots
    \, ,
\end{equation}
where~$\sigma_c^{(n)} \!=\! d^n \sigma/df^n (f_c)$. 
Substituting this expansion into Eq.~(\ref{P(X) profile equation}) and expanding the equation 
itself to third order yields
\begin{align}
\MoveEqLeft[1]
0 =
    \frac{6}{\sigma_c} \biggl[
        \Bigl( \sigma_c \sigma_c' + \frac{1}{3f_c^2} \Bigr)^{\!2}
        - \frac{1}{9f_c^4}
        + \frac{\sigma_c^4}{2(1\!-\!f_c)^2}
        \biggr] (f \!-\! f_c)^2
    +
    6 \biggl\{
        \Bigl( \sigma_c \sigma'_c + \frac{1}{3f_c^2} \Bigr) \sigma_c''
\nonumber \\
&
        -
        \frac{1}{6\sigma_c^3}
        \biggl[
        4 \Bigl( \sigma_c \sigma_c' + \frac{1}{3f_c^2} \Bigr)^{\!3}
        +
        \frac{2}{f_c^2} \Bigl( \sigma_c \sigma_c' + \frac{1}{3f_c^2} \Bigr)^{\!2}
        +
        3 \Bigl( \frac{4}{3f_c^2} - \frac{3 \sigma_c^2}{1\!-\!f_c} \Bigr)
            \Bigl( \frac{4}{3f_c^2} - \frac{\sigma_c^2}{1\!-\!f_c} \Bigr)
            \Bigl( \sigma_c \sigma_c' + \frac{1}{3f_c^2} \Bigr)
\nonumber \\
&
    -
    \frac{58}{27 f_c^6}
    +
    \frac{48 \sigma_c^2}{9f_c^4(1\!-\!f_c)}
    +
    \frac{9 \sigma_c^4}{3f_c^2(1\!-\!f_c)^2}
    -
    \frac{10 \sigma_c^6}{(1\!-\!f_c)^3}
        \biggr]
        \biggr\} (f \!-\! f_c)^3
    +
    \mathcal{O}\bigl[(f\!-\!f_c)^4\bigr]
    \, .
\label{CriticalExpandedUprime}
\end{align}
No derivatives higher than~$\sigma''$ appear at cubic order, which is a direct 
consequence of expanding around the critical point. The coefficients of the higher 
derivatives vanish due to the conditions~(\ref{critical f sigma}) and~(\ref{critical beta}).

The coefficient of the quadratic term determines the first derivative at the critical 
point. Because the two solution curves intersect there (see Fig.~\ref{phase diagrams}), 
the equation determining this derivative is quadratic, accounting for both branches. 
The curve that interpolates between the boundary conditions relevant for our analysis 
has the following derivative
\begin{equation}
\sigma_c' =
    \frac{ -1 + \sqrt{ 1 - \frac{18f_c^2}{(1+3f_c)^2}} }{ 3 f_c^2 \sigma_c }
    \, .
\end{equation}
The second derivative is then obtained from the linear equation given by the 
coefficient of the cubic term in~(\ref{CriticalExpandedUprime}). 
Both derivatives depend only on~$\mu^2$, through the critical quantities defined 
in~(\ref{critical f sigma}) and~(\ref{critical beta}).

Having determined the derivatives at the critical point, we can explicitly compute 
the correction to the dimensionless flux~(\ref{beta_1}),
\begin{equation}
\beta_1 = \frac{f_c \sigma_c^3}{4\beta_c^3} 
    \bigl( f_c \sigma_c'' + \sigma_c' \bigr)
    \, .
\end{equation}
This, in turn, determines the bracketed coefficients 
in~(\ref{Inequality1}),~(\ref{LargeXiFoneEpsilonPasymp}), and~(\ref{dw2})
that govern the behavior of the corrections to the field profile, energy density, pressure, and equation-of-state parameter far from the horizon:
\begin{align}
\text{coeff}(\sigma_1/\sigma_0) 
    ={}&
    1 - 2\mu^2 - 16 \beta_0^3 \beta_1
    \, ,
\\
\text{coeff}(\epsilon_1/\epsilon_0) 
    ={}&
    3 - 5 \mu^2 - 16( 3 \!-\! \mu^2 ) \beta_0^3 \beta_1
\\
\text{coeff}(p_1/p_0) 
    ={}&
    - 1 - 48\beta_0^3 \beta_1
    \, ,
\\
\text{coeff}(w_1/w_0) 
    ={}&
    \bigl[ - 6 + 19\mu^2 - 16\mu^4 - 48 \mu^2  \beta_0^3 \beta_1 \bigr]
        \theta(-\mu^2)
    +
    \bigl[ -6 + 7\mu^2 - 48 \mu^2 \beta_0^3 \beta_1 \bigr]
        \theta(\mu^2)
    \, .
\end{align}
It is found that only the third of these coefficients is always positive, while 
the remaining three flip sign close to~$\mu^2\!=\!1$, as 
 shown in Fig.~\ref{CoeffFigure} below.
\begin{figure}[h!]
\vskip+1mm
\centering
\includegraphics[width=15cm]{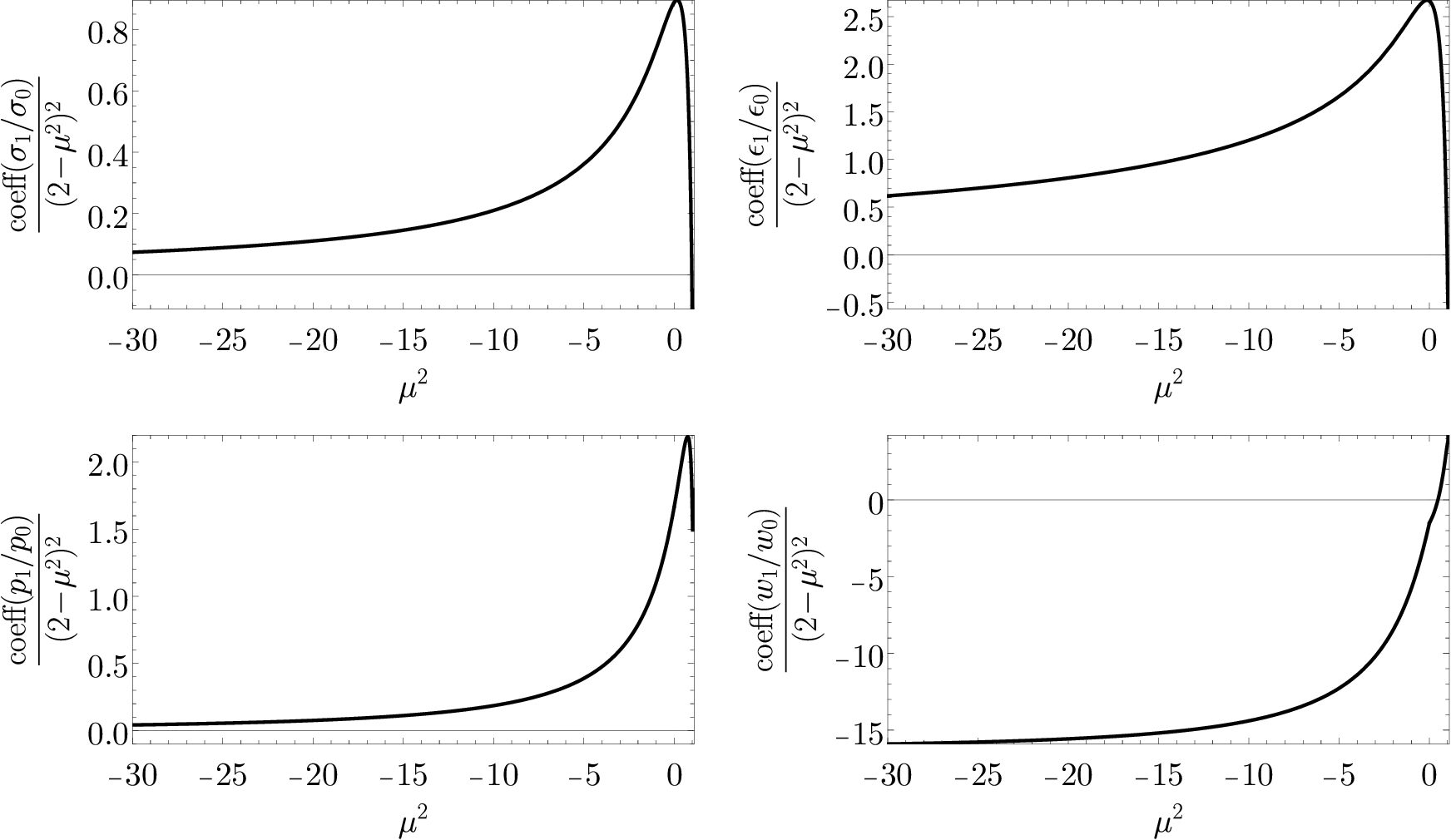}
\vskip-3mm
\caption{Coefficients appearing in~(\ref{Inequality1}),~(\ref{LargeXiFoneEpsilonPasymp}), 
and~(\ref{dw2}), that determine whether, in the large-$\xi^2$ limit and far from the 
horizon, the complex-scalar gradient corrections raise the modulus-field profile 
({\it top left}), energy density~({\it top right}), pressure~({\it bottom left}), 
and equation-of-state parameter~({\it bottom right}) above or below the
corresponding~$P(X)$ curve. The plots show that sign flips are possible in all cases 
except pressure for values~$\mu^2$ close to unity. The normalization~$(2\!-\!\mu^2)^2$ 
of the coefficients is always positive, and is chosen for convenience. }
\label{CoeffFigure}
\vskip-4mm
\end{figure}


\end{document}